\documentclass[notitlepage, twocolumn, nofootinbib,british,balancelastpage,preprintnumbers]{revtex4-1}
\usepackage{flushend}
\usepackage{graphicx}
\usepackage{grffile}
\usepackage{rotate}
\usepackage{rotating}
\usepackage{relsize}
\usepackage{slashed}
\usepackage{balance}
\usepackage{url}
\usepackage{amsmath,mathrsfs}
\usepackage{amssymb}
\usepackage{longtable}
\usepackage{multirow}
\usepackage{tabularx}
\usepackage{units}
\usepackage[usenames,dvipsnames]{xcolor}
\usepackage{subfigure}
\usepackage{xr}

\graphicspath{{images/}{/scratch/Programs/NMSSM/images}{/hdd/results/natural_NMSSM/images/resultPlots/}{/Users/jsk/tex/all_new_images/}{/afs/atlass01.physik.uni-bonn.de/user/jsk/tex/results_NMSSM/}}

\newcommand{\mueff}{\mu_\text{eff}}
\newcommand{\Beff}{(A_\lambda + \kappa s)}
\newcommand{\tanb}{\tan \beta}
\newcommand{\cotb}{\cot \beta}
\newcommand{\mHusq}{m_{H_u}^2}
\newcommand{\mHdsq}{m_{H_d}^2}
\newcommand{\mSsq}{m_S^2}
\newcommand{\mQthreesq}{m_{Q_3}^2}
\newcommand{\mUthreesq}{m_{U_3}^2}
\newcommand{\mDthreesq}{m_{D_3}^2}
\newcommand{\Ala}{A_{\lambda}}

\newcommand{\mgl}{$m_{\tilde{g}}$}
\newcommand{\mst}{$m_{\tilde{t}}$}
\newcommand{\mstone}{$m_{\tilde{t}_1}$}
\newcommand{\msttwo}{$m_{\tilde{t}_2}$}

\newcommand{\ms}{$m_{\tilde{S}}$}

\newcommand{\Mq}{M_{\tilde{q}3}}
\newcommand{\At}{A_{\tilde{q}3}}

\newcommand{\Checkmate}{\texttt{Check\textsc{mate}} }
\usepackage{soul}

\begin{document}
\preprint{IFT-UAM/CSIC-15-105}

\title{Naughty or Nice? The Role of the `N' in the Natural NMSSM for the LHC}%

\author{Jong Soo Kim}%
\email{jong.kim@csic.es}
\affiliation{Instituto de Fisica Teorica UAM/CSIC, Madrid, Spain}
\preprint{TTK-15-18}

\author{Daniel Schmeier }%
\email{daschm@th.physik.uni-bonn.de}
\affiliation{Physikalisches
  Institut and Bethe
  Center for Theoretical Physics, University of Bonn, Bonn, Germany}

\author{Jamie Tattersall }%
\email{tattersall@physik.rwth-aachen.de}
\affiliation{Institut f\"ur Theoretische Teilchenphysik und Kosmologie, RWTH Aachen, Aachen, Germany}

\begin{abstract}

In this work, we present mass limits on gluinos and stops in a {\it natural} Next-to-Minimal Supersymmetric 
Standard Model (NMSSM) with a singlino as the lightest supersymmetric particle. Motivated by naturalness, we consider 
spectra with light higgsinos, sub-TeV third generation sparticles and gluinos well below the multi-TeV regime while the 
electroweak gauginos, the sleptons and the first and second generation squarks are decoupled.  We check that our natural supersymmetry 
spectra satisfy all electroweak precision observables and flavour measurements as well as theoretical constraints. By reinterpreting
the results from the 8~TeV ATLAS supersymmetry searches we present the 95$\%$ CL exclusion limits on the model. The results
show that the presence of a singlino LSP can lengthen decay chains and soften the final state particle energies. 
Whilst this does reduce the strength of the bounds in some areas of parameter space, the LHC still displays good sensitivity to the model.

\end{abstract}

\keywords{NMSSM, Collider, Natural, Light Stops, Decoupled}

\date{\today}%
\maketitle

%
%
%
\section{Introduction}
The discovery of a Higgs Boson at the Large Hadron Collider in 2012 \cite{Aad:2012tfa,Chatrchyan:2012xdj,Aad:2015zhl}
was a triumph for experimental particle physics. Its measured 
mass of $\approx 125$ GeV fits perfectly into the framework of the Standard Model (SM), which required --- and 
hence predicted --- a scalar particle with mass of the order $\leq$ 2 TeV from unitarity constraints \cite{PhysRevD.16.1519}. Moreover, the 
observed Higgs mass falls into the narrow mass window $m_h=96^{+31}_{-24}$ GeV predicted in global fits of the SM model 
to precision electroweak observables \cite{Baak:2011ze}. However, this 
theory suffers from a well known Hierarchy Problem due to the quadratic sensitivity 
of the Higgs mass to new physical 
scales. Supersymmetry (SUSY) \cite{Wess:1974tw,Nilles:1983ge,Drees:1996ca,Martin:1997ns} is able to ameliorate this problem and to stabilise 
the Higgs mass at the electroweak scale by cancelling the quadratic divergences. 
However, the minimal incorporation into the Standard Model, called the Minimal Supersymmetric Standard Model (MSSM), leads to the prediction that 
the CP--even Higgs should be lighter than the Z--Boson at tree-level.
Under the assumption that the LHC has measured the lightest supersymmetric Higgs boson, we thus require
significant radiative corrections in order to raise the mass to the experimentally measured value. These corrections are often provided by the supersymmetric partners of the top quark, called stop squarks, since their large Yukawa couplings dictate that they provide the leading one-loop correction.

In the MSSM, the Higgs is expected to have a mass between 113 and 135 GeV \cite{Degrassi:2002fi}. However, the problem is that 
these corrections only reproduce the correct Higgs mass when both stop masses have very large masses for negligible mixing in the 
stop sector, or the trilinear $A_t$ term is very large with at least one heavy stop (e.g. \cite{Bechtle:2012jw}). This is an 
issue since a large separation between the electroweak and SUSY breaking scale introduces the {\it little hierarchy 
problem} \cite{Giusti:1998gz}. Thus, in this case the model is deemed to be `unnatural' since fine-tuned cancellations are still
required. To confront this issue in the MSSM, extended models that already predict a heavier Higgs mass
at the tree level have become popular. The simplest example of such a model is known as the Next-to-Minimal Supersymmetric 
Standard Model (NMSSM) where an extra gauge singlet chiral superfield is added to the 
spectrum \cite{Ellwanger:2009dp}. Since the singlet superfield couples both to the up- and the down-type 
Higgs superfield, the singlet scalar components contribute to the Higgs potential and can 
thus raise the tree-level Higgs mass, reducing the need for heavy stops. 

In addition, the only dimensionful 
SUSY conserving parameter, $\mu$, of the MSSM can be dynamically generated in the NMSSM by a non-vanishing vacuum expectation 
value $s$ of the extra singlet scalar. To get a phenomenologically acceptable scenario of electroweak symmetry 
breaking, $|\mu|$ should lie within $M_Z$ and $M_\text{SUSY}$, the scale where Supersymmetry is broken. In the MSSM, 
the scale of $\mu$ is in principle arbitrary and no theoretical reasoning binds it to low scales, which leads to 
the so-called $\mu$-problem \cite{Kim:1983dt}. In the NMSSM, however, the effective $\mu$ parameter is determined by the scale
of the vev $s$, which is automatically of the right order.

The large Higgs mass is not the only experimental evidence from the LHC that puts the idea of SUSY solving the
hierarchy problem under strain. The fact that no SUSY particles have yet been seen pushes the limits on SUSY gluons, called gluinos,
and SUSY quarks, called squarks, to masses $\geq1.5$~TeV (see e.g. \cite{Aad:2015iea}) that would already be deemed unnatural in constrained models like the 
CMSSM. These results have motivated a deeper study of exactly which pieces of the SUSY spectrum are required 
to be light for a theory to be considered natural \cite{Papucci:2011wy}. Firstly, since the singlet itself now generates the $\mu$ parameter 
that sets the Higgs(ino) masses, all of these particles, including the fermionic partner of the singlet (singlino), can be expected to 
have masses of the same order. Furthermore, the dominant one-loop corrections to the Higgs sector come from the stops and consequently
these cannot be too heavy. Also, since the gluino yields a sizeable correction to the stop masses at one loop, we also 
have another, looser constraint on the mass of this particle for the same reason. Finally due to the weak isospin
symmetry, the partners of the left handed bottom quarks (sbottoms), must have a mass similar to that of the 
left handed stops. 

Consequently we are drawn to a SUSY spectrum with light singlinos and higgsinos, stops and sbottoms
that may be a little more massive and a gluino that can be heavier still. Since none of the other SUSY partners are 
required by naturalness principles to be light enough to be seen at the LHC, we simply decouple these 
from our spectrum in this study.

In the context of the MSSM, naturalness is now used as a guiding principle for many LHC searches for 
gluinos, stops, sbottoms. These studies set bounds on the gluino of $m_{\tilde{g}} \geq 1150$~GeV in the case of a 
light ($m_{\tilde{\chi}^0_1}\lesssim$ 100~GeV) LSP, but this can be reduced to, $m_{\tilde{g}} \geq 500$~GeV in the limit that the gluino
becomes degenerate with the LSP \cite{Aad:2015iea,Chatrchyan:2013iqa,Khachatryan:2015vra,CMS:2014wsa,Dreiner:2012gx,Dreiner:2012sh}. For 
stops, the bounds can reach up to $m_{\tilde{t}}\geq 700$~GeV, if the dominate decay mode 
is $\tilde{t}\to t \tilde{\chi}^0_1$, and $m_{\tilde{t}}\gtrsim 600$~GeV for $\tilde{t}\to b \tilde{\chi}^+_1$.
Again, if the spectrum is compressed, the bounds weaken significantly and the limit is 
only $m_{\tilde{t}}\gtrsim 255$~GeV for $m_{\tilde t_1}-m_{\widetilde{\chi}_1^0}\approx m_b$ \cite{Aad:2014nra}. 
In addition there 
are regions of parameter space ($m_{\tilde{t}} \sim m_t + m_{\tilde{\chi}^0_1}$) where no limit can be set at 
all since the kinematics very closely resemble the SM $t\overline{t}$ background but with 
a substantially smaller production cross-section \cite{Aad:2015pfx,CMS:2014wsa,Chatrchyan:2013xna,CMS:2014yma,Rolbiecki:2015lsa}. Sbottom limits are similar
to those of stops (up to $m_{\tilde{b}}\geq650$~GeV for light $\tilde{\chi}^0_1$ and $m_{\tilde{b}}\geq250$~GeV in compressed
regions) but are more robust and do not contain holes as we move across the 
mass plane\cite{Aad:2015pfx,CMS:2014nia,CMS:2013ida,Chatrchyan:2013fea}.

Since the LHC direct production constraints still allow for relatively light gluinos and have no model
independent limit on the stop mass, the question of naturalness is driven by the Higgs mass in
the MSSM. In the NMSSM however, the reduced need for heavy stops to contribute to the Higgs means
that the direct production constraints become far more relevant. In addition, the limits can be expected
to be different since a light singlino will be present in the spectrum. However, as the singlino does 
not couple directly to the squarks the state does not normally contribute to LHC phenomenology unless
it is the lightest particle in the spectrum (LSP). The effect of a singlino LSP has now been examined 
in a number of studies and it has been claimed that it generally weakens the LHC limits since the longer
decay chains softens the $p_T$ spectra and reduces the 
$E_T^{\text{miss}}$ \cite{Ellwanger:2014hia,Ellwanger:2014hca}. Other studies have also looked at purely
Higgsino-singlino spectra \cite{Ellwanger:2013rsa,Kim:2014noa}, direct 
stop \cite{Chakraborty:2015xia,Beuria:2015mta} or gluino \cite{Cheng:2013fma} production and the 
possibility that the singlino may be light \cite{Dermisek:2005ar,Dermisek:2006wr,Dermisek:2007yt,Gunion:2008dg,Domingo:2009tb,Potter:2015wsa}.
A comprehensive list of the expected signatures of the NMSSM is given in \cite{Dreiner:2012ec} 
whilst \cite{Moortgat-Pick:2014uwa} has explored possible methods to distinguish the NMSSM 
from the more commonly discussed MSSM.

In this study we wish to explore in detail the claim that a singlino LSP generally
weakens the LHC bounds. As stated above this is expected and seen \cite{Ellwanger:2014hia,Ellwanger:2014hca}
because the longer decay chain produce soften particles for similar LSP masses. However if we examine
the particles produced in the extra NMSSM decay we see that this may not always be true. In
particular the decays that may occur are $\tilde{\chi}^0_2\to\tilde{\chi}^0_1 X^0$ where $X^0$ is either 
a $Z^0$ or Higgs and $\tilde{\chi}^{\pm}_2\to\tilde{\chi}^0_1 X^{\pm}$ where $X{\pm}$ is either 
a $W^{\pm}$ or a charged Higgs. In the case of $W^{\pm}$ or $Z^0$ production we can expect increased 
production of leptons over the MSSM that may improve the bounds but the branching ratio suppression 
makes it unlikely that this will result in a large change. However a bigger difference can be expected
when a Higgs is produced that will decay to a $b\overline{b}$ final state. If the mass splitting 
$m_{\tilde{\chi}^0_2}-m_{\tilde{\chi}^0_1} > m_h$, then this decay dominates in a large portion 
of the natural NMSSM parameter space. The reason that this final state can be so important for 
LHC phenomenology is that many SUSY searches use $b$-tags as a way to suppress the SM background (e.g. \cite{Aad:2014mfk,Aad:2014kra,ATLAS:2013cma,Khachatryan:2015wza})
and some even search for the presence of on-shell Higgs bosons e.g(\cite{Aad:2015jqa,TheATLAScollaboration:2013tha,Khachatryan:2014mma,Khachatryan:2014doa,Khachatryan:2015pwa}). Both of these strategies give the 
possibility that the natural NMSSM may be even more constrained than the MSSM.

In order to fully test the effect of the additional $b$-quarks we require many LHC searches to be simultaneously
checked. For this reason we use the \Checkmate tool \cite{Drees:2013wra} which now contains over 40 analyses
implemented via \texttt{AnalysisManager} \cite{Kim:2015wza}. In addition, we also test various theoretical and 
experimental constraints via \texttt{NMSSMTools} \cite{Ellwanger:2004xm,Ellwanger:2005dv,Ellwanger:2006rn,Das:2011dg,Muhlleitner:2003vg}, 
\texttt{HiggsSignals}\cite{Bechtle:2013xfa} and \texttt{HiggsBounds} \cite{Bechtle:2013wla}.

We begin the paper by describing the Lagrangian of the natural NMSSM in Sec.~\ref{sec:Lagrangian} and the spectrum 
that we decide to investigate along with the LHC signatures this will lead to. In Sec.~\ref{sec:methodology} we describe
exactly how the model parameters are chosen and the experimental and theoretical constraints that are applied. Here
we also introduce how we perform the LHC phenomenology in this paper. Sec.~\ref{sec:results} displays the results of 
our study, concentrating on the LHC bounds now present on the natural NMSSM. Finally in Sec.~\ref{sec:conclusion} we conclude.
\vfill
\section{A Phenomenologically Natural NMSSM}\label{sec:Lagrangian}
\subsection{Lagrangian, Masses and Parameters}
\label{subsec:params}
In the following we present the Lagrangian formulation of our model, the resulting mass matrices and the features 
that motivate our spectra. Most definitions and relations are taken from \cite{Ellwanger:2009dp} and we 
refer readers to check this source and references therein for more information.

The Next-to-Minimal Supersymmetric Standard Model (NMSSM) extends the well-known Minimal Supersymmetric Standard Model (MSSM) by an 
additional chiral superfield $\hat{S}$ which is uncharged under the Standard Model gauge groups. In this work we consider a 
simplified, natural version of a $\mathbb{Z}_3$--invariant NMSSM. Here, only terms involving exactly three fields are allowed to appear in the superpotential, for reasons explained below. Furthermore, only the scalar partners of the gluon and the third generation quarks plus the fermionic components of the three Higgs superfields $\hat{H}_u, \hat{H}_d$ and $\hat{S}$ are assumed to be phenomenologically observable among all supersymmetric particles. This setup can be described by the following superpotential:
\begin{align}
\mathcal{W} &=  h_t (\hat{Q}_3 \cdot \hat{H}_u) \hat{t}_{R}^c + h_b (\hat{Q}_3 \cdot \hat{H}_u) \hat{b}_{R}^c  \nonumber \\
&+ \lambda  (\hat{H}_u \cdot \hat{H}_d)\hat{S} + \frac{\kappa}{3} \hat{S}^3,
\end{align}
where the $\cdot$ symbol denotes the usual SU(2) invariant antisymmetric product of the respective isospin 
doublets $\hat{H}_u \equiv (\hat{H}_u^+, \hat{H}_u^0), \hat{H}_d \equiv (\hat{H}_d^0, \hat{H}_d^-)$ 
and $\hat{Q}_3 \equiv (\hat{t}_L, \hat{b}_L)$. Here, $h_t$ and $h_b$ are the dimensionful Yukawa couplings, 
while $\lambda$ and $\kappa$ correspond to dimensionless Yukawas. Note that the assumed additional $\mathbb{Z}_3$ symmetry prohibits the term $\mu (\hat{H}_u \cdot \hat{H}_d)$ usually present in the MSSM and hence provides a superpotential without any dimensionful parameters. A vacuum expectation value (vev) of the scalar singlet $\langle{}S\rangle \equiv s $ of electroweak scale order reintroduces this term after expanding the scalar field $S$ around its minimum and thus generates an effective $\mu$ term $\lambda s (H_u \cdot H_d) \equiv \mueff (H_u \cdot H_d)$ of naturally the correct scale, evading the known $\mu$--problem of the MSSM:

In addition to the terms derived from this superpotential, the following dimensionful `soft' parameters have to be added to the Lagrangian of the theory:
\begin{align}
- \mathcal{L}_\text{soft}^{\text{mass}} &= \mHusq | H_u |^2 + \mHdsq | H_d |^2 + \mSsq | S |^2 +  \frac{1}{2} M_3 \tilde{g} \tilde{g} \nonumber \\
&+ \mQthreesq | \tilde{Q}_3 |^2 + \mUthreesq | \tilde{t}_R  |^2 + \mDthreesq |  \tilde{b}_R |^2,  
\end{align}

\begin{align}
- \mathcal{L}_\text{soft}^{\text{trilinear}}&= h_t A_t (\tilde{Q}_3 \cdot H_u) \tilde{t}_{R}^* + h_b A_b (\tilde{Q}_3 \cdot H_b) \tilde{b}_{R}^* \nonumber\\
&\qquad+\lambda  \Ala  (H_u \cdot H_d)S  + \frac{\kappa}{3} A_\kappa S^3  + \text{h.c.}\ .
\end{align}
Here, $\tilde{g}$ denotes the gluino, i.e. the fermionic part of the vector superfield associated to the SU(3) gauge group. All other field names denote the scalar component of the respective chiral superfield in the superpotential. We assume all couplings to be real--valued to simplify the discussion.  

The other SUSY particles, namely the squarks of the first two generations, the sleptons as well as 
the SU(2)$\times$U(1) gauginos are assumed to be decoupled from the experimentally accessible spectrum as 
explained in Sec.~\ref{subsec:natspec}. Consequently they are not listed here.

Adding  $\mathcal{L}_\text{soft}$ to the supersymmetric F-- and D--terms yields the full scalar potential from which three minimisation conditions for the non-vanishing singlet vevs and the doublet vevs  $\langle H_{u/d} \rangle \equiv v_{u/d}$ can be derived:
\begin{align}
\mHusq + \mueff^2 + \lambda^2 v_d^2 + \frac{(g_1^2+g_2^2)}{4(v_u^2-v_d^2)} &= \mueff  B_{\text{eff}} \cotb , \\
\mHdsq + \mueff^2 + \lambda^2 v_u^2 + \frac{(g_1^2+g_2^2)}{4(v_d^2-v_u^2)} &= \mueff   B_{\text{eff}} \tanb, \\
\mSsq + \kappa s (A_\kappa  + 2 \kappa s) + \lambda^2 v^2 - \lambda \kappa &v_u v_d = \lambda \frac{v_u v_d}{s} B_{\text{eff}} 
\end{align}
with $B_{\text{eff}} \equiv \Beff, v^2 \equiv v_u^2 + v_d^2$. The first two of these can be reformulated as follows:
\begin{align}
\frac{M_Z^2}{2} &= \frac{2}{\tanb^2-1} (\mHdsq - \tanb^2 \mHusq) - \mueff^2, \label{eqn:minimisation}\\
\frac{\sin 2 \beta}{2} &= \frac{\mueff B_{\text{eff}}}{\mHusq + \mHdsq + 2 \mueff^2 + \lambda^2 v^2}.
\end{align}
 Here we have used the fact that $M_Z^2 = v^2 (g_1^2+g_2^2)/2 $ is fixed by the known mass of the $Z$ boson. The above equations allow one to choose the parameters in the set $\{\lambda, \kappa, A_\lambda, A_\kappa, \mueff, \tanb\equiv v_u/v_d \}$ to be independent, where $\mueff$, $\tan \beta$ and the known Standard Model parameter $M_Z$ replace the Lagrangian parameters $\mHusq, \mHdsq$ and $\mSsq$.
After expanding $H_u, H_d$ and $S$ around their minima, one gets the following symmetric mass matrix for their CP--even components $\{h_u, h_d, h_s\}$ at tree level:
\begin{widetext}
\begin{align}
\mathcal{M}^2_{\text{scalar}} =\begin{pmatrix}
g^2 v_d^2 + \mueff B_{\text{eff}} \tanb & (2 \lambda^2 - g^2) v_u v_d - \mueff B_{\text{eff}} & \lambda (2 \mueff v_d - (B_{\text{eff}} + \kappa s) v_u) \\
 \ldots & g^2 v_u^2 + \mueff B_{\text{eff}} \cotb & \lambda (2 \mueff v_u - (B_{\text{eff}}+ \kappa s) v_d) \\
 \ldots & \ldots  & \lambda A_\lambda v_u v_d/s + \kappa s A_\kappa + 4 \kappa^2 s^2
\end{pmatrix}, \label{eqn:cpevenmass}
\end{align}
\end{widetext}
with $g^2 \equiv M_Z/(v_u^2+v_d^2)$ given by the Standard Model gauge sector. We call the diagonalised mass eigenstates $h, H$ and $H_3$ which have increasing mass from left to right. 

Similarly, the matrix of the respective CP--odd components $\{a_u, a_d, a_s\}$ reads
\begin{widetext}
\begin{align}
\mathcal{M}^2_{\text{pseudoscalar}} = \begin{pmatrix}
\mueff B_{\text{eff}} \tanb & \mueff B_{\text{eff}} & \lambda v_u (A_\lambda - 2 \kappa s) \\
\ldots & \mueff B_{\text{eff}} \cotb & \lambda v_d(A_\lambda - 2 \kappa s) \\
\ldots & \ldots & \lambda(A_\lambda + 4 \kappa s) v_u v_d/s - 3 \kappa A_\kappa s
\end{pmatrix},
\end{align}
\end{widetext}
which yields one massless Goldstone mode and two CP--odd mass eigenstates $A_1$ and $A_2$. 

Finally, the charged components $\{h_u^+, h_d^{-*}\}$ have the mass matrix:
\begin{align}
\mathcal{M}^2_{\pm} &= m^2_\pm \cdot \begin{pmatrix} \cot \beta & 1 \\ 1 & \tanb \end{pmatrix} \\
\text{with }m_\pm^2 &\equiv \mueff \Beff + v_u v_d (g_2^2/2-\lambda^2),
\end{align}
resulting in one massless Goldstone mode and one massive charged Higgs boson $H^\pm$.

The three neutral fermionic partners of these fields, $\tilde{h}^0_u$, $\tilde{h}^0_d$ and $\tilde{s}$, mix to three neutralinos $\widetilde{\chi}_{1,2,3}^0$ after diagonalising the matrix
\begin{align}
\mathcal{M}_{\text{neutralinos}} = 
\begin{pmatrix}
0 & -\mueff & -\lambda v_u \\
\ldots & 0 & - \lambda v_d \\
\ldots & \ldots & 2 \kappa s
\end{pmatrix}.
\label{eq:higgsinos}
\end{align} 
The two charged higgsino components combine to a single Dirac chargino $\widetilde{\chi}_1^\pm$ with mass term $\frac{1}{2} \mueff \tilde{h}_u^+ \tilde{h}_d^- + \text{h.c.}$. 

In the following, we will use the collective term `higgsino' ($\tilde{h}$) for the two higgsino-like neutralinos and the chargino. Furthermore, for the sake of simplicity, we will 
use  `electroweakino' ($\widetilde{\chi}$) collectively for all three neutralinos and the chargino, even though strictly speaking $\tilde{s}$ does not have any electroweak charge.

The stop and sbottom tree level mass matrices in the bases $(\tilde{t}_R, \tilde{t}_L)$  and $(\tilde{b}_R, \tilde{b}_L)$ read
\begin{widetext}
\begin{align}
\displaystyle
\mathcal{M}^2_{\text{stops}} &= \begin{pmatrix}
\mUthreesq + h_t^2 v_u^2 - (v_u^2 - v_d^2) g_1^2/3 & h_t  v_u (A_t - \mueff \cot \beta) \\
\ldots & \mQthreesq + h_t^2 v_u^2 + (v_u^2-v_d^2) (g_1^2/12 - g_2^2/4) 
\end{pmatrix}, \\
\mathcal{M}^2_{\text{sbottoms}} &= \begin{pmatrix}
\mDthreesq + h_b^2 v_d^2 - (v_u^2 - v_d^2) g_1^2/6 & h_b v_d (A_b - \mueff \tan \beta) \\
\ldots & \mQthreesq + h_b^2 v_d^2 + (v_u^2-v_d^2) (g_1^2/12 + g_2^2/4) 
\end{pmatrix},
\end{align}
\end{widetext}
with eigenstates $\tilde{t}_{1/2}, \tilde{b}_{1/2}$.

Even though we haven't shown the full NLO corrections to these tree level masses, it can be understood that the whole model is 
fixed by Standard Model parameters plus the set $\{\lambda, \kappa, A_\lambda, A_\kappa, \mueff, \tanb, \mQthreesq, \mUthreesq, \mDthreesq, A_t, A_b, M_3 \}$
\subsection{Natural Spectrum}
\label{subsec:natspec}
\allowdisplaybreaks
Naturalness comes into play in the context of Eq.~(\ref{eqn:minimisation}). For a model to be natural all of the individual terms should be of order $M_Z^2$ and no fine-tuned cancellations
should be present. 
In contrast to the $\mu$ parameter in the MSSM, which is a free parameter of the superpotential without 
any \emph{a priori} relation to the electroweak scale, the $\mueff$ parameter in the NMSSM is itself 
induced by electroweak symmetry breaking and the vacuum expectation value of $S$.  Thus, it is naturally of 
right order and determines the expected mass scale of the higgsinos which are mainly determined by $\mueff$, 
see Eq.~(\ref{eq:higgsinos}) and below.
In the limit of vanishing mixing, the tree level singlino mass 
reads $ \kappa s = \mueff\ (\kappa / \lambda) \lesssim \mueff$, where we have used that the stability 
of the $s \neq 0$ vacuum usually requires $\kappa / \lambda < 1$ \cite{Ellwanger:1996gw}. We therefore 
expect a singlino that is lighter than the higgsinos in a natural setup.

 This tree level relation 
is affected by loop corrections to the respective parameters. As an example, the large Yukawa coupling to the 
stops and their $\mathcal{O}(\alpha_s)$ correction from gluino loops induces a sizeable effect on $\mHusq$ in Eq.~(\ref{eqn:minimisation}) while running  from the SUSY breaking scale $\Lambda_S$ down to the TeV scale. In the leading log approximation \cite{Papucci:2011wy}, these corrections read
\begin{align}
\left. \Delta \mHusq \right|_{\tilde{t}}  &\approx  -\frac{3  y_t^2}{8 \pi^2} (\mQthreesq + \mUthreesq + |A_t|^2)\ \text{ln} \Big(\frac{\Lambda_S}{\text{TeV}}\Big), \\
\left. \Delta \mHusq \right|_{\tilde{g}}  &\approx  -\frac{2 y_t^2}{\pi^3}  \alpha_s |M_3|^2\  \text{ln}  \Big(\frac{\Lambda_S}{\text{TeV}}\Big).
\end{align}
Naturalness requires these corrections to be moderately small which translates into mass bounds \mbox{\mst{} $\lesssim$ \mgl{} $\lesssim \mathcal{O}(1 \text{ TeV})$}. Note that the naturalness bound on $\mQthreesq$ also sets the scale of the $\tilde{b}_L$-like scalars, as they lie in the same SU$(2)_L$ doublet as the $\tilde{t}_L$ field. Though no equivalent naturalness constraint applies to the $\tilde{b}_R$ scalar, we assume that there is no \emph{a priori} reason why the SUSY breaking mechanism should induce large splittings $\mUthreesq - \mDthreesq$ or $A_t - A_b$ and thus we assume the soft breaking parameters to be degenerate (see Sec.~\ref{sec:methodology}). The appearance of a second light sbottom, however, does not affect the collider results significantly.

Parameters related to the SU$(2)\times$U$(1)$ gauginos and the squarks of the first two generations have negligible 
effect on the parameters in Eq.~(\ref{eqn:minimisation}) and thus are not constrained by naturalness arguments. They 
can therefore safely be set to experimentally inaccessible scales while keeping the electroweak breaking scale small.

Note that the above consideration of naturalness is only performed on the qualitative level and solely serves 
as a motivation for the hierarchies and mass scales of our following collider study.  More quantitative analyses 
in terms of so--called \emph{fine tuning} are possible but require a more specific formulation of the 
decoupled supersymmetric sector to get a valid dependence of low-scale observables on independent high-scale 
parameters (see e.g. \cite{Baer:2013gva}). 

While trying to keep the parameters natural, our model should still not violate experimental observation, i.e.\ a Standard 
Model like scalar boson with mass of order 125 GeV should emerge. In the limit where the lightest CP-odd Higgs boson
decouples (\mbox{$M_A \equiv 2 \mueff (A_\lambda + \kappa s) / \sin 2 \beta \rightarrow \infty$}), the lightest SM-like eigenvalue of Eq.~(\ref{eqn:cpevenmass}) including leading order tree--level correction\footnote{We have set $\kappa = 0$ in Eq.~(\ref{eqn:higgsmass}) due to the relation $\kappa < \lambda$ and the expansion in orders of $\lambda$.} in $\lambda < 1$ and dominant radiative corrections of tops and stops reads
\begin{align}
  m_{h, \text{SM}}^2 &\approx M_Z^2 \cos^2 2 \beta + \frac{\lambda^2}{g^2} \frac{\sin^2 2 \beta}{\cos^2 2 \beta(1+\tan^2 \beta)} \frac{M_A^4}{\mueff^2} \nonumber \\
&+ \frac{3 m_t^4}{4 \pi^2 v^2} \Big( \text{ln} \Big( \frac{\Mq}{m_t^2} \Big) + \frac{A_t^2}{\Mq^2} \Big(1-\frac{A_t^2}{12 \Mq^2} \Big) \Big) \label{eqn:higgsmass}
\end{align}
assuming degenerate soft breaking stop masses $\Mq^2 \equiv  \mQthreesq = \mUthreesq \gg m_t^2$. While the MSSM contribution shows 
an upper limit $M_Z^2$ in the decoupling limit and thus requires a significant contribution from heavy stops, the NMSSM 
contribution of $\mathcal{O}(\lambda^2)$ can lead to a sizeable enhancement of the Higgs boson mass itself. Consequently 
no heavy stops are needed, making it easier to acquire a natural spectrum as explained above.

\begin{figure}
\includegraphics[width=0.4\textwidth]{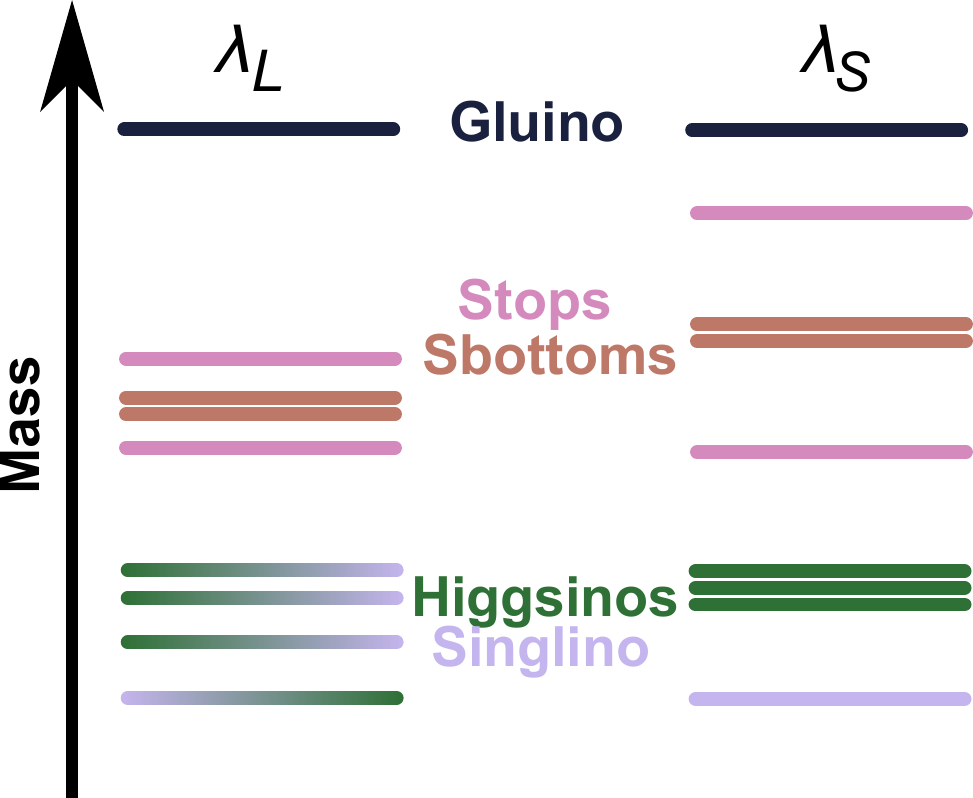}
\caption{Schematical setups of the considered benchmark models $\lambda_S$ and $\lambda_M$, their hierarchies 
and the respective expected mass splittings.}
\label{fig:spectrum}
\end{figure}

The benchmark spectra we are going to consider in the upcoming analysis are sketched in Fig.~\ref{fig:spectrum}. We 
distinguish two main limits of the NMSSN, steered by the size of the dimensionless coupling parameter $\lambda$:
\begin{description}
\item[large $\lambda \equiv \lambda_L$]
When the coupling $\lambda$ is large, Eq.~(\ref{eqn:higgsmass}) suggests that we can reach a large enough Higgs 
mass if $\sin 2 \beta$ is large. In our analysis we choose $(\lambda = 0.7, \tan \beta = 2)$, with the value 
for $\lambda$ chosen at the maximum possible value which does not run into Landau poles at higher scales. In this 
setup, no large radiative corrections are required and as such it is expected that one can keep both stops 
(and the respective sbottoms) rather light while still being able to reach the correct Higgs mass. As a consequence
all third generation scalars may be kinematically accessible at the LHC. The neutralinos can mix largely in this 
scenario and direct decays of coloured scalars into singlets and singlinos are possible \cite{Kraml:2005nx}.
\item[small $\lambda \equiv \lambda_S$ ]
In case of a very small $\lambda$, the Higgs mass is very MSSM like. To maximise 
the tree level value one needs a larger $\tan \beta$, which is why we define this point 
via $(\lambda = 0.01, \tan \beta = 15$). Large radiative corrections are needed, which asks for at least 
one heavy stop. The sparticles of the MSSM sector decouple from the singlet states and experimentally, the 
only difference between the MSSM and the NMSSM would be sparticle decays into the singlino LSP. For very 
small $\lambda$,  the scalar vev $s$ must be large in order to have a sufficiently large $\mu$ term. This 
generally translates into $s\gg v_u,v_d$.  Contrarily to the previous case, this scenario will come along with a rather 
split sector of third generation squarks, mostly degenerate higgsinos and a mostly decoupled singlino and 
a singlet scalar sector.
\end{description}
In both scenarios, to avoid having an LSP--like 
chargino we always require the singlino to be lighter than the higgsinos. Furthermore, we always require the 
gluino to be heavier than the stops to avoid the consideration of loop-induced 2--body or off-shell 3--body 
decays. Whilst the relative hierarchy between stops/sbottoms and the higgsinos is not fixed by our setup, we will
nevertheless find that it is often as depicted in Fig.~\ref{fig:spectrum}.
\subsection{Signatures of Interest}
\label{subsec:signatures}
The spectrum described in the previous section leads to interesting signatures for the LHC: due to the 
light $\tilde{t}$ and $\tilde{b}$ scalars we expect final states with many $t$ and $b$ quarks. The hadronised jets 
originating from $b$--quarks have a high probability of being correctly tagged as so--called \emph{b-jets} and 
many analyses from both ATLAS and CMS have been designed to specifically tag final states with these objects, see Sec.~\ref{sec:collpheno}. In 
the following we only focus on final states with these objects and neglect other signatures:

In this work, we consider third generation squark and gluino hadro-production via the strong interaction. In general, 
the gluon fusion diagrams will be the dominant production channel for not too heavy gluino and third generation 
scalar masses and the cross section is only determined by the respective mass and spin of the respective final state 
sparticle at leading order,
\begin{align}
    p p &\rightarrow \tilde{g} \tilde{g},\quad \tilde{t}_i \tilde{t}^*_i,\quad \tilde{b}_i \tilde{b}^*_i \qquad i \in \{1, 2\}\nonumber. \\    
\intertext{Here, we have omitted the production of electroweakino pairs since the cross section is negligible compared 
to the production of coloured sparticles unless the higgsino and the singlino are the only kinematically accessible 
sparticles at the LHC \cite{Ellwanger:2013rsa,Kim:2014noa} --- a scenario we are not going to assume. In addition, 
we have not considered compressed spectra where a hard initial state radiation jet has to be taken into account.
\endgraf
Let us now turn to the discussion of the decay modes. The decay chains can be very complicated in 
natural SUSY and typically depends on the details of the mass spectrum and the mixing angles.  Since we define
that the gluino is always the heaviest sparticle in our setup, the following strong two body decay 
modes are the dominant gluino decay channels}
\tilde{g} &\rightarrow \tilde{t}_i t_i,\quad \tilde{b}_i b_i \qquad i \in \{1, 2\}.\nonumber
\end{align}
There is no tree level coupling between the squarks and the singlino in the NMSSM and thus
the decays to neutralino states with a significant singlino component are suppressed. As the $\lambda_S$
scenario contains an almost pure singlino LSP, direct decays of the squarks to the LSP very rarely occur. Even
in the $\lambda_L$ scenario, which can contain an LSP that is a higgsino-singlino mix, the large singlino
component significantly suppresses the direct decay to this state. This situation is different to the 
natural MSSM where stop decays into the LSP are common and consequently we expect longer and more complicated decay chains
in the natural NMSSM. A by-product of such longer decay chains is that the individual particles produced are
necessarily softer and $E_T^{\text{miss}}$ can be expected to be reduced. 

However in common with the natural MSSM, the sparticles will decay in final states with third generation SM 
particles which will give rise to a high $b$ jet multiplicity,
\begin{align}
\tilde{t}_i &\rightarrow \widetilde{\chi}^0_{2/3} t,\quad \widetilde{\chi}^{+}_1 b, \nonumber  \\
\tilde{b}_i &\rightarrow \widetilde{\chi}^0_{2/3} b,\quad \widetilde{\chi}^{-}_1 t. \nonumber  \\
\intertext{In addition, the large expected squark mass splittings in the $\lambda_S$ scenario can lead 
to the following squark-to-squark decays 
with additional gauge bosons and Higgs scalars}
\tilde{t}_2 &\rightarrow \tilde{t}_1 X^0,\quad  \tilde{b}_i X^{+},   \nonumber  \\
\tilde{b}_2 &\rightarrow \tilde{b}_1 X^0,\quad  \tilde{t}_1 X^{-},  \nonumber  \\
\intertext{with $X^0 \in \{Z^0, h, H, H_3, A_1, A_2\}$ and  $X^{\pm} \in \{H^{\pm}, W^{\pm}\}$. As the 
production rate of the  heavy squark in such a case is largely suppressed compared 
to $\tilde{t}_1/\tilde{b_1}$, this decay is however not expected to contribute significantly to the observed event rates.
\endgraf
The biggest 
difference between the natural MSSM and the natural NMSSM is that we can now have 
a singlino LSP. This leads to additional decays of the (now NLSP) higgsinos $\chi^0_{2/3}$
such as,}
\widetilde\chi_{2/3}^0 &\rightarrow\widetilde{\chi}_1^0 X^0 \nonumber \\
\widetilde\chi_1^{\pm} &\rightarrow\widetilde{\chi}_1^0 X^{\pm}. \nonumber
\end{align}

Generally, a light singlino is accompanied by relatively light singlet scalars. Depending on the mass 
difference between the NLSP and the LSP and the mass of the decay products $X$, differences 
between the MSSM and the NMSSM will arise, which may modify the decay patterns of the higgsino in a MSSM scenario.

Of course, for each of the above listed decays there exists a mode with all involved particles 
charge conjugated. Obviously the listed decay
modes are only possible subject to kinematic constraints and all decay modes 
mentioned above can have a related three (or four) body decay mode 
if one (or more) of the final state particles are virtual. 

We have not listed the tediously large list of possible decays for the neutral scalars $\{h, H, H_3, A_1, A_2\}$: They generally 
involve Standard Model like Higgs decays, decays of heavy into light scalars and decays of heavy scalars into pairs of lighter 
squarks or electroweakinos. However, in most cases the heavy scalars $H, H_3$ and $A_{2}$ do not appear in the observed decay chains and thus their 
decay modes are of no relevance in the following. It is mostly the Standard Model like Higgs and the 
singlet like scalars which are of 
importance and their decays are practically Standard Model like after having applied the experimental constraints as explained in 
upcoming Sec.~\ref{sec:scan}.  

%
\section{Model Test Methodology}
\label{sec:methodology}

As described at the end of Sec.~\ref{subsec:params}, our model of interest can be described by 12 free 
parameters. To simplify the discussion, we assume a degeneracy\footnote{To be more precise, the 
degeneracy is assumed to hold at the scale $Q_{\text{SUSY}} = 5$ TeV with the exact choice being 
of minor relevance for the numerical results. Note that this is also the scale to which we put the 
decoupled SUSY particles.} of the soft parameters in the third generation, i.e. 
\begin{align}
\At &\equiv A_t = A_b, \\
\Mq^2 &\equiv  \mQthreesq = \mUthreesq = \mDthreesq.
\end{align}
This assumption always fixes the mass of the bottom squarks for given stop masses in a way 
as depicted in Fig.~\ref{fig:spectrum}. In the following we explain how we fix the free 
parameters of our model
\begin{align}
\lambda, \kappa, A_\lambda, A_\kappa, \mueff, \tanb, \At, \Mq, M_3 \label{eqn:parameters}
\end{align}
with respect to the hierarchies of the models we want to consider. We follow with a
discussion on how we test the respective parameter combination.
\subsection{Scan Setup and Definitions}
\label{sec:scan}
Each of the data points that we are going to analyse is solely defined by the following set of information:
\begin{enumerate}
\item The NMSSM scenario $\lambda_S$ or $\lambda_L$,
\item the mass \mgl{} of the gluino,
\item the mass \mstone{} of the lightest stop,
\item the higgsino mass parameter $\mueff$ and
\item the singlino mass parameter \ms{} $ \equiv 2 \kappa s$
\end{enumerate}
This fixes the following parameters in Eq.~(\ref{eqn:parameters}):
\begin{alignat}{2}
\lambda &= 0.7\  (0.01) &&\quad \text{for $\lambda_L$\ ($\lambda_S$)}, \nonumber \\
\tanb   &= 2\ (15) &&\quad \text{for $\lambda_L$\ ($\lambda_S$) },  \\
\kappa  &= \nicefrac{\lambda}{2} \cdot \nicefrac{\text{\ms}}{\mueff} &&\quad \text{due to $\mueff = \lambda s$}. \nonumber
\end{alignat}
The remaining five parameters are found as follows: We require a natural, realistic particle content, that is we aim for 
a spectrum with as light as possible stops while having a Higgs boson at the correct mass.
In addition we demand that the Higgs boson passes the most relevant 
theoretical and phenomenological constraints. Such a spectrum is found by using the public
tool \texttt{NMSSMTools} \cite{Ellwanger:2004xm,Ellwanger:2005dv,Ellwanger:2006rn,Das:2011dg,Muhlleitner:2003vg}. This allows us to specify the above mentioned parameters at scale $Q_t$ to get the corresponding physical particle masses,
mixing matrices, branching ratios and test against a variety of observational tests (see below).

In order to find a parameter combination with a viable, natural spectrum, we perform the following chain of actions:
\begin{table*}[t]
\begin{tabular}{l@{\hskip 0.3cm}l@{\hskip 0.4cm}l@{\hskip 0.4cm}l}
\hline
\hline
 Ref. & \Checkmate{} identifier & Sensitive to which decay scenario(s) \\
\hline
 \cite{ATLAS:2013cma} & \texttt{atlas\_conf\_2013\_024} & stop/sbottom decay chains leading to purely hadronic final states\\
 \cite{TheATLAScollaboration:2013tha} & \texttt{atlas\_conf\_2013\_061} & $\tilde{g}\tilde{g}\rightarrow t\bar t \tilde{t} \tilde{t}^*, b \bar b \tilde{b} \tilde{b}^*$ and/or decays involving $h \rightarrow b\bar b$. \\
 \cite{TheATLAScollaboration:2013uha} & \texttt{atlas\_conf\_2013\_062} & stop/sbottom decay chains with 1 isolated lepton from $W/Z$\\
 \cite{Aad:2013ija} & \texttt{atlas\_1308\_2631} & $\tilde{t} \rightarrow b \tilde{\chi}^\pm, t \tilde{\chi}^0$ with a purely hadronic final state \\
 \cite{Aad:2014qaa} & \texttt{atlas\_1403\_4853} & $\tilde{t} \rightarrow b \tilde{\chi}^\pm, t \tilde{\chi}^0$ with an OS isolated lepton pair in the final state\\
 \cite{Aad:2014pda} & \texttt{atlas\_1404\_2500} & $\tilde{g}\tilde{g}$ with decays into stop/sbottom producing 2 SS or 3 isolated leptons\\
 \cite{Aad:2014kra} & \texttt{atlas\_1407\_0583} & stop/sbottom decay chains with 1 isolated lepton from $W/Z$ \\
\hline
\hline
\end{tabular}
\caption{Summary of the expected most sensitive analyses within \Checkmate{} to the considered natural model, listed in 
alphabetical order. All analyses require a significant amount of missing transverse momentum in the final state and have at 
least one signal region which requires b-tagged jets. All other ATLAS analyses implemented 
in \Checkmate{} are tested in parallel, but are always found to be less sensitive than those listed.}
\label{tbl:analyses}
\end{table*}
\begin{description}
\item[Loop over the heavy stop mass \msttwo{}]
We are interested in stops that are as light as possible, i.e. we aim 
to find the lightest spectrum that passes the most important phenomenological constraints. For that purpose, 
with \mstone{} set above, we perform a loop over \msttwo: Starting from \mstone+25 GeV and using a step-size 
of 5 GeV, we steadily increase the heavy stop mass and try to find a valid parameter point according to the 
steps described next. As soon as a valid point is found, that one is taken for the further collider study.
\item[Fix the strong sector $M_3$, $\At$, $\Mq$]
The masses of the stops and the gluino are mostly determined by  these three parameters. Given the target 
values \mstone, \mgl{} and the looped value for \msttwo, we use \texttt{NMSSMTools}\footnote{\texttt{NMSSMTools} 
has been modified to allow scanning over $\At$ and $\Mq$ which the public version does not allow} to scan 
over $M_3$, $\At$ and $\Mq$ and find the combination that reproduces the desired masses\footnote{The mass 
calculation performed by \texttt{NMSSMTools} first uses 2-loop RGEs to run the parameters from $Q_{\text{SUSY}}$ 
down to $Q_{\tilde{t}} = \Mq$ and then evaluates the pole mass at $Q_{\tilde{t}}$ using next-to-leading 
order corrections in $\mathcal{O}(\alpha_s)$} best. For this scan, the values of $A_\lambda$ and $A_\kappa$ 
are barely of relevance as they have only a minor impact on the third generation stop masses. 
Consequently they are therefore fixed to the central values of the ``scalar sector scan'' described below. 
Note that at this stage we use \texttt{NMSSMTools} solely to find the correct mapping 
of physical masses to parameters. No phenomenological constraints are applied at this stage.
\item[Explore the scalar sector $A_\lambda, A_\kappa$]
Having the strong sector fixed we start a new grid scan over the scalar trilinear parameters $A_\lambda$, $A_\kappa$ in order to find a phenomenologically allowed scalar sector. We test $A_\lambda$ uniformly in the range 0 to \mbox{$2 ( \mueff / \sin 2 \beta-  \text{\ms})$}, which is chosen such that the central value minimises the higgsino-singlino mixing in Eq.~(\ref{eqn:cpevenmass}) and hence maximises the SM-like Higgs boson mass \cite{Ellwanger:2006rm}. $A_\kappa$ is uniformly scanned in the range $[-550$ GeV, 450 GeV$]$.

For each point, \texttt{NMSSMTools} tests
\begin{itemize}
\item the absence of tachyonic masses and charge or colour breaking minima in the scalar potential, 
\item that there is a SM-like Higgs boson in the mass window 121 to 129 GeV\footnote{The window for $m_h$ is motivated by theory uncertainties and the fact that the decoupled sector, most importantly the electroweakinos, can influence the Higgs mass by higher order corrections if they are of order $\mathcal{O}($few TeV$)$, see e.g. \cite{Staub:2015aea}. The exact details of the heavy electroweakino sector would not affect our collider analysis at all and thus are incorporated by a looser constraining on the light Higgs boson mass.},
\item consistency with all other implemented collider constraints (mostly LEP limits on the Higgs sector, neutralinos and charginos)
\item consistency with all other implemented low energy observables. (e.g. $b \rightarrow s \gamma, B_s \rightarrow \mu^+ \mu^-, \ldots$) apart from
$(g-2)_\mu$ where our natural model will reproduce the SM expectation.
\end{itemize}
To consider more recent collider results from LEP, Tevatron and the LHC that constrain the scalar sector, we further 
use \texttt{HiggsBounds 4.1.2}\cite{Bechtle:2013wla} and \texttt{HiggsSignals 1.2.0}\cite{Bechtle:2013xfa} 
to perform final tests on the scalar sector of the considered parameter points. For that purpose we fix the mass uncertainty for all Higgs bosons to be 4 GeV. \texttt{HiggsBounds} is used with the \verb@LandH@ setup. A parameter combination is discarded if \texttt{HiggsBounds} returns ``excluded''. In \texttt{HiggsSignals}, the \verb@both@ setting is used that performs both a mass centred and a peak centred method using \verb@latestresults@. A point is discarded if it produces a p-value smaller than 0.05. 
\item[Exit \msttwo{} scan]
If at the end of this stage no allowed $A_\lambda, A_\kappa$ combination is left, the \msttwo{} loop starts with the next iteration. If however a parameter combination of $M_3$, $\At$, $\Mq$, $A_\lambda$ and $A_\kappa$ passes all the aforementioned constraints, this parameter point is used for collider phenomenology part described next. 
\end{description}

For completeness it should be noted that the 5 parameters mentioned at the beginning of this section are closely related 
to the physical electroweakino masses. Firstly due to the decoupled wino, the mass of the chargino, $m_{\widetilde{\chi}_1^\pm}$, is 
practically identical to the input parameter $\mueff$ and we will therefore use both variables synonymously in the following. As 
depicted in Fig.~\ref{fig:spectrum}, $\mueff$ (or $m_{\widetilde{\chi}_1^\pm}$) is also very close to the mass of the two neutral 
higgsinos within $\lambda_S$. Likewise, the singlino mass parameter $m_{\tilde{S}}$ sets the mass of the lightest 
neutralino, $m_{\widetilde{\chi}_1^0}$. Within $\lambda_L$ however, 
large mixing in the neutralino sector will lead to deviations from these identities. In the following, instead of the 
input variable $m_{\tilde{S}}$ we will only show the physical mass of the lightest neutralino, $m_{\widetilde{\chi}_1^0}$, which 
by construction is predominantly singlino like.

\subsection{Collider Phenomenology}
\label{sec:collpheno}

As explained in Sec.~\ref{subsec:signatures}, we assume that pair production of the light $\tilde{g}, \tilde{t}_i$ and $\tilde{b}_i$ dominates 
the expected signal. Production cross sections for these particles are calculated 
using \texttt{NLLFast 2.1}\cite{Beenakker:1996ch,Beenakker:1997ut,Kulesza:2008jb,Kulesza:2009kq,Beenakker:2009ha,Beenakker:2010nq,Beenakker:2011fu} 
using \texttt{CTEQ6.6NLO} PDF \cite{Nadolsky:2008zw}. Uncertainties due to scale variations, parton density 
functions and $\alpha_s$ are provided and we take the quadratic sum of these to set the total theory 
error $\Delta \sigma$.
For each production mode, 50 000 signal events are generated 
using \texttt{Herwig++ 2.7.0} \cite{Bahr:2008pv,Bellm:2013lba}  with the \verb@NMSSM@ model 
setting. 
For practical reasons, decay tables of all relevant particles are calculated within Herwig++, which 
contains all tree level 2-- and 3--body decays and effective implementations of the loop-induced decays $h_i \rightarrow \gamma \gamma, g g$.

To test the model against a variety of LHC results, we use \Checkmate\cite{Drees:2013wra,Kim:2015wza}: This tool applies an ATLAS 
tuned version of the \texttt{Delphes 3} \cite{deFavereau:2013fsa} detector simulation which uses 
\texttt{FastJet} with the anti-$k_T$ jet algorithm \cite{Cacciari:2005hq,Cacciari:2008gp,Cacciari:2011ma}. Reconstructed events 
are tested against various ATLAS analyses and the derived number of signal events is tested against observation and the 
Standard Model expectation. The compatibility of signal and observation is tested by comparing the predicted 
signal $S \pm \Delta S$ to the model independent 95\% CL limit $S95$, determined by using the CL$_\text{S}$ 
method \cite{0954-3899-28-10-313}. Here, $\Delta S$ considers both the MC error on our statistics as 
well as the theory error on the total cross sections.
\Checkmate considers a large list of ATLAS analyses, however due to the signatures described in Sec.~\ref{subsec:signatures} it is 
expected that only a subset of these will be sensitive to the characteristics of our model. We list these 
analyses in Tbl.~\ref{tbl:analyses}. They all require a significant amount of missing transverse momentum due to the 
expected undetected LSP in the final state and have signal regions that check for b--jets. They mainly differ by 
the final state jet multiplicities and the total amount and relative charge of final state isolated leptons (i.e. electrons and muons). 
The analyses also differ in the kinematics of the respective signal regions that are designed and tuned for 
particular final states. As we expect different final state signatures in our model, it is highly favourable to 
check all these possibilities in parallel and filter out the most sensitive one for each case. Fortunately, \Checkmate{} 
allows for an easy comparison of that kind.

\section{Results}
\label{sec:results}

In the following we show exclusion lines in the parameter space of the model explained above. Since 
we still have $m_{\tilde{g}}, m_{\tilde{t}_1}, m_{\widetilde{\chi}_1^\pm}$ and $m_{\widetilde{\chi}_1^0}$ as 
continuous degrees of freedom, we choose to present results for specific chosen benchmark scenarios. 

As one of our considered decay chains in Sec.~\ref{subsec:signatures} starts with the production 
of gluinos and ends with the decay into the singlino LSP, we first choose to show exclusion lines in 
the plane spanned by the masses of these two particles. We do so for various 
choices of $m_{\tilde{t}_1}, m_{\widetilde{\chi}_1^\pm}$ and always compare the results 
for $\lambda_L$ and $\lambda_S$. As it will turn out, light gluinos mostly lead to severely 
constrained models. Thus we will follow with a scenario in which the gluino 
is decoupled from the spectrum as well. We then show exclusion lines 
in the $m_{\tilde{t}_1}$--$m_{\widetilde{\chi}_1^0}$--plane for different 
chargino masses, again putting the results for $\lambda_L$ and $\lambda_S$ side by side. For 
the specific case of a light LSP, we also present results in the 
 $m_{\tilde{t}_1}$--$m_{\widetilde{\chi}_1^\pm}$--plane to illustrate the dependence on the 
 chargino mass for both $\lambda$ scenarios.

To keep the discussion compact, we only show  $95\%$ exclusion lines in different parameter 
planes within this section. An exhaustive list of plots showing distributions of masses, 
cross sections and branching ratios can be found in the appendix.

\subsection{Gluino--LSP--Plane}
\label{sec:results:glu_limits}

\begin{figure*}
\includegraphics[width=0.4\textwidth]{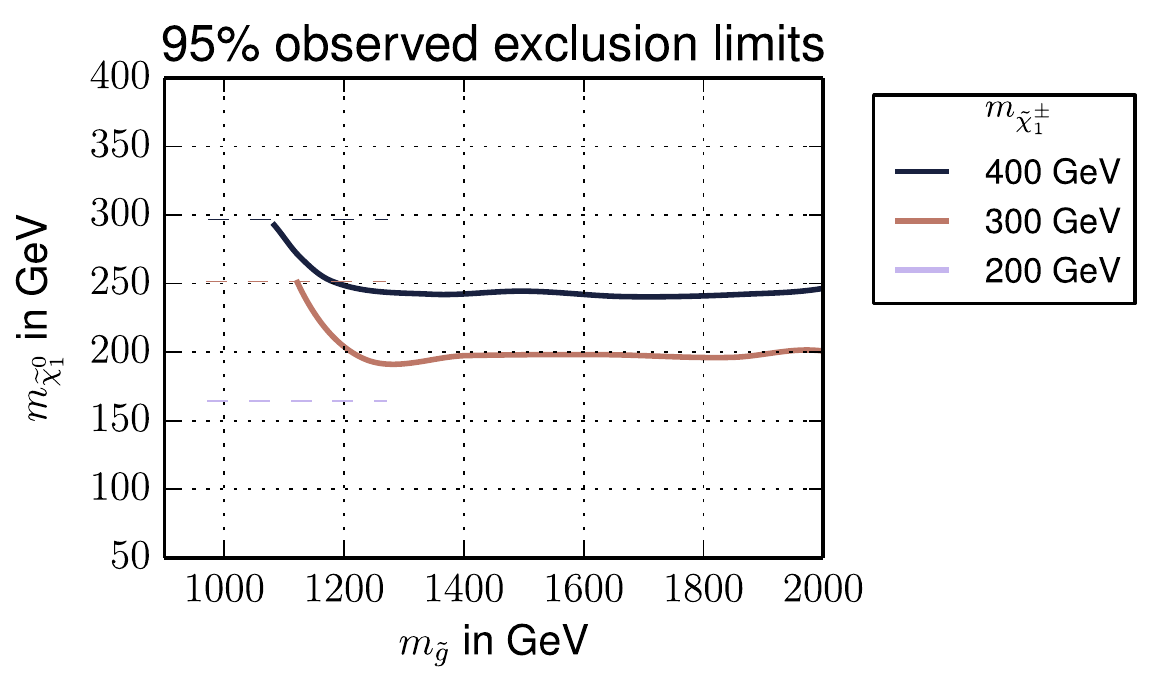}\qquad
\includegraphics[width=0.4\textwidth]{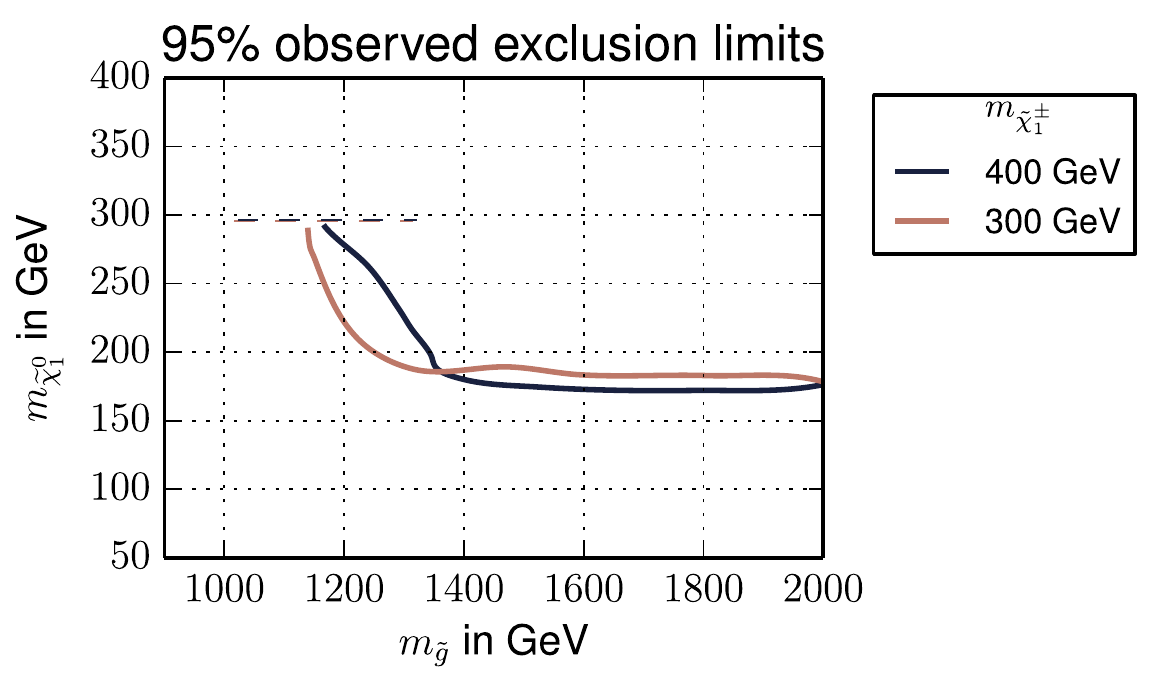}
\vspace{-0.5cm}
\caption{Observed 95\% C.L. exclusion limits for $m_{\tilde{t}_1} = 400$ GeV. Left: $\lambda_L$. Right: $\lambda_S$}
\label{fig:exclusion_gluino_singlino_mstop400}
\includegraphics[width=0.4\textwidth]{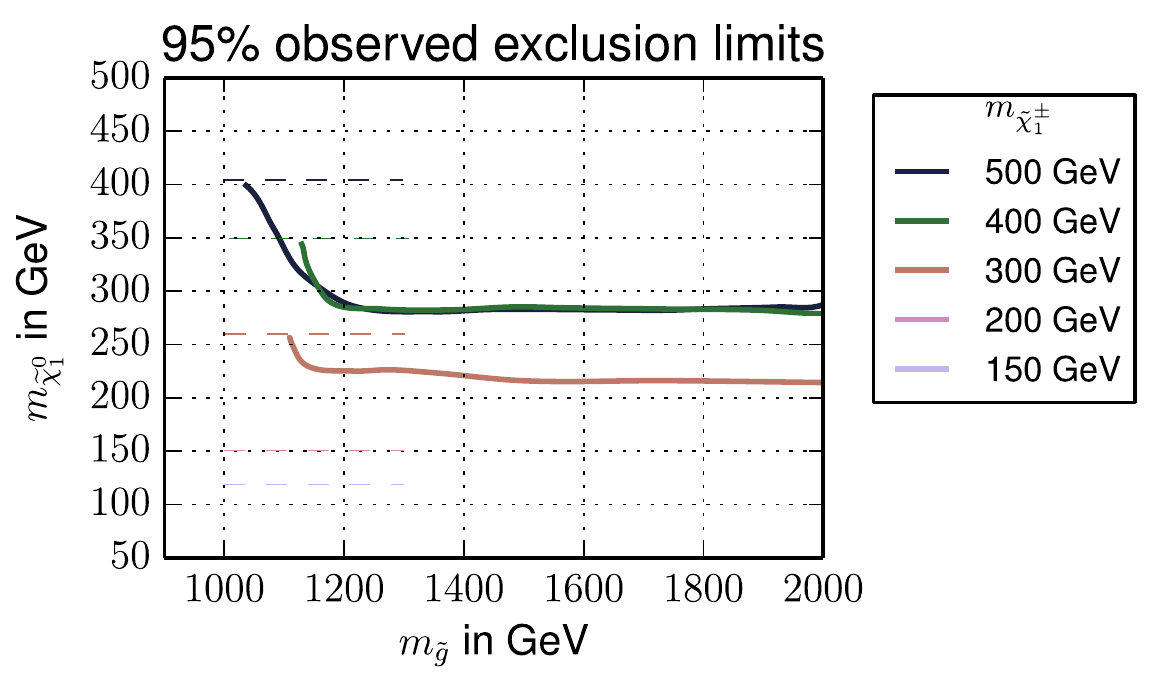}\qquad
\includegraphics[width=0.4\textwidth]{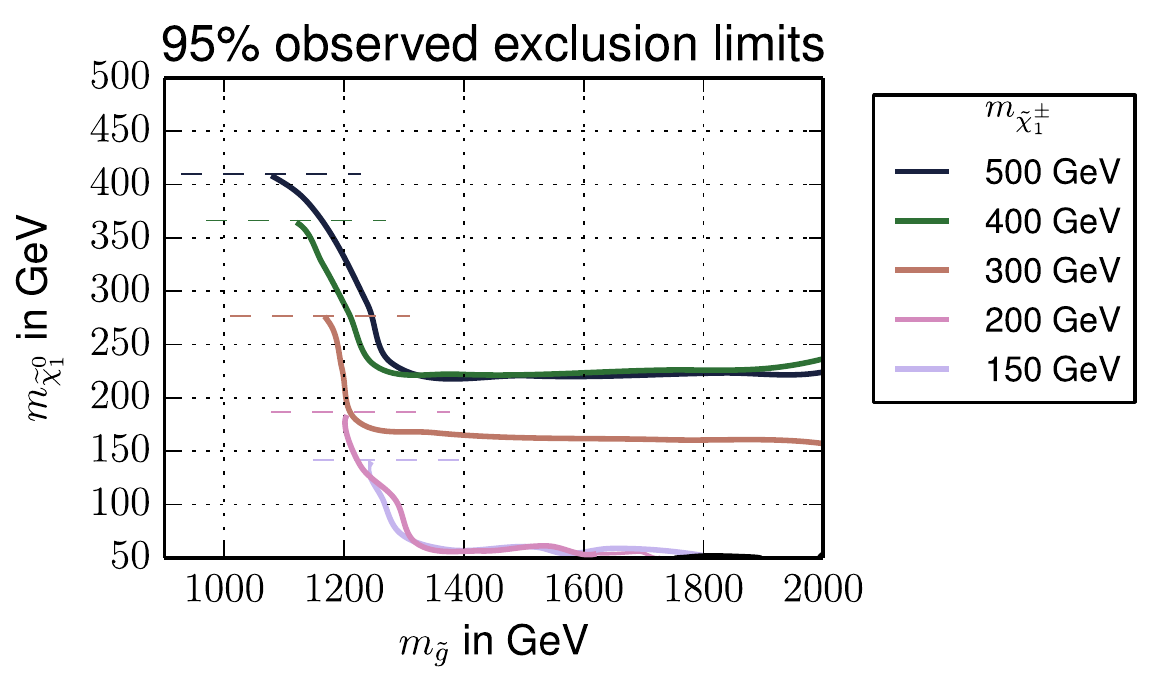}
\vspace{-0.5cm}
\caption{Observed 95\% C.L. exclusion limits for $m_{\tilde{t}_1} = 500$ GeV. Left: $\lambda_L$. Right: $\lambda_S$}
\label{fig:exclusion_gluino_singlino_mstop500}
\includegraphics[width=0.4\textwidth]{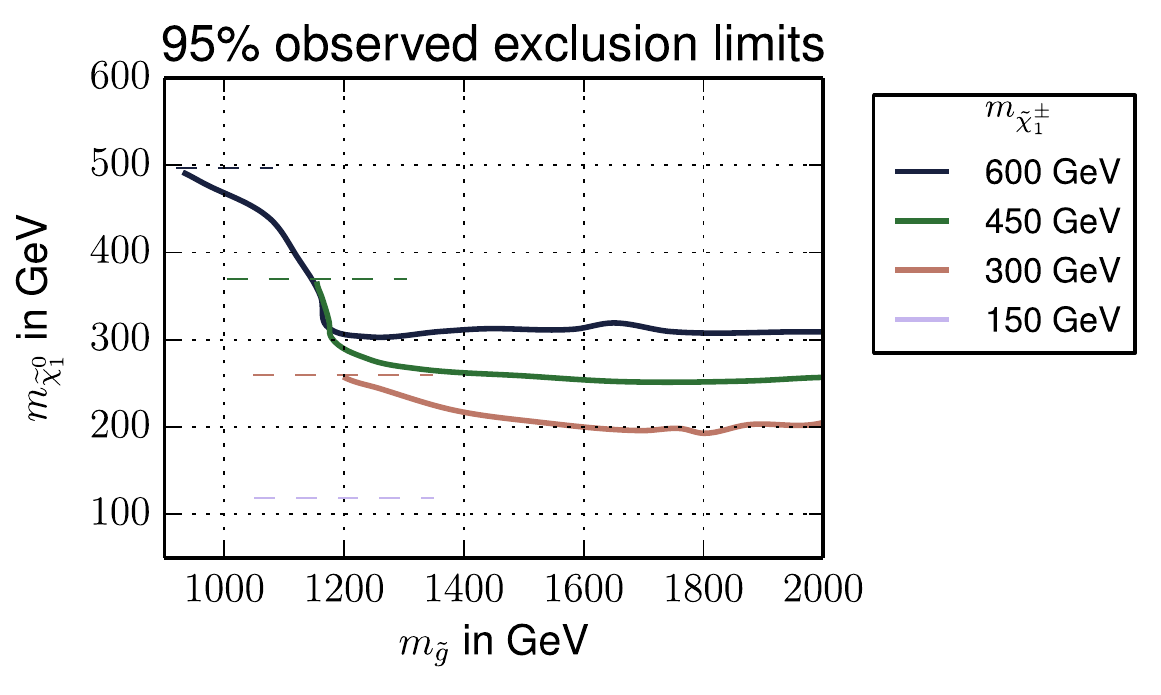}\qquad
\includegraphics[width=0.4\textwidth]{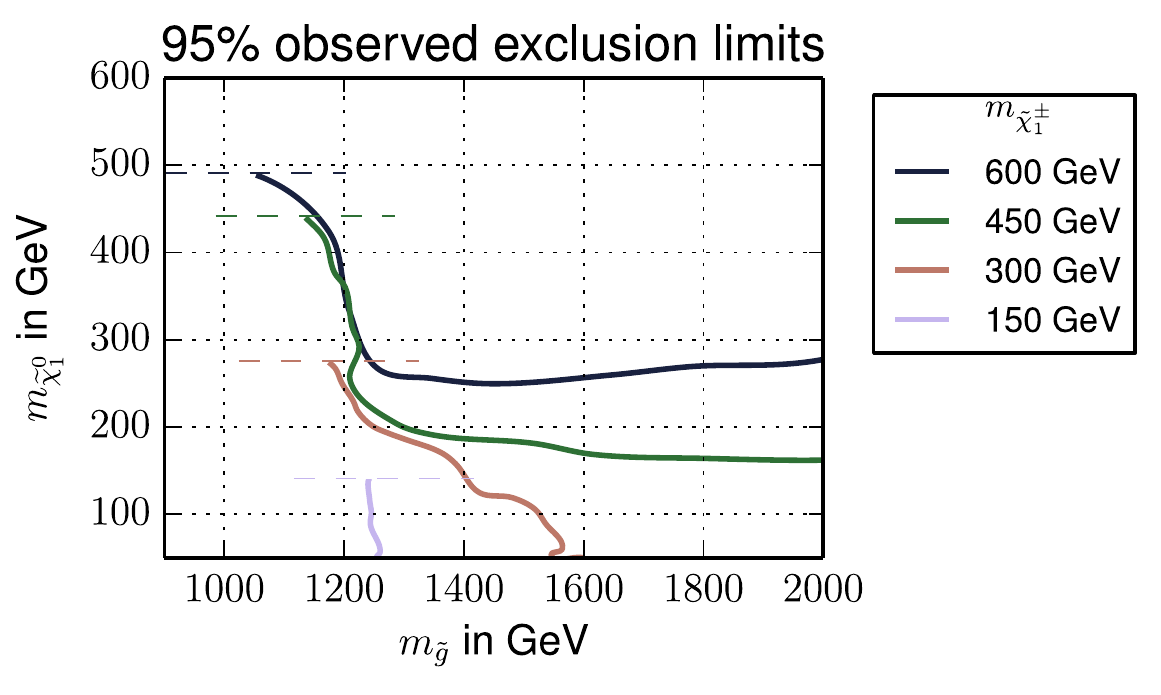}
\vspace{-0.5cm}
\caption{Observed 95\% C.L. exclusion limits for $m_{\tilde{t}_1} = 600$ GeV. Left: $\lambda_L$. Right: $\lambda_S$}
\label{fig:exclusion_gluino_singlino_mstop600}
\includegraphics[width=0.4\textwidth]{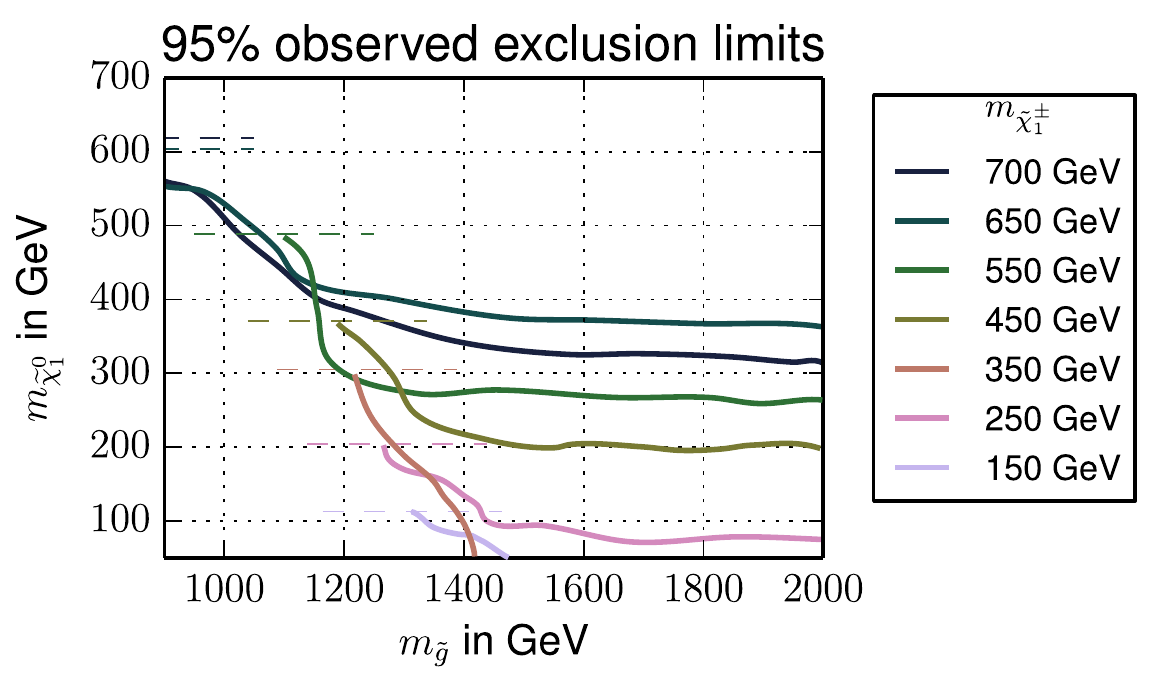}\qquad
\includegraphics[width=0.4\textwidth]{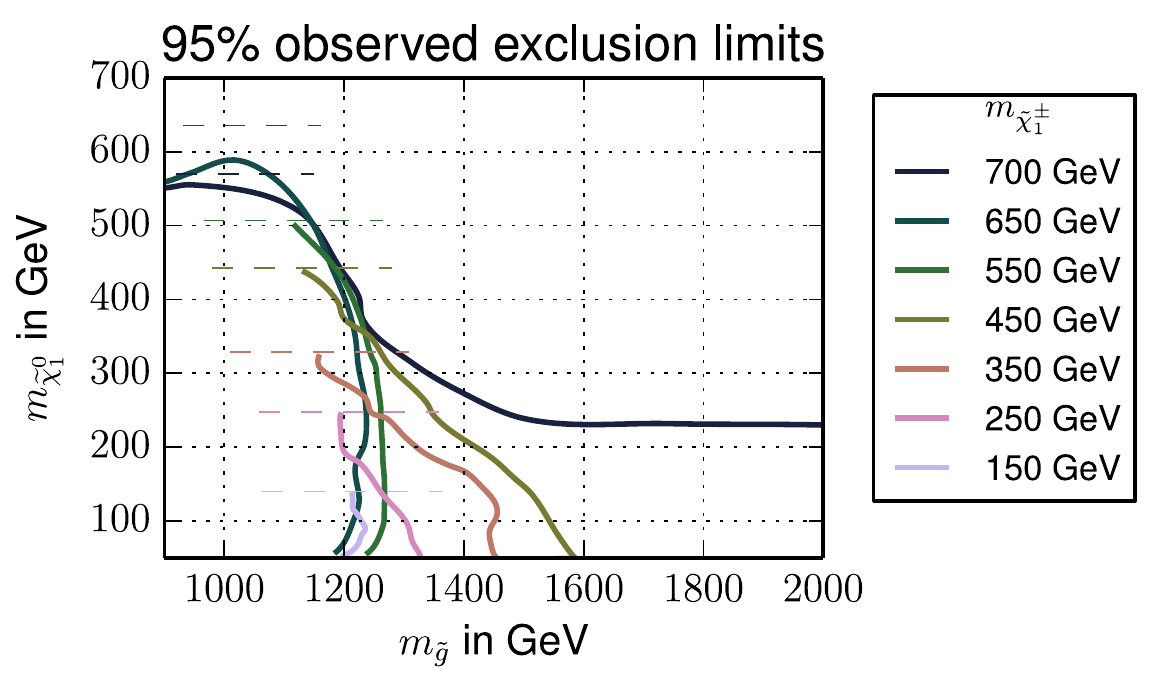}
\vspace{-0.5cm}
\caption{Observed 95\% C.L. exclusion limits for $m_{\tilde{t}_1} = 700$ GeV. Left: $\lambda_L$. Right: $\lambda_S$}
\label{fig:exclusion_gluino_singlino_mstop700}
\includegraphics[width=0.4\textwidth]{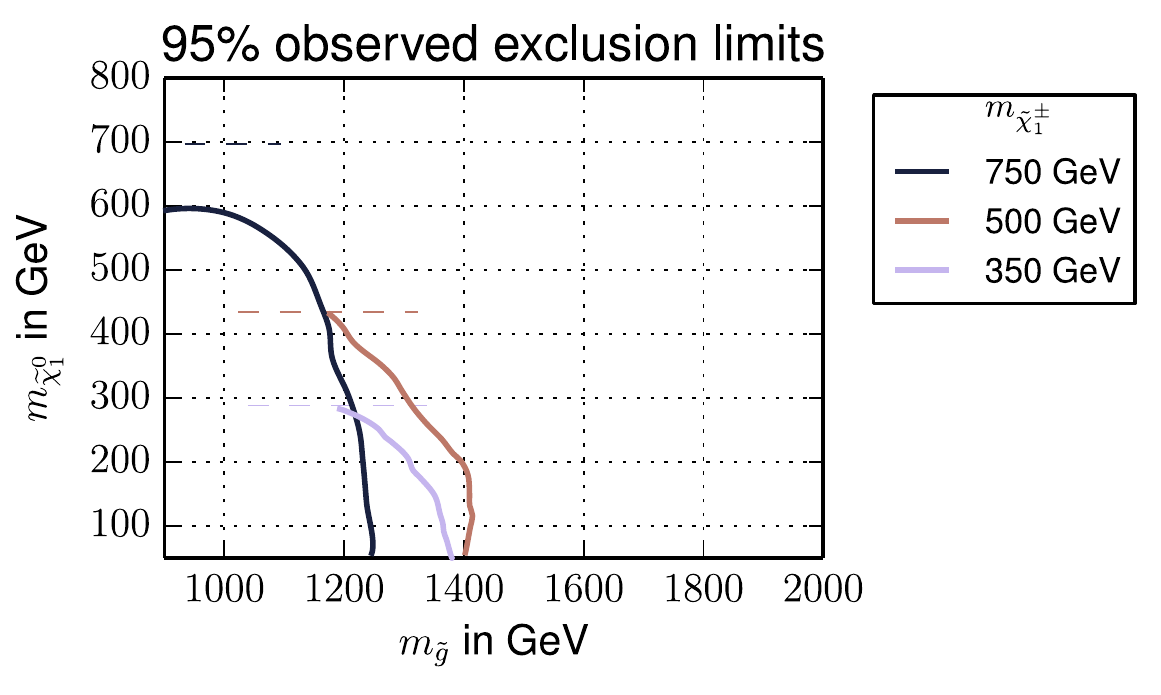}\qquad
\includegraphics[width=0.4\textwidth]{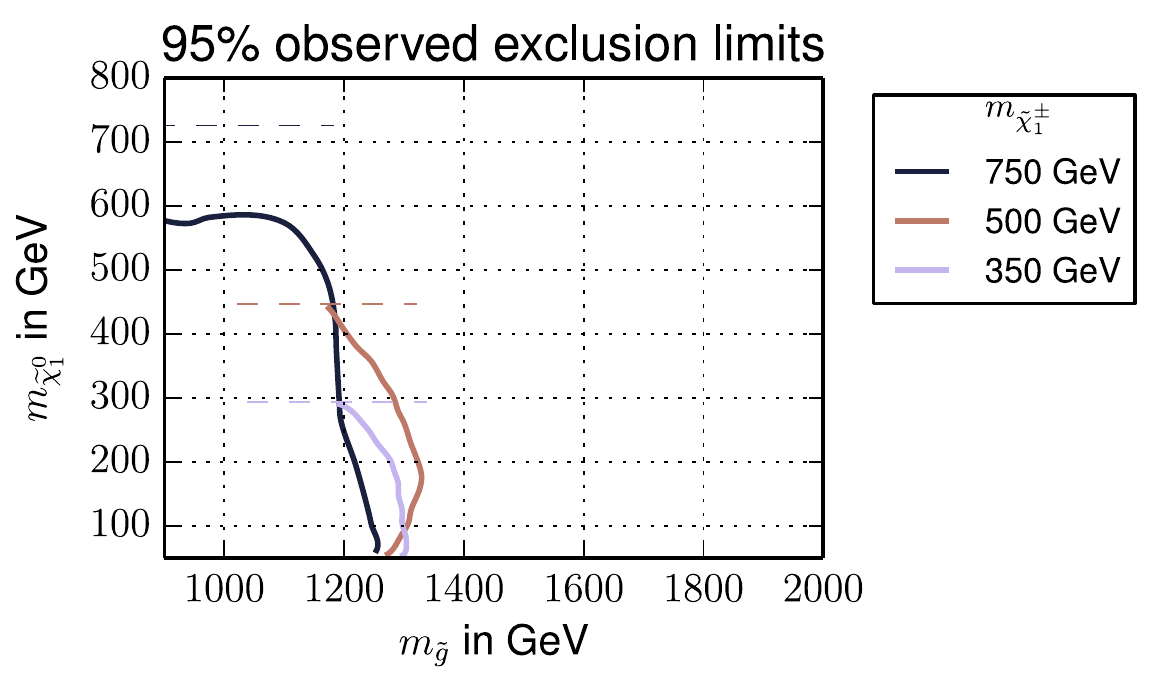}
\vspace{-0.5cm}
\caption{Observed 95\% C.L. exclusion limits for $m_{\tilde{t}_1} = 800$ GeV. Left: $\lambda_L$. Right: $\lambda_S$}
\label{fig:exclusion_gluino_singlino_mstop800}
\end{figure*}

In Figs.~\ref{fig:exclusion_gluino_singlino_mstop400}-\ref{fig:exclusion_gluino_singlino_mstop800} we show the 95$\%$ exclusion 
region in the gluino-LSP mass plane, using fixed stop masses in the range $m_{\tilde t_1}=400$ to $800$ GeV. For 
each case, the $\lambda_L$ and $\lambda_S$ scenarios are compared in the
left and right panel, respectively. Within each panel we compare the exclusion regions for different 
chargino mass values that obey $m_{\widetilde{\chi}_1^\pm} < m_{\tilde t_1}$. Since the chargino must not be lighter 
than the LSP, each exclusion line has an individual upper limit on the $m_{\chi^0_1}$ axis, drawn by dashed 
horizontal lines\footnote{For given $m_{\widetilde{\chi}_1^\pm} \approx \mueff$, this theoretical upper limit 
should appear for $m_{\widetilde{\chi}_1^\pm} = m_{\widetilde{\chi}_1^0}$. However, since $\mueff$ also sets the scale of the neutral 
higgsinos in our setup, mixing in the neutralino sector does not allow for points which fulfil the equality. Therefore 
the dashed horizontal lines appear slightly below the $m_{\widetilde{\chi}_1^\pm} = m_{\widetilde{\chi}_1^0}$ line, namely at the 
heaviest singlino-like $\widetilde{\chi}_1^0$ that can be achieved for given $\mueff$.}. Chargino mass values that are 
listed in the legend but do not appear in the plot should be interpreted as being entirely excluded
across the whole mass plane.

Generally, the exclusion lines split the parameter space into two regions of interest and we discuss these regions separately:

\subsubsection{Light Gluinos}
\begin{figure*}
\includegraphics[height=0.2\textheight]{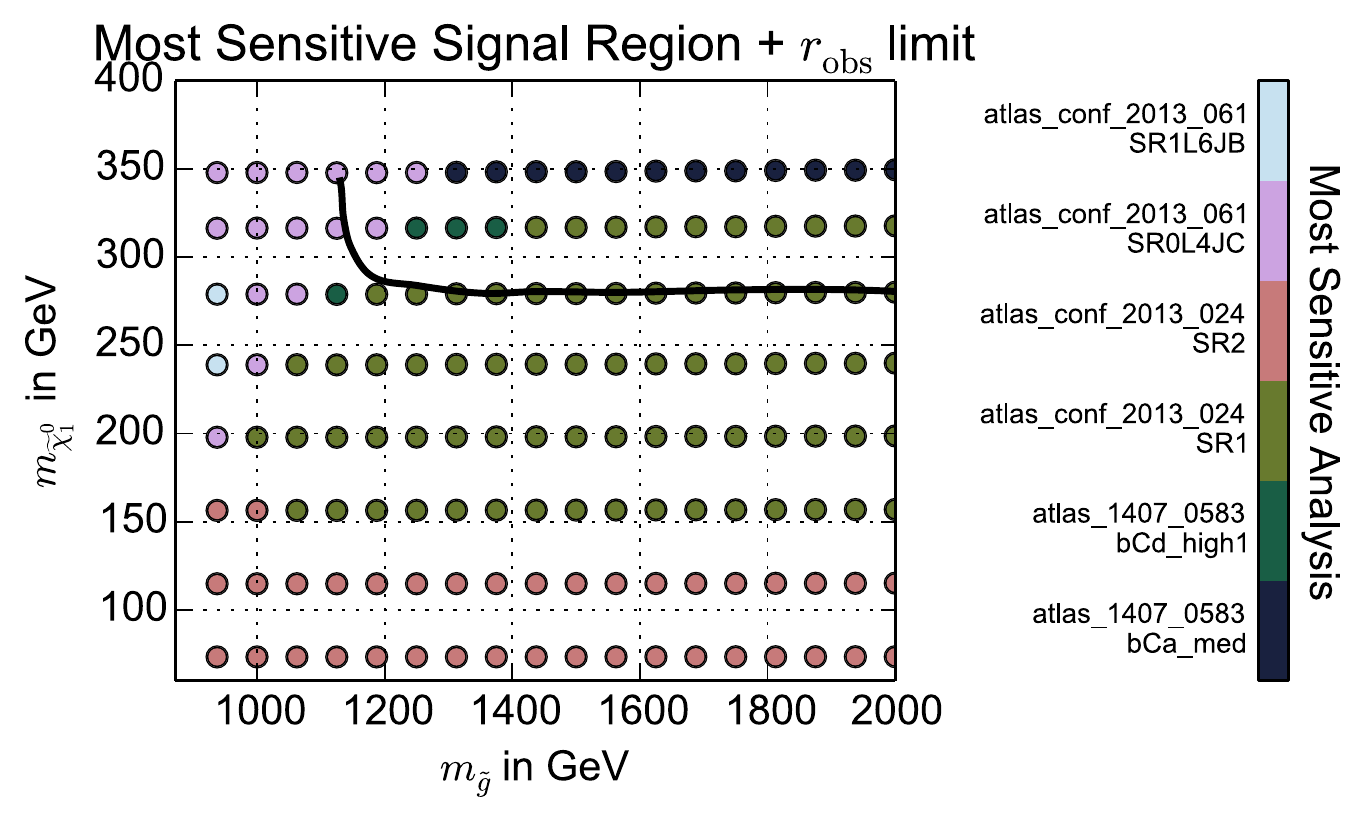}\qquad \qquad \qquad
\includegraphics[height=0.2\textheight]{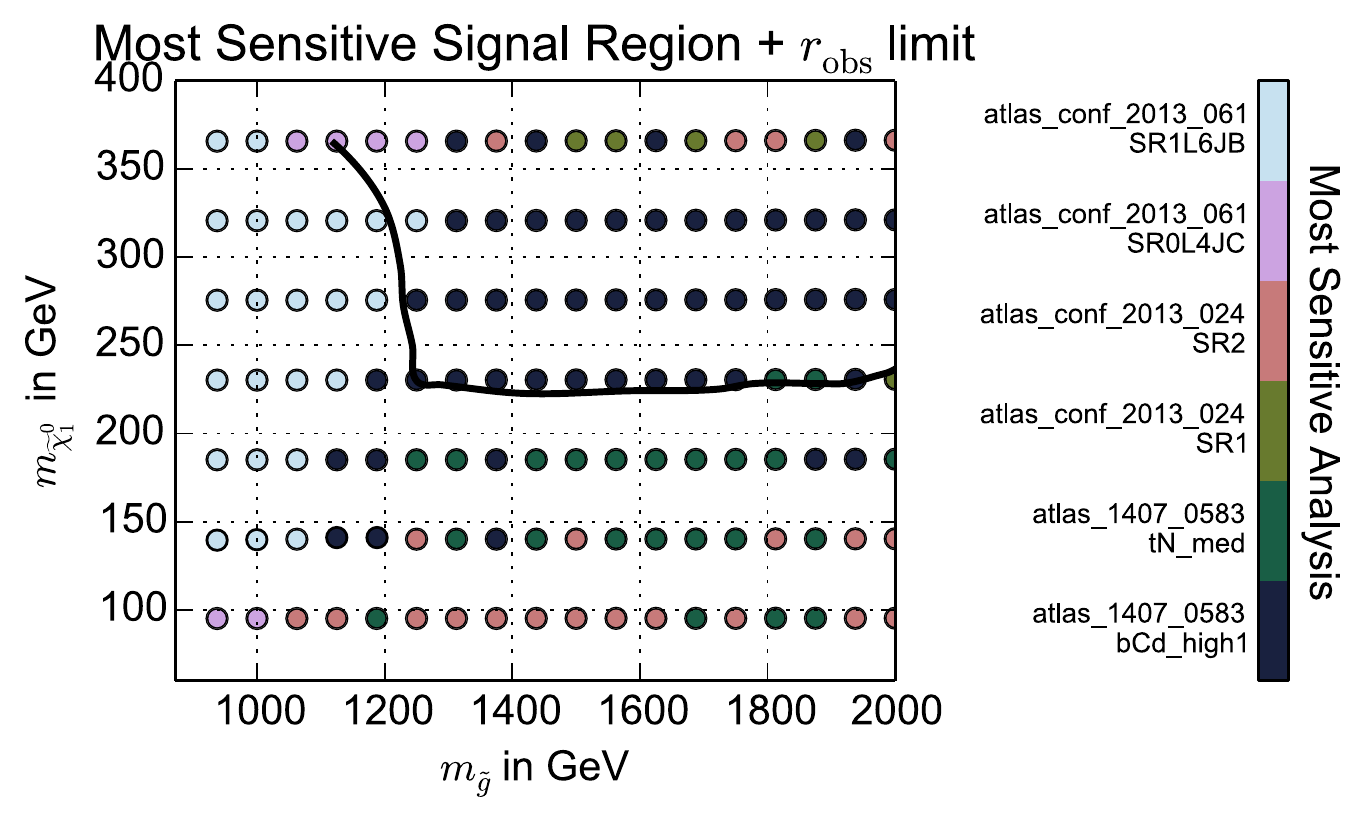}
\includegraphics[height=0.2\textheight]{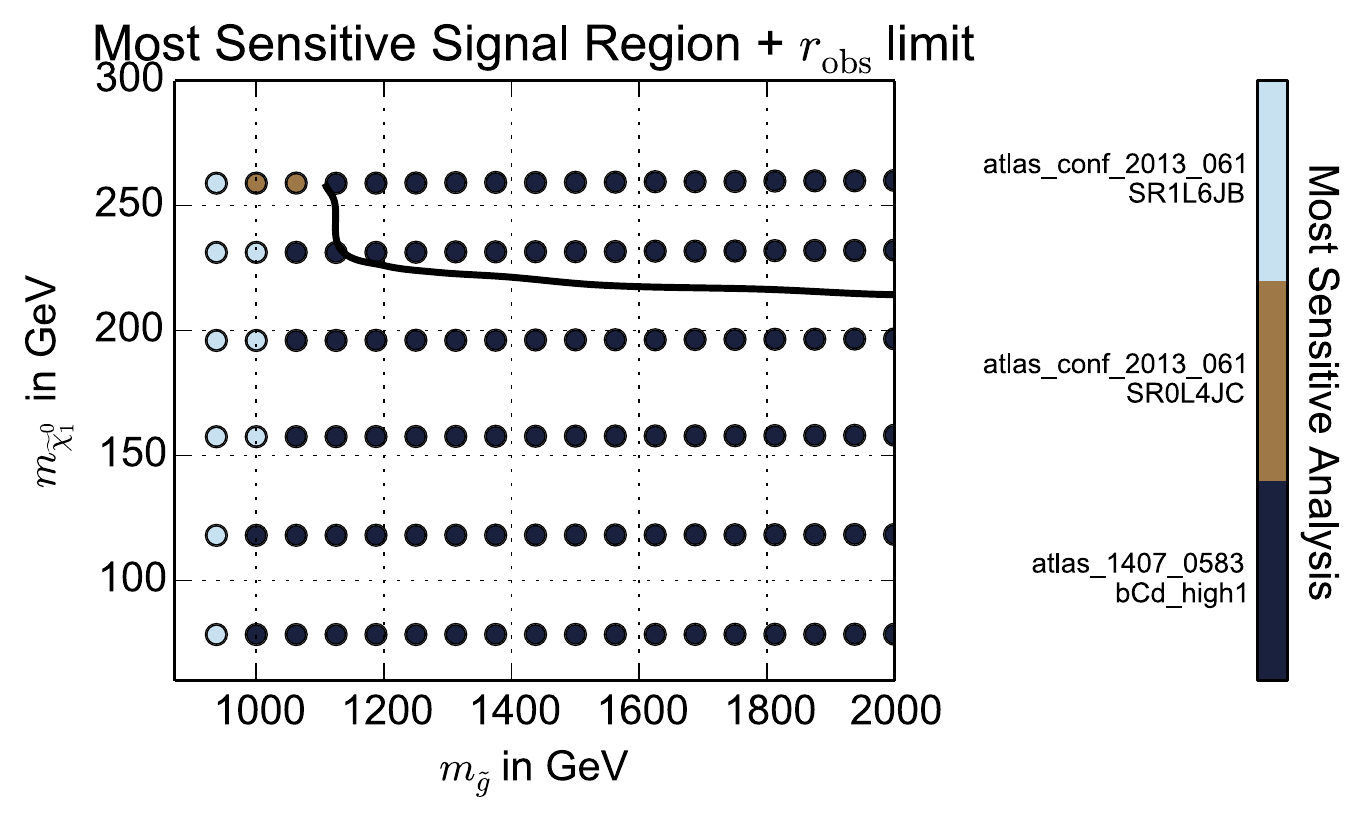}\qquad \qquad \qquad
\includegraphics[height=0.2\textheight]{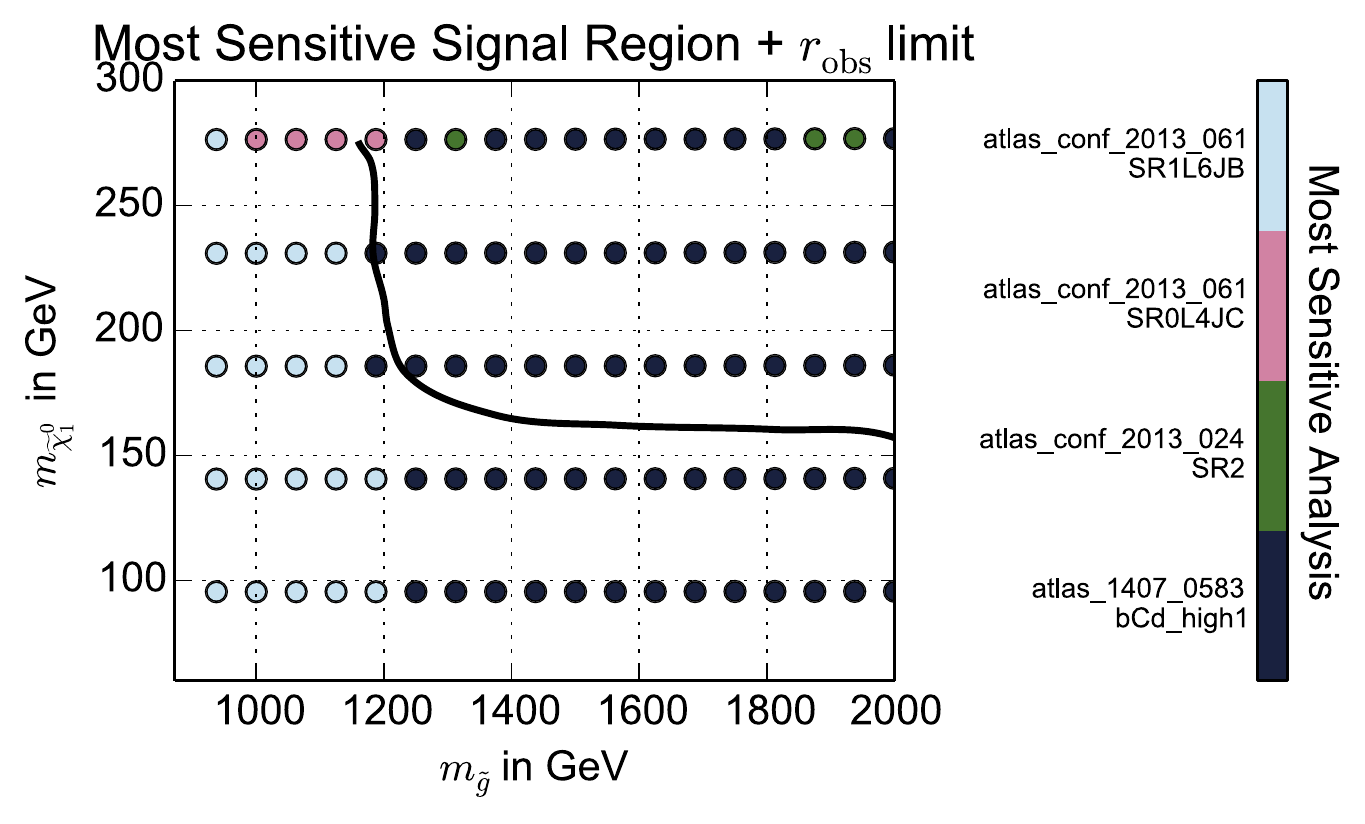}
\caption{Most sensitive signal region for each individual point in the gluino-LSP scan, 
using $m_{\tilde t_1} = 500$ GeV, $m_{\widetilde{\chi}_1^\pm} = 400$ GeV (top) and $300$ GeV (bottom). Left: $\lambda_L$. Right: $\lambda_S$.}
\label{fig:gluinobestsr}
\end{figure*}
For $m_{\tilde g} \lesssim 1100 $ GeV, 
Fig.~\ref{fig:exclusion_gluino_singlino_mstop400}--\ref{fig:exclusion_gluino_singlino_mstop800} show that the limits
are mostly independent of the chargino mass and apparently primarily driven by the gluino decay products in the 
decays $\tilde g\rightarrow \tilde b b, \tilde t t$. 
 
As the bounds 
in that region do not seem to 
vary significantly as we change the mass of the electroweakinos (and only barely if we change the mass of the lightest 
stop), we conclude that the 
details of the decay chain of the third generation scalars into the LSP is almost irrelevant when setting limits
on the model. The only exception is if very small mass splittings occur in the decay chain, for example between the gluino
and the stop or the stop and the higgsinos. We can see the effect in the
left parts of Fig.~\ref{fig:exclusion_gluino_singlino_mstop800} and also can be observed for all 
scenarios with $m_{\widetilde{\chi}_1^\pm} \lesssim m_{\tilde{t}_1}$ in 
Figs.~\ref{fig:exclusion_gluino_singlino_mstop600}-\ref{fig:exclusion_gluino_singlino_mstop800}.

When we compare the $\lambda_S$ and $\lambda_L$ limits we also see that the limits are stable between the two scenarios once 
gluino production is dominant. Consequently, we again conclude that the precise decay modes of 
the $\tilde {t} (\tilde{b})$ and the various $\widetilde{\chi}_1^0, \widetilde{\chi}_1^{\pm}$ do not effect the LHC phenomenology in this region
of parameter space.

The above conclusions may be different to the thoughts we had before commencing this study. In fact we may have guessed that the 
additional decay step present due to the singlino would have made the model
more difficult to see at the LHC. The reason is that the extra decay can reduce the individual final state particle energies and also the 
total missing energy (e.g. \cite{Fan:2012jf}). We believe the reason that this does not occur here is the number of
studies and therefore signal regions contained within the \Checkmate{} program. For example, in Fig.~\ref{fig:gluinobestsr}, we can compare the respective most constraining signal regions 
in the gluino dominated region for a specific benchmark scenario. We see
that the signal regions used to constrain the models are different between the two scenarios. In particular the $\lambda_S$ scenario 
which generically contains longer decays is better constrained by signal regions that have a larger final state particle multiplicity.
For instance, in the gluino dominated region, the ATLAS search with at least 3 $b$-jets \cite{TheATLAScollaboration:2013tha} (\texttt{atlas\_conf\_2013\_061}) is the 
most powerful but whilst the $\lambda_L$
scenario is best constrained with the 4-jet signal region, the 6-jet + 1-lepton region dominates for $\lambda_S$.
In addition, the multi $b$-jet ATLAS search demands moderate missing transverse momentum and hence the 
reduction of the total missing energy in the NMSSM does not significantly change the efficiency in the signal 
regions. The demands of the signal region therefore translates into the necessity of a  sufficiently large  gluino production 
cross section and a sizeable mass splitting of gluinos and squarks as well as squarks and electroweakinos. It is thus 
expected that limits should not depend significantly on the $\lambda$ scenario and only on the masses if they are 
close to threshold, as can be seen in our results.
\subsubsection{Heavy Gluinos}
For gluinos with mass above the production threshold of about 1.2 TeV, the 
exclusion sensitivity will be dominantly driven by the production of the third generation 
sparticles $\tilde{t}_{1/2}, \tilde{b}_{1/2}$ if they are sufficiently light. To illustrate this, 
we show the total production cross section for gluinos in Fig.~\ref{fig:gluinoxsect} of the appendix 
and third generation squark production for fixed $m_{\tilde{t}_1}$ in Fig.~\ref{fig:stopplotstopxsect}. Comparing 
the cross section values in regions with $m_{\tilde{g}} > 1.2$ TeV, $m_{\tilde{t}_1} < 800$ GeV, one 
expects far more $\tilde{t}_1$ than gluinos to be produced. Depending on the $\lambda$ scenario large numbers of events with sbottoms and heavier 
stops are expected in addition. Therefore, beyond the gluino threshold 
we observe a gluino-independent upper limit on the mass of the lightest neutralino. 

However, contrarily to the gluino-dominated 
region, one now finds significant dependencies of the limits on the chargino mass parameter and 
the $\lambda$ scenario in Fig.~\ref{fig:exclusion_gluino_singlino_mstop400}--\ref{fig:exclusion_gluino_singlino_mstop800}. In general, 
we observe that for a fixed mass of the lightest stop, limits on the LSP mass become weaker the lighter we chose the 
intermediate chargino. Also, throughout all cases we find consistently better limits in the $\lambda_L$ scenario than 
for $\lambda_S$.  

To understand these differences, we first have to shed light on the analyses and signal regions which define our 
exclusion limits in this part of parameter space. In Fig.~\ref{fig:gluinobestsr}, we take the specific example of a 
light stop mass of $500$ GeV and show the most sensitive signal regions for chargino masses of 400 and 300 GeV, 
comparing $\lambda_L$ on the left to $\lambda_S$ on the right. One finds two main classes of final states to be of importance here:
\begin{enumerate}
\item Signal regions from \texttt{atlas\_conf\_2013\_024} and `\texttt{tN}-type' regions in \texttt{atlas\_1407\_0583} focus on 
final states that originate from direct $\tilde{t} \rightarrow t \widetilde{\chi}_1^0$ decays. That is, they 
require missing transverse momentum, b-jets and final state objects whose invariant mass lie close to the top-quark mass. 
\item `\texttt{bC}-type' regions in \texttt{atlas\_1407\_0583} have been designed to tag events of type $\tilde{t} \rightarrow b \widetilde{\chi}^\pm, \widetilde{\chi}^\pm \rightarrow W^\pm \widetilde{\chi}_1^0$ by using kinematic variables that are sensitive to intermediate decay steps.
\end{enumerate}

In the following, we will refer to these as `$\texttt{tN}$-like' and `$\texttt{bC}$-like' analyses and signal regions, respectively.

In our model setup, the choice of the Higgs mass parameter $\mu$ (which sets the $m_{\widetilde{\chi}^{\pm}_1}$ and $m_{\widetilde{\chi}^{0}_{2,3}}$ )
is crucial to determine how many events are expected to be 
counted for the above most sensitive signal regions. Its value sets the kinematically open channels from the full 
list in Sec.~\ref{subsec:signatures}, fixes the branching ratios and determines the energy distribution among the final state particles.

For $m_{\widetilde{\chi}_1^\pm} \geq m_{\widetilde{\chi}_1^0}$, the branching ratio for $\tilde{t}_1 \rightarrow t \widetilde{\chi}_1^0$ is almost 100\% --- 
regardless of $\lambda$ --- and thus the upper LSP mass limits in both scenarios are determined by 
results from \texttt{tN}-like signal regions. If the $\tilde{t}_1$ was the only squark kinematically 
available, the limits of $\lambda_L$ and $\lambda_S$ would be expected to coincide. Comparison of the 
corresponding $m_{\tilde{t}_1} = m_{\widetilde{\chi}_1^\pm}$ lines in  Figs.~\ref{fig:exclusion_gluino_singlino_mstop400}--\ref{fig:exclusion_gluino_singlino_mstop800} 
however shows that $\lambda_L$ yields stronger limits, with the difference being larger for 
lighter $m_{\tilde{t}_1}$. The reason here is that $\lambda_L$ can allow for additional lighter 3rd generation squarks while 
still being able to get the right SM Higgs mass, as in Eq.~(\ref{eqn:higgsmass}). These lighter squarks have a 
larger production cross section and thus contribute more to the observable events, e.g. 
via decays $\tilde{b}_1 \rightarrow t \widetilde{\chi}_1^\pm$ which can also pass the signal region cuts. If a light $\tilde{t}_1$ is present in a $\lambda_S$ scenario however,
the additional 3rd generation squarks are required to be much heavier.

For lighter chargino masses, the decay $\tilde{t}_1 \rightarrow b \widetilde{\chi}_1^\pm$ opens kinematically. Within the $\lambda_S$ scenario we 
have an almost purely singlino LSP which causes the branching ratio for $t \widetilde{\chi}_1^0$ final states to become almost immediately 
disfavoured below the chargino threshold. Thus in this scenario almost all stops have to decay via intermediate electroweakinos. 
Interestingly, $\texttt{tN}$-like analyses are still most significant to set the limit if the charginos are not too 
light (see Fig.~\ref{fig:gluinobestsr} top right). The reason is that events with intermediate charginos can lead 
to $b W^+ \widetilde{\chi}_1^0$ final states misidentified as top quarks within $\texttt{tN}$-like signal region selections 
if the neutralino is light enough (the top mass window is very large in this analysis, as wide as $130<m_t<250$~GeV). In 
addition one expects a significant contribution of sbottoms 
decaying into $t \widetilde{\chi}_1^\pm$ final states which also look $\texttt{tN}$-like. 

For even lighter charginos, the limit 
is however only set by $\texttt{bC}$-like analyses (see Fig.~\ref{fig:gluinobestsr}). Decreasing the chargino mass further 
leads to softer decay products in the decay $\widetilde{\chi}_1^\pm \rightarrow \widetilde{\chi}_1^0 X^\pm$, which weakens the resulting upper limits on 
the $\widetilde{\chi}_1^0$ mass. Finally, decays into $t \widetilde{\chi}_{2/3}^0$ can reduce the branching ratio into the above mentioned decays once 
the chargino becomes light enough (see Fig.~\ref{fig:stopplotbrstoplsp}). 

It should also be mentioned that the branching ratios of the stop into neutral and charged higgsinos are fixed by the stop mixing 
matrix and $\tan\beta$ \cite{Rolbiecki:2009hk,Brooijmans:2014eja}. This results in a significant number of events 
displaying an `asymmetric' topology in which each of the initially produced sparticles decays
differently. However, the signal 
regions within the analyses that we use are mainly designed for symmetric decay scenarios, which leads to a reduction of the overall sensitivity.

Most of the explanations in the above discussion apply similarly to the  $\lambda_L$ scenario. However, a 
distinctive feature is the strong mixing in the neutralino sector which allows for the LSP to have a large higgsino 
component and thus  $t\widetilde{\chi}_1^0$ decays still having a large branching fraction below the chargino threshold. For example one finds that for $m_{\tilde{t}_1} - m_{\widetilde{\chi}_1^\pm} \lesssim 150$ GeV direct 
stop-to-top decays still happen with more than $20 \%$ probability (see Fig.~\ref{fig:stopplotbrstoplsp}). We therefore expect, 
and observe, that also within $\lambda_L$ the $\texttt{tN}$ signal regions set the limit for 
charginos within that mass region (see Fig.~\ref{fig:gluinobestsr}, top left). 

For lighter charginos, the limits 
become weaker due to the decreasing branching ratio of the `golden channel' $\tilde{t}_1 \to t \widetilde{\chi}_1^0$ and eventually 
the \texttt{bC} signal regions dominate and sets the limits thereafter (see Fig.~\ref{fig:gluinobestsr}, bottom left). The overall 
stronger exclusions within the $\lambda_L$ scenario can therefore be attributed to two different reasons. Firstly, the other 
3rd generation squarks will again be lighter in the $\lambda_L$ scenario due to the additional singlet contributions to 
the Higgs mass. Secondly, the increased branching ratio of $\tilde{t}_1 \to t \widetilde{\chi}_1^0$ which the LHC analyses are particularly
sensitive to also helps.

Interestingly, in both $\lambda$ scenarios, $\mueff$ lighter than $m_{\tilde{t}_1} - m_t$ opens decay channels of the 
type $\tilde{t}_1 \rightarrow t \widetilde{\chi}_{2/3}^0$. These could lead to NMSSM specific final states as 
discussed in Sec.~\ref{subsec:signatures}. However, we do not observe any improvement on the LSP limits in these cases. Quite 
the contrary, the reduction of the branching ratio into $b \widetilde{\chi}_1^\pm$ final states resulting from the new decay channel 
and asymmetric final states mentioned above
weakens the limits even more as can be observed when comparing the limits in 
Fig.~\ref{fig:exclusion_gluino_singlino_mstop400}--\ref{fig:exclusion_gluino_singlino_mstop800} above or below this 
threshold. We investigate the impact of this more closely in the upcoming section.

\subsection{Stop-Electroweakino-Plane}
\label{sec:results:stop_limits}
\begin{figure*}
\includegraphics[width=0.45\textwidth]{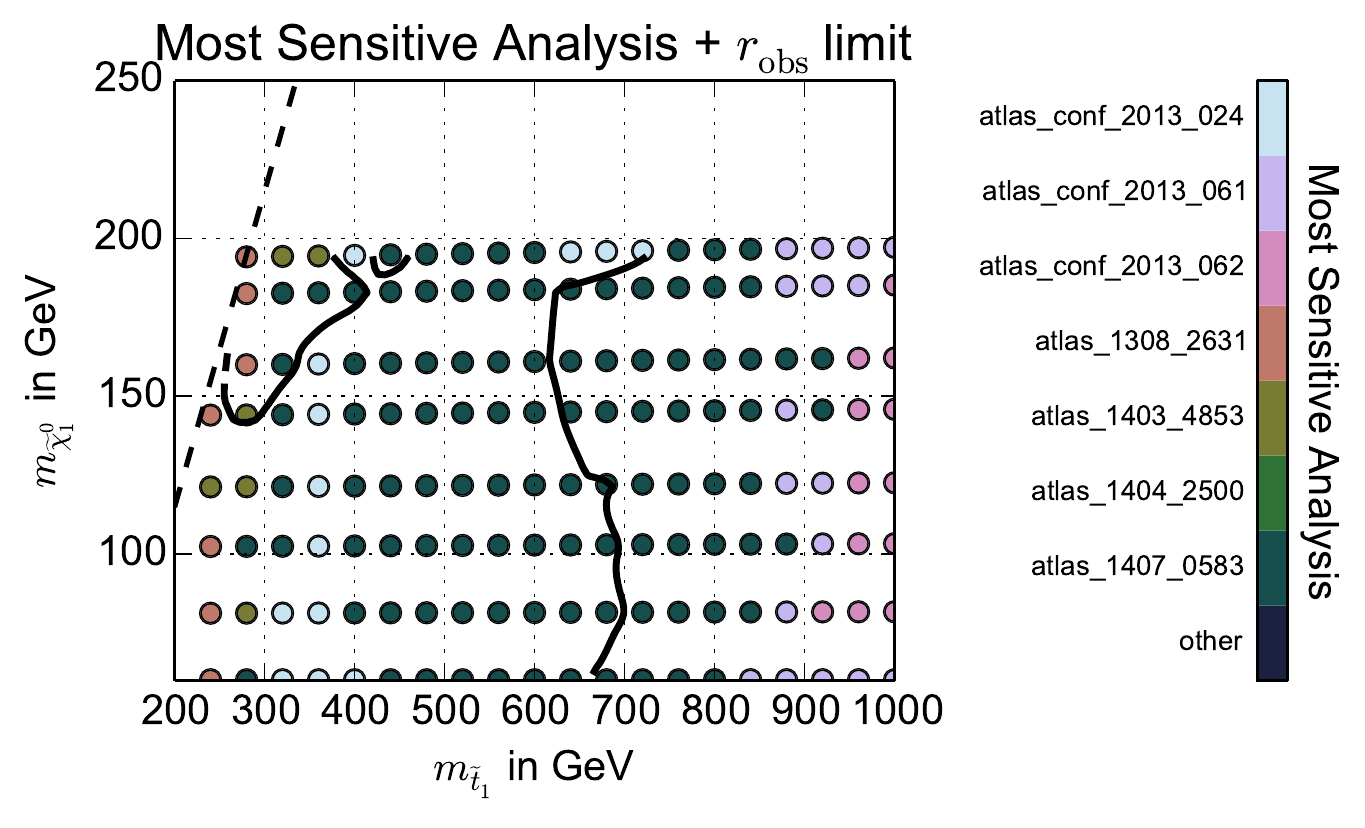}\qquad
\includegraphics[width=0.45\textwidth]{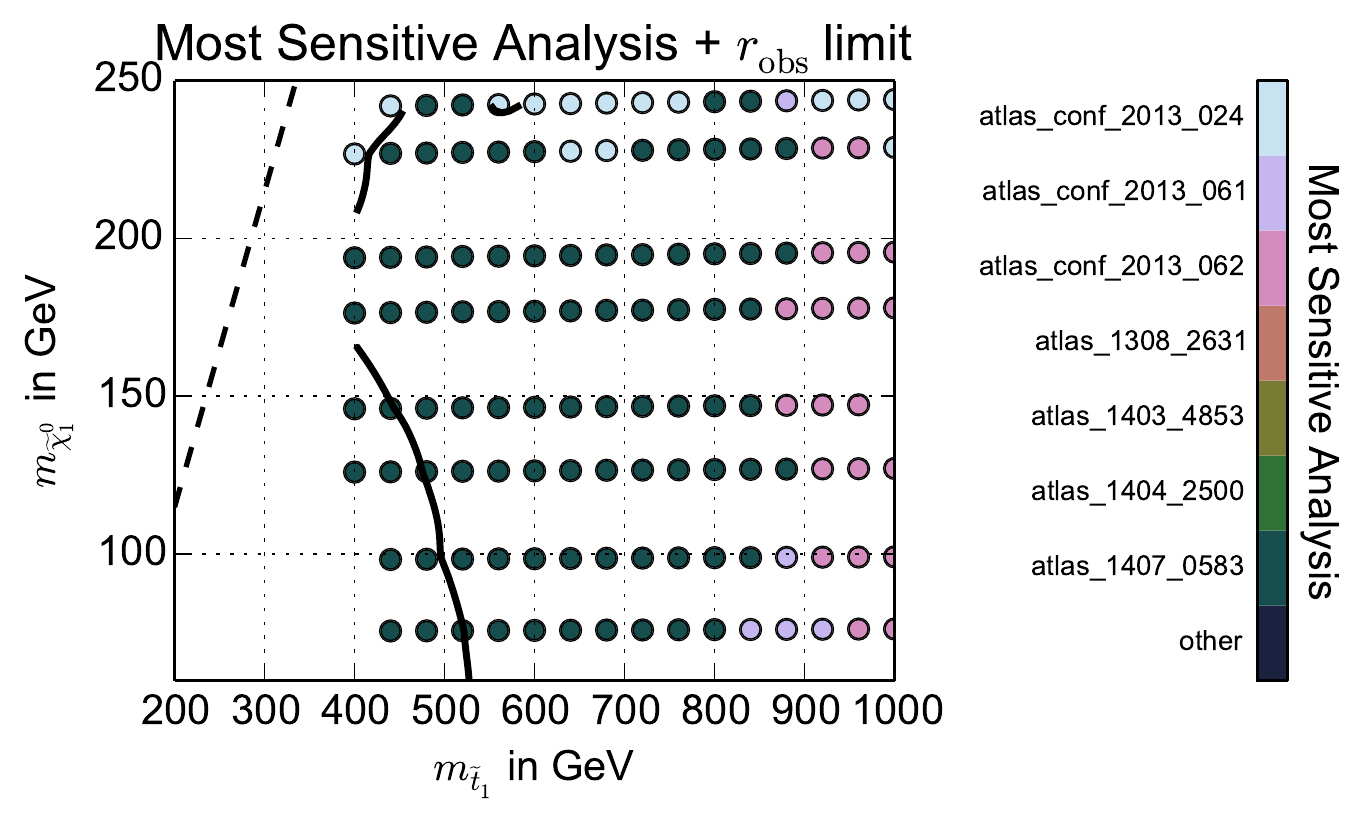}
\caption{Observed 95\% C.L. exclusion limit and most sensitive analysis per point for $m_{\widetilde{\chi}_1^\pm} = 250$ GeV. Left: $\lambda_L$. Right: $\lambda_S$}
\label{fig:stopplot250}
\includegraphics[width=0.45\textwidth]{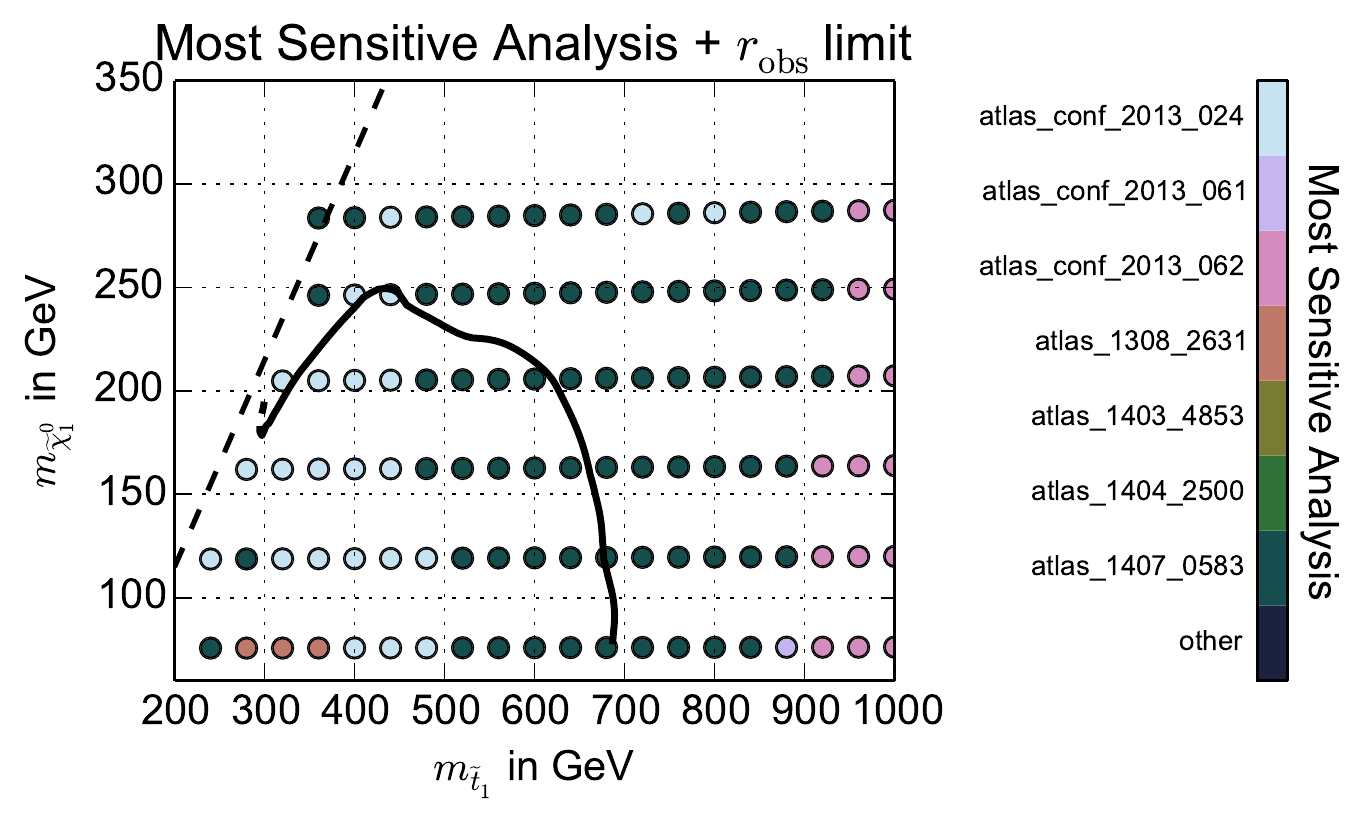}\qquad
\includegraphics[width=0.45\textwidth]{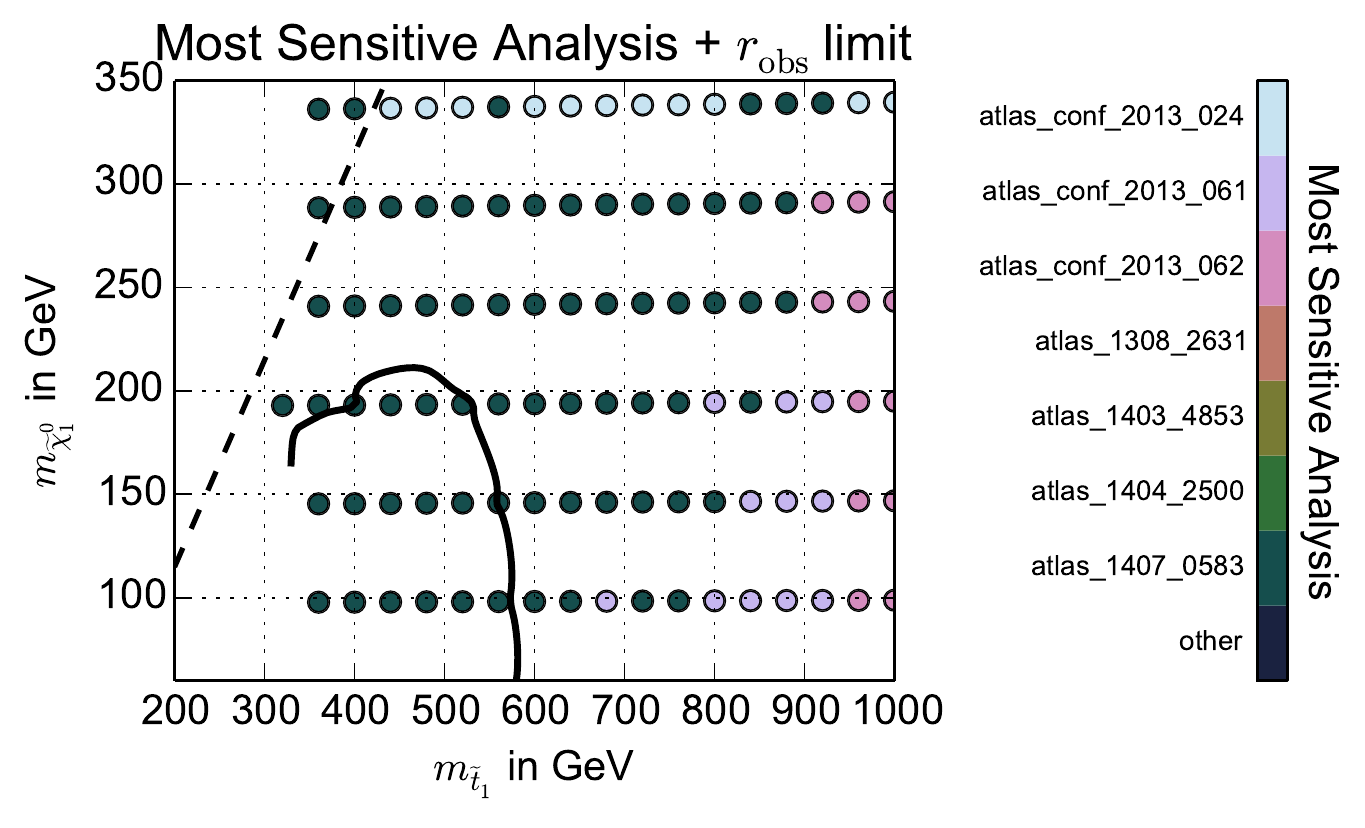}
\caption{Observed 95\% C.L. exclusion limit and most sensitive analysis per point for $m_{\widetilde{\chi}_1^\pm} = 350$ GeV. Left: $\lambda_L$. Right: $\lambda_S$}
\label{fig:stopplot350}
\includegraphics[width=0.45\textwidth]{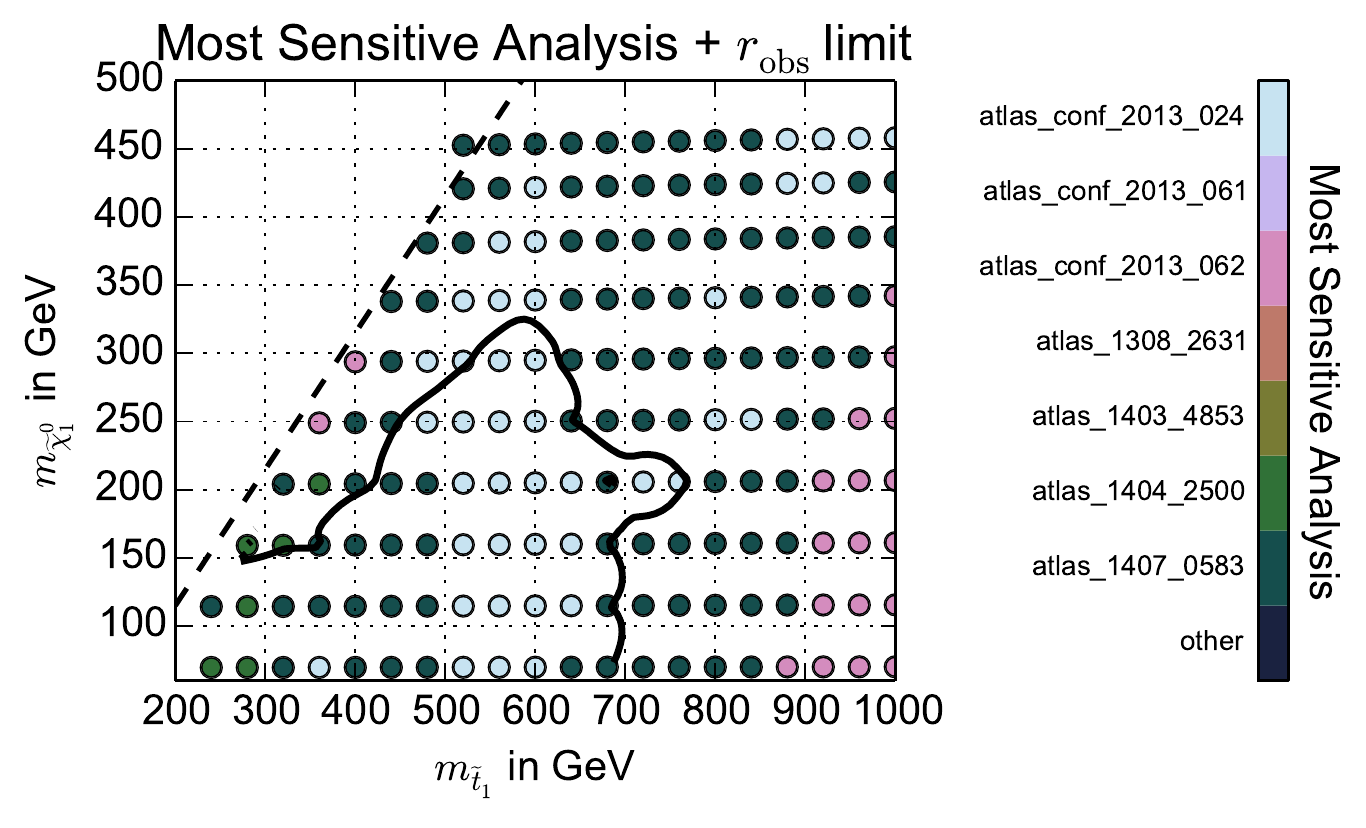}\qquad
\includegraphics[width=0.45\textwidth]{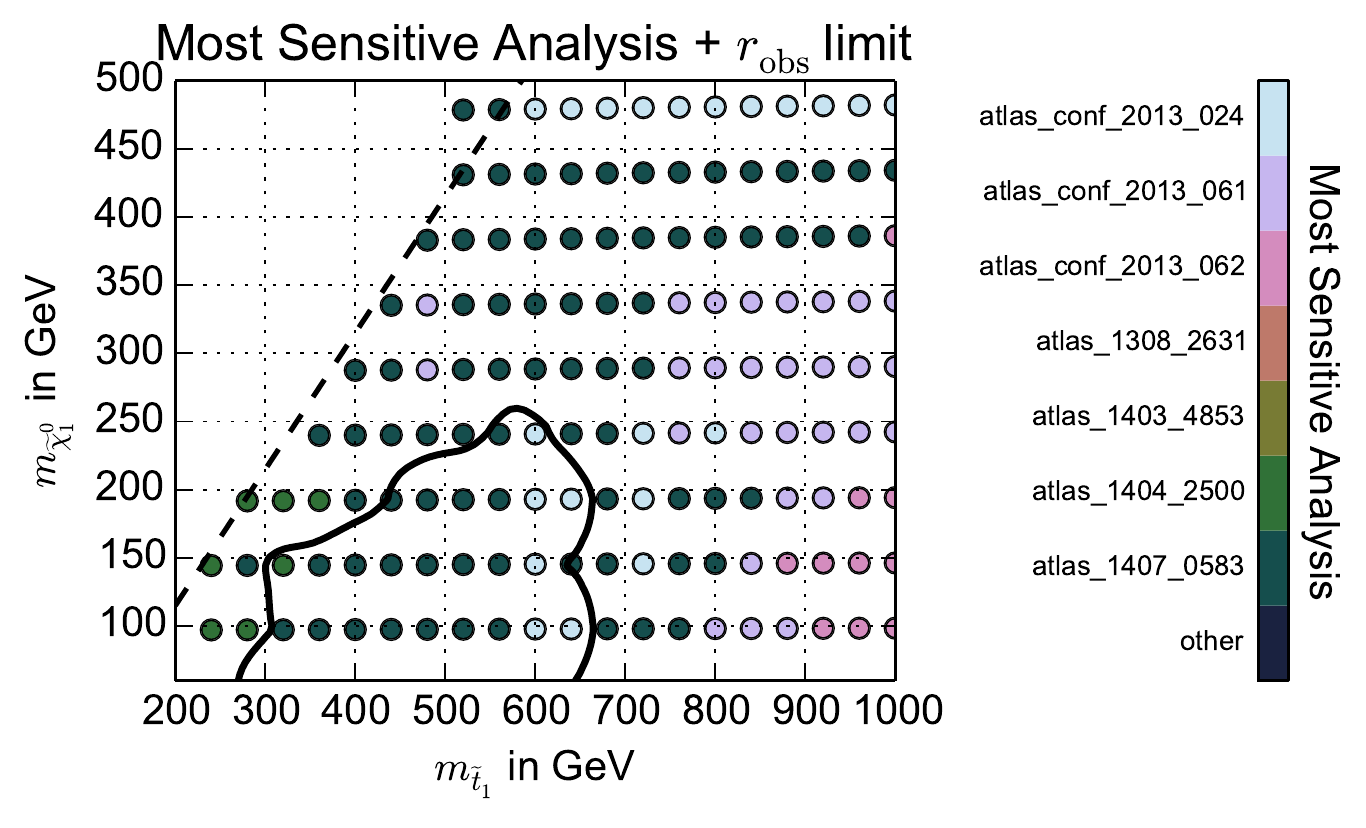}
\caption{Observed 95\% C.L. exclusion limit and most sensitive analysis per point for $m_{\widetilde{\chi}_1^\pm} = 500$ GeV. Left: $\lambda_L$. Right: $\lambda_S$}
\label{fig:stopplot500}
\includegraphics[width=0.45\textwidth]{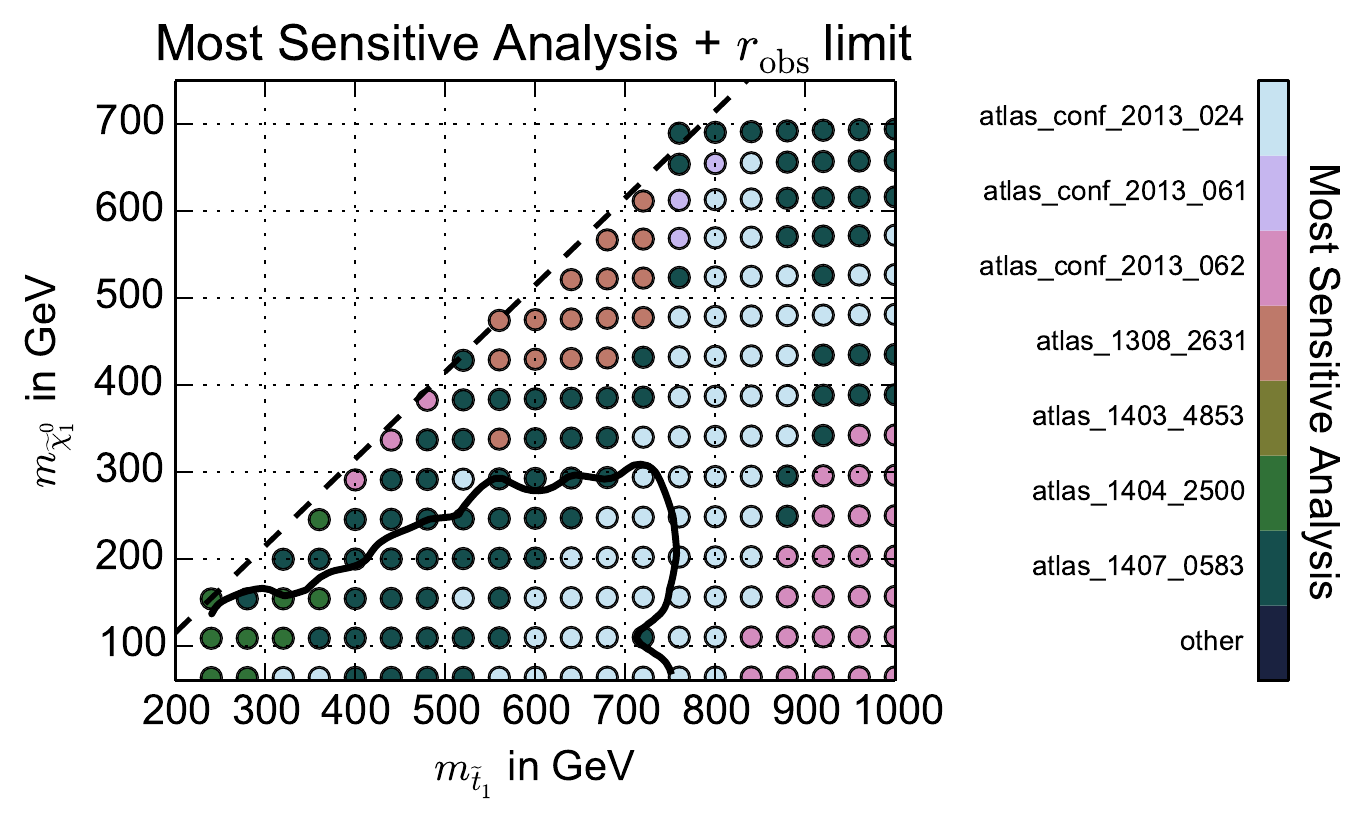}\qquad
\includegraphics[width=0.45\textwidth]{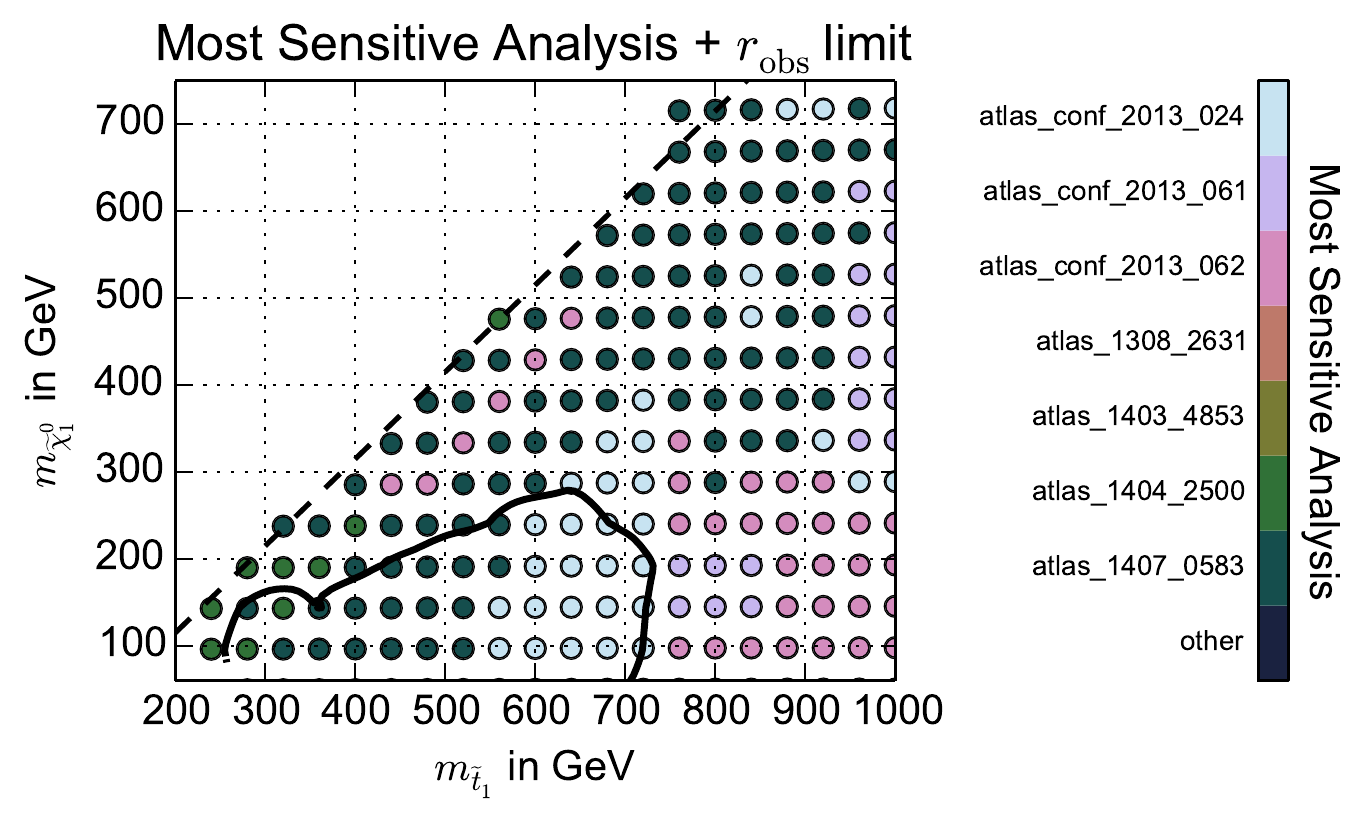}
\caption{Observed 95\% C.L. exclusion limit and most sensitive analysis per point for $m_{\widetilde{\chi}_1^\pm} = 750$ GeV. Left: $\lambda_L$. Right: $\lambda_S$}
\label{fig:stopplot750}
\end{figure*}
As shown in the last set of results, below the gluino production threshold, the LHC limits only have a small
dependence on the details of the natural spectrum. However, as we decouple the 
gluino, the masses and couplings of the electroweakino sector become more important.
For that reason we also show results 
in the $m_{\tilde{t}_1}$-$m_{\widetilde{\chi}_1^0}$-plane for a decoupled gluino of mass 2 TeV in Figs.~\ref{fig:stopplot250}-\ref{fig:stopplot750}. With one degree of 
freedom less, we are now able to show one exclusion limit per plot for specific values of $m_{\widetilde{\chi}^\pm}$, 
again comparing $\lambda_L$ (left) to $\lambda_S$ (right). 
The parameter space that we investigate does not include the region where $m_{\tilde{t}_1}$ becomes 
close to $m_{\tilde{\chi}^0_1}$. This is shown by the diagonal dashed line within each plot which shows the kinematic range 
for which $m_{\tilde{t}_1} < m_b + m_W + m_{\widetilde{\chi}_1^0}$ and only 
4-body final states or flavour changing neutral current decays 
such as $\tilde t_1\rightarrow c \widetilde{\chi}_1^0$ are possible. Given the 
small mass difference, initial state radiation searches provide the most constraining limits in this region \cite{Drees:2012dd,Bornhauser:2010mw}. These searches 
are relatively insensitive to the details of the decay chain in question and thus we expect the results to be very similar 
to those of the MSSM.
\begin{figure*}
\includegraphics[width=0.45\textwidth]{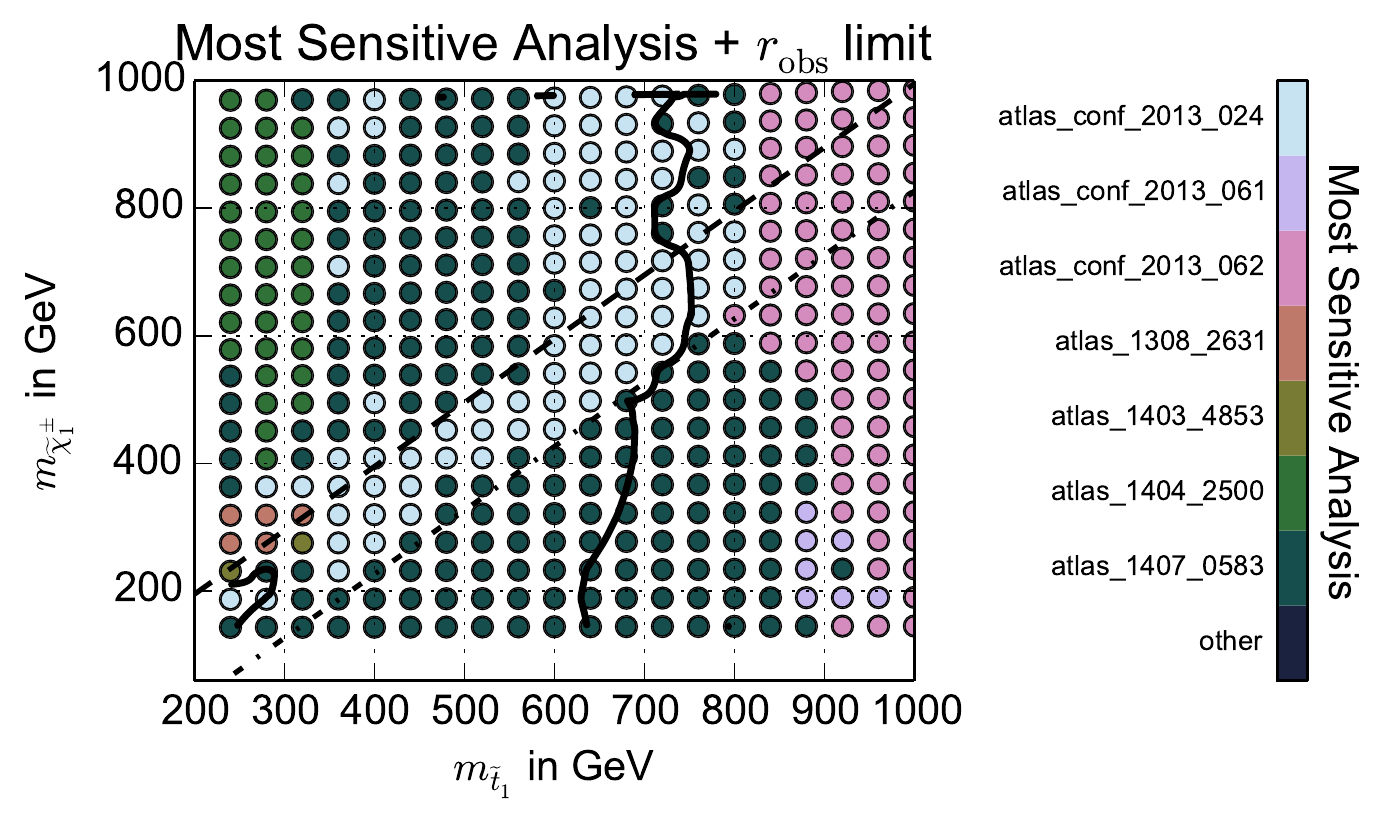}\qquad
\includegraphics[width=0.45\textwidth]{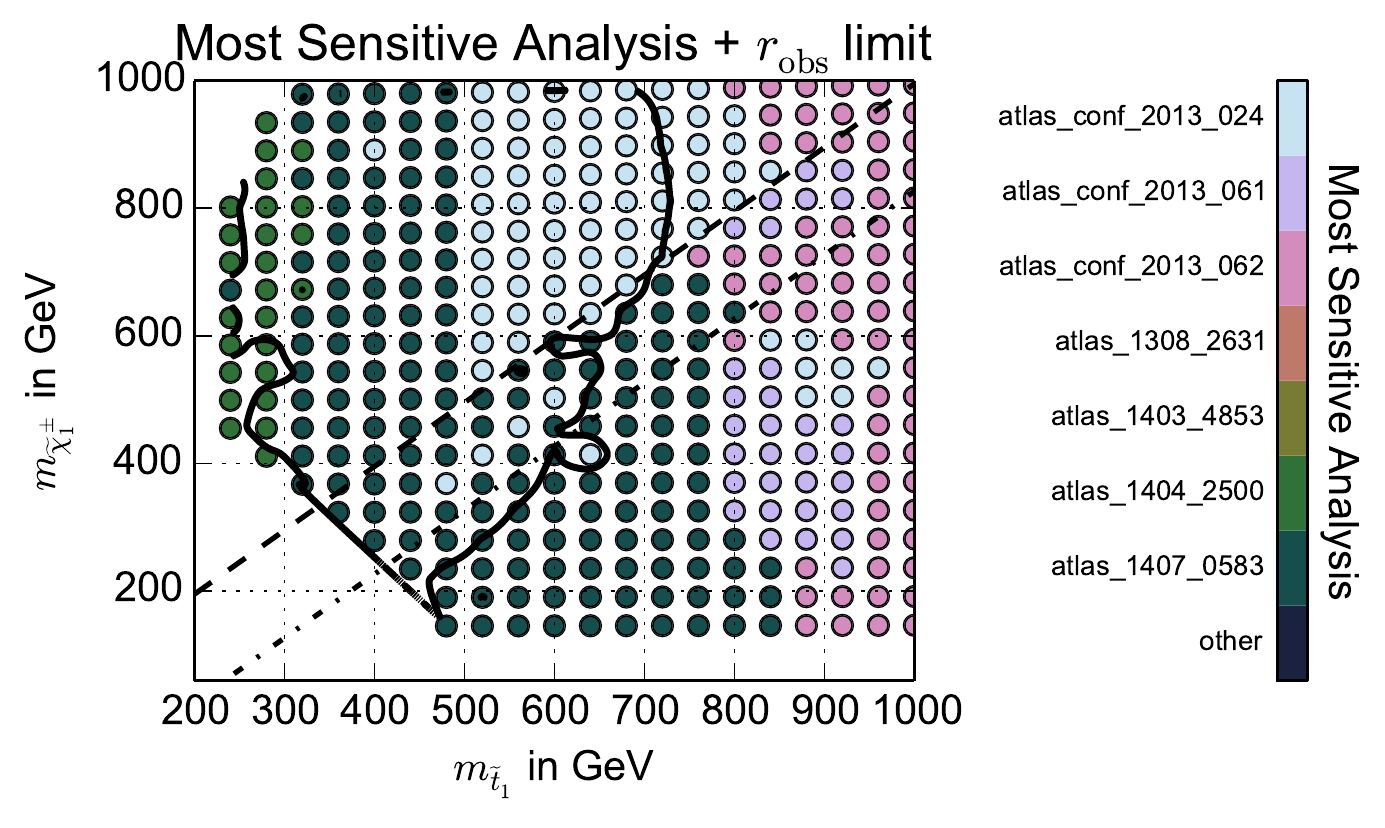}
\caption{Observed 95\% C.L. exclusion limit and most sensitive analysis per point for $m_{\chi^0_1} = 100$ GeV. Diagonal 
dashed (dashed-dotted) lines show shows the threshold 
for $\tilde{t}_1 \rightarrow b \widetilde{\chi}^\pm_1$ ($\tilde{t}_1 \rightarrow t \widetilde{\chi}^0_{2/3}$). Left: $\lambda_L$. Right: $\lambda_S$}
\label{fig:stopplotchargino}
\end{figure*} 

Similarly to the gluino-LSP scan, the upper limit on the LSP mass is set by 
requiring $m_{\widetilde{\chi}_1^0} < m_{\widetilde{\chi}^\pm}$. For the $\lambda_L$ case, mixing in the 
neutralino sector leads to a maximum achievable value of $m_{\widetilde{\chi}_1^0}$ which lies somewhat 
below $m_{\widetilde{\chi}^\pm}$. In the $\lambda_S$ scenario, realistic
parameter points are not possible with $m_{\tilde{t}_1} \lesssim 400$~GeV  
 if $m_{\widetilde{\chi}^\pm} \approx \mueff$ 
is small. The reason is that for small $\mueff$, the radiative corrections to the Higgs boson mass are not large enough
to correctly reproduce the LHC measurement (see e.g. Eq.~(\ref{eqn:higgsmass})). 

In all plots we again show, for each individual considered data point, the most sensitive analysis that has been 
used to calculate the confidence level of that particular point. However, we do not show the numerous individual 
signal regions (as we did for Fig.~\ref{fig:gluinobestsr}) to keep the amount of different values to a reasonable level.

We again observe that the choice of analysis responsible for the limit setting is strongly correlated with the 
branching ratio of the lightest stop and from Fig.~\ref{fig:stopplotbrstoplsp} we expect four main regions of interest. These are respectively, 
direct decays of the stop into the LSP and an a) on- or b) off-shell top, c) intermediate decays via charginos 
or d) via neutral higgsinos. The thresholds for these regions often coincide with similar threshold for sbottom decays, as 
can be seen in Fig.~\ref{fig:sbottomplotbrstoplsp}. As an example the $\tilde{b}_1 \rightarrow t \widetilde{\chi}^\pm$ and 
the $\tilde{t}_1 \rightarrow t \widetilde{\chi}_{2/3}^0$ lie very close in the $\lambda_L$ scenario.

Using the branching ratio information, we can closely follow the explanations from the last section to understand the 
limits in Figs.~\ref{fig:stopplot250}-\ref{fig:stopplot750}. For stops lighter than the given chargino, only 
direct decays $\tilde{t}_1 \rightarrow t^{(*)} \widetilde{\chi}_1^0$  are kinematically allowed. $\texttt{tN}$-like analyses 
are therefore the most sensitive and lead to similar limits for $\lambda_L$ and $\lambda_S$, with the former 
being slightly stronger than the latter due to the lighter sbottoms in this model. In $\lambda_S$, a 
strip for $m_{\tilde{t}_1} - m_{\widetilde{\chi}_1^0} < m_t$ cannot be excluded as the final state with an off-shell top is not 
observed by \texttt{tN}-like analyses and hard to distinguish from the SM background. Within $\lambda_L$, this 
region can still be explored since it is possible that the spectrum also contains a light $\tilde{b}_1$. This 
can be excluded via $\tilde{b} \rightarrow t \widetilde{\chi}^\pm$ specific selections 
in \texttt{atlas\_1404\_2500} (see e.g. Figs.~\ref{fig:stopplot500}, \ref{fig:stopplot750}).

For kinematically allowed chargino decays, a transition from `\texttt{tN}' into `\texttt{bC}' signal regions can be observed for 
increasing $m_{\tilde{t}_1}$, that is for larger stop-chargino splitting. As in the previous setup, $\lambda_L$ profits from 
the Higgsino fraction of the LSP and the generally lighter 3rd generation squarks. The highest sensitivities are 
reached via `\texttt{tN}' final states in \texttt{atlas\_conf\_2013\_024}. The highest sensitivity to the LSP 
mass can be reached when these final states set the limit, which can reach up 
to $(m_{\widetilde{\chi}_1^0}^{\text{max}} \approx 325$ GeV. In $\lambda_S$, \texttt{bC} signal regions dominate the limit earlier, 
which require lighter neutralinos to observe the intermediate chargino decay step. The 
experimental reach to the LSP mass is therefore smaller in these scenarios and of order 250 GeV.

As we further increase the stop masses, a maximum value of $m_{\tilde{t}_1}$ is reached. This stop sensitivity 
limit seems to depend on the chosen chargino mass and the considered $\lambda$ scenario and is rather 
independent of the LSP mass as long as it is light enough, that is for $m_{\widetilde{\chi}_1^0} \lesssim 150$~GeV.

To better understand the parameter dependence, we chose to show results in the $m_{\tilde t_1}$-$m_{\widetilde \chi_1^\pm}$-plane for a 
fixed, light LSP mass of $100$ GeV in Fig.~\ref{fig:stopplotchargino}.  We show the previously discussed thresholds 
for $\tilde{t}_1 \rightarrow b \widetilde{\chi}^\pm$  and $\tilde{t}_1 \rightarrow t \widetilde{\chi}^0_{2/3}$ and it 
can be seen that they can have an important impact on the sensitivity of the experimental analyses to the 
stop mass. Within $\lambda_L$, the upper limit on $m_{\tilde{t}_1}$ is almost constant at $\approx 700$~GeV 
for charginos above the $t \widetilde{\chi}^0_{2/3}$ threshold. This corresponds to similar limits 
from simplified $\tilde{t} \rightarrow t \widetilde{\chi}_1^0$ topologies as in \cite{ATLAS:2013cma,Aad:2014kra}. The limit 
gets slightly weaker if the chargino threshold is passed, dropping by at most 50 GeV as 
soon as \texttt{bC} signal regions dominate the limit. In Figs.~\ref{fig:stopplotbrstoplsp2}, \ref{fig:sbottomplotbrstoplsp2} 
and \ref{fig:stopplot22} we show the branching ratio distributions in the same plane and the same LSP mass as the 
results in Fig.~\ref{fig:stopplotchargino}. One observes that the mass values in our spectrum are such that the above 
behaviour coincides with the threshold for $\tilde{b} \rightarrow t \widetilde{\chi}^\pm$, which 
also explains why the $\texttt{bC}$-like signal regions become important within this region of parameter space.

As long as the higgsinos do not appear in the squark decay chains, $\lambda_S$ returns similar limits as 
the $\lambda_L$ scenario, for the same reasons discussed in the previous section. However, within this model one 
observes a sizeable weakening of the limits as soon as the intermediate chargino and NLSP higgsino decays open 
kinematically. Interestingly, the latter has a particularly negative impact on the result, as the experimental analyses are only weakly sensitive to parameter regions 
in $\lambda_S$ where $\tilde{t}_1 \rightarrow t \widetilde{\chi}^0_{2/3}$ is kinematically allowed. As discussed in 
Sec.~\ref{subsec:signatures}, it is this decay chain which yields NMSSM-specific features in the final state topology: the 
decay of the higgsino NLSPs into the singlino LSP should create a sizeable excess of $h/H/A_1 \rightarrow \bar{b} b$ final states. It 
seems, however, that none of the many distinct final states within the numerous analyses that \Checkmate{} contains 
is sufficiently sensitive to this topology. Thus, the existing \texttt{bC}-like limits are weakened due to reduced 
branching ratios after passing the NLSP higgsino threshold.

We therefore conclude that not only can many limits on natural NMSSM scenarios be derived from very similar 
topologies in natural MSSM studies, but we also find that regions of parameter space which produce 
NMSSM-exclusive final state features are not sufficiently covered by existing studies. Therefore, only 
weak limits on the NMSSM can be set within this region of parameter space which suffer under branching-ratio penalties.

\section{Conclusion}
\label{sec:conclusion}

In this study we explore the natural NMSSM to determine how the additional singlino can effect the LHC searches compared to the more 
studied MSSM case. To do this we explored a number of different scenarios, mostly notably examining the difference between a small-$\lambda$ case,
where the LSP is dominantly a singlino and the large-$\lambda$ case, where the LSP can contain a substantial higgsino component. We also study
in detail the differences that occur when the gluino is light enough that it dominates the SUSY production cross-sections and 
what happens when the gluino is pushed to a mass where LHC production is no longer copious.

We find that, when constructing a realistic phenomenological model, the NMSSM-specific decay chains via intermediate 
heavy neutralinos often create an MSSM-like topology, $\tilde{q}_3 \to q_3 \tilde{\chi}^0_1$ which can be preceded by $\tilde{g} \to \tilde{q}_3 q_3$
if the gluino is light. If the branching ratio to these decay chains are large, the limits very closely follow those often studied as 
simplified models in the MSSM. However, the branching ratio depends on the size of the NMSSM-coupling $\lambda$. If it is 
large, all neutralinos have a sizeable higgsino fraction and direct decays into the lightest neutralino are significant. However, in case of 
small $\lambda$, the coupling of the squarks to the LSP is made small since it has a large singlino content. Therefore decays via intermediate 
charged and neutral higgsinos are preferred if kinematically allowed which lengthens the decay chains seen. In addition, since different
decay modes may be competing with similar branching ratios, `asymmetric' decay chains can often occur.

These longer decay chains can lead to weaker LHC bounds for two particular reasons. First of all, the ATLAS searches have more focussed 
of the MSSM specific signatures and consequently not been designed with these final states in mind. Secondly, the longer decay chains
lead to a higher final state particle multiplicity but with each individual particle carrying smaller $p_T$. In addition the same effect
reduces the final state $E_T^{\text{miss}}$ as observed in other studies with more complicated decay topologies e.g. \cite{Fan:2012jf}. On the 
other hand, additional
final states, namely jets and leptons, can improve the sensitivity even though the invisible transverse momentum is reduced. Therefore 
an important conclusion of this study is that it is not obvious if the efficiency is smaller or larger in a particular NMSSM
scenario simply by looking at the spectrum and decays. Instead it is crucial to test the model against a large 
number of searches covering various final state topologies.

Within this study we do test a large variety of different analyses but still only use one signal region to define the overall limit. In the 
models with extended and asymmetric decay chains (where we observe a weakening of the LHC limit), we expect the signal to populate a more
varied number of signal regions than if the model predicted a single dominating decay chain. Therefore it may be expected that a combination 
of the sensitivities across all analyses can significantly enhance the limits but this is beyond the scope of this study.

\vfill

\section*{Acknowledgements}
JT would like to thank Prof. Herbi Dreiner and the Bethe Centre for Theoretical Physics at Bonn University for 
hospitality and support while part of this work was completed.
The work of JSK has been partially supported by the MICINN, Spain, under contract FPA2013-44773-P, Consolider-Ingenio CPAN CSD2007-00042 and the Spanish MINECO Centro de excelencia Severo Ochoa Program under grant SEV-2012-0249. 
\onecolumngrid
\pagebreak
\appendix
\onecolumngrid
\section{Mass Distributions}
\begin{figure}[h!]
\includegraphics[height=0.2\textheight]{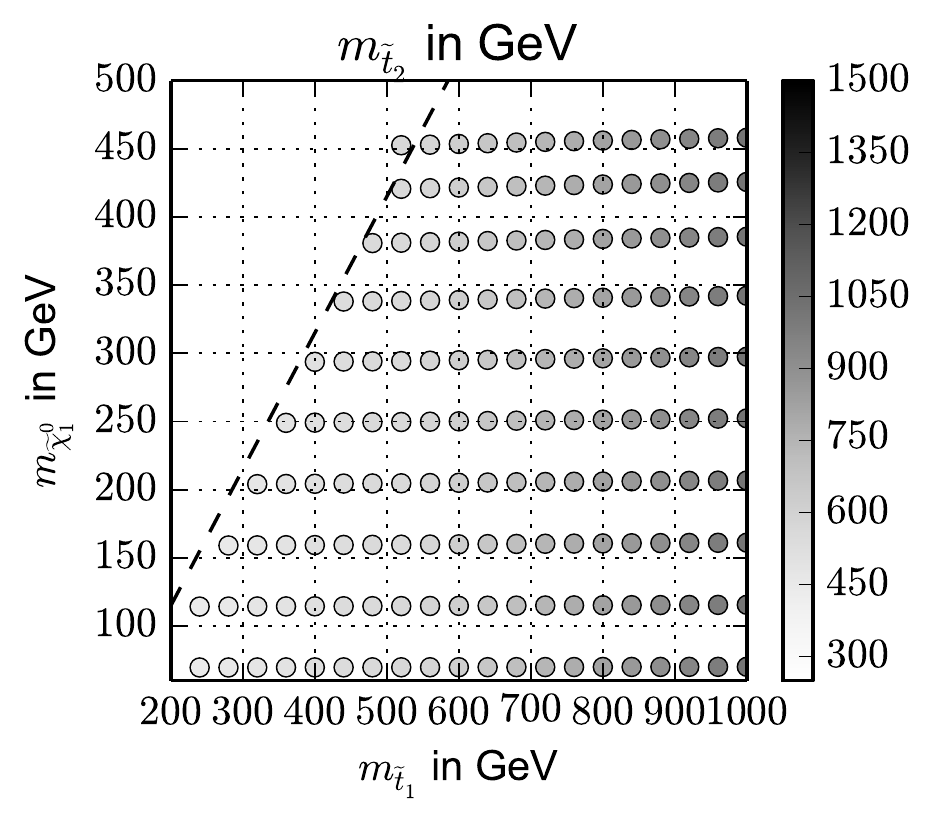}\qquad \qquad \qquad
\includegraphics[height=0.2\textheight]{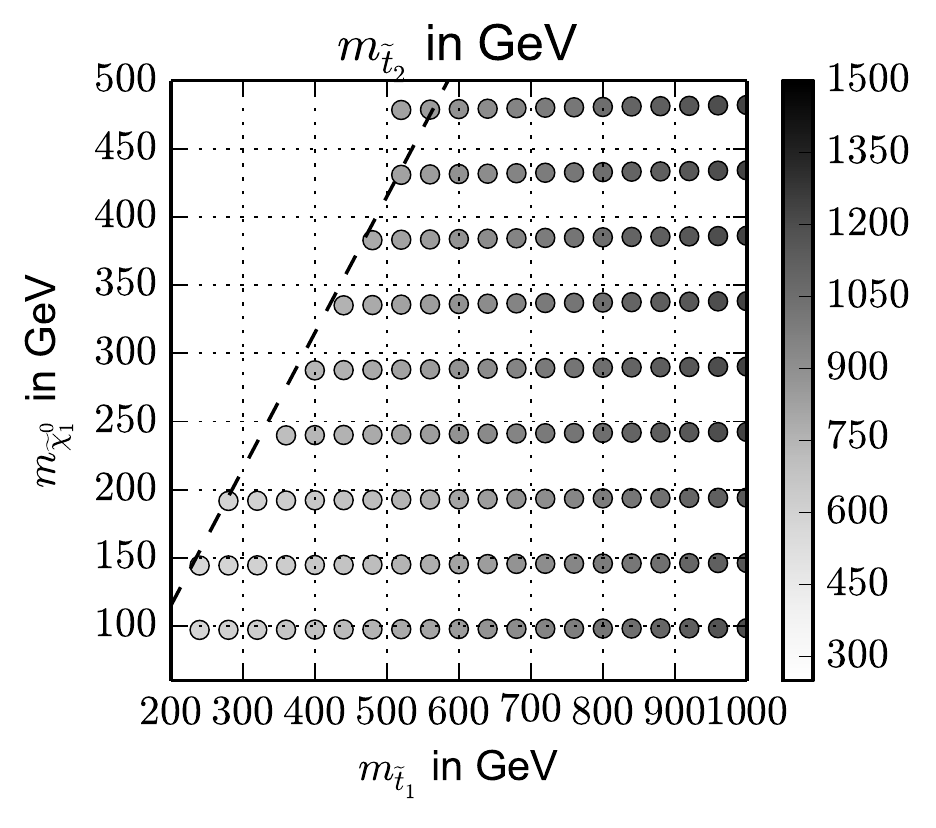}
\includegraphics[height=0.2\textheight]{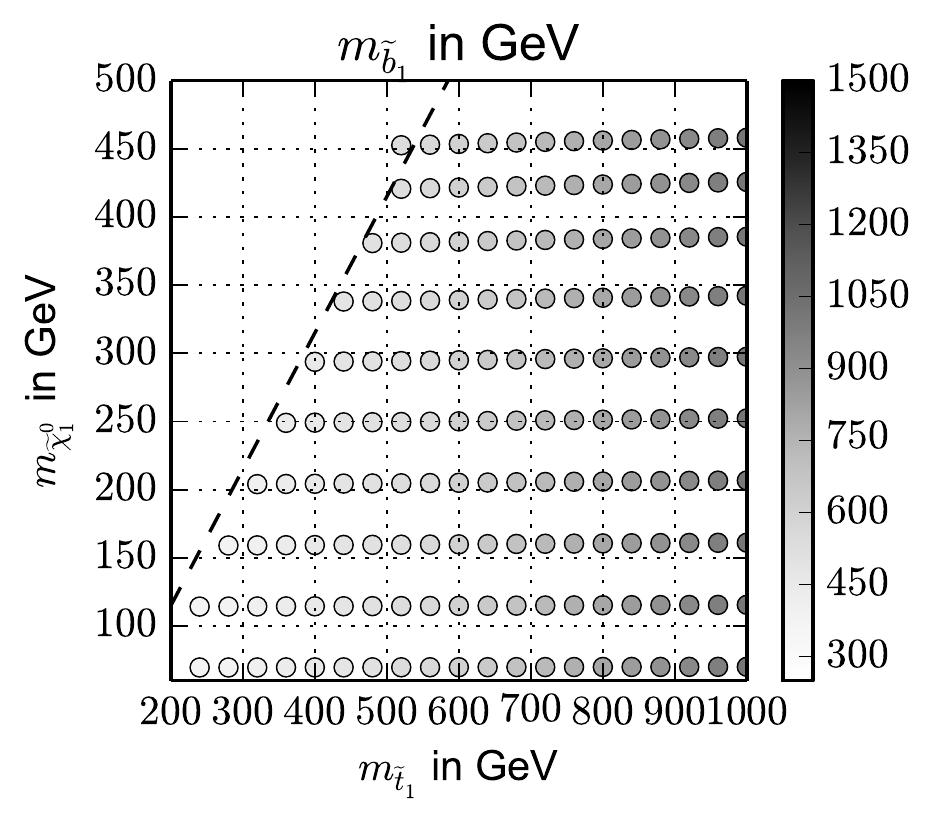}\qquad \qquad \qquad
\includegraphics[height=0.2\textheight]{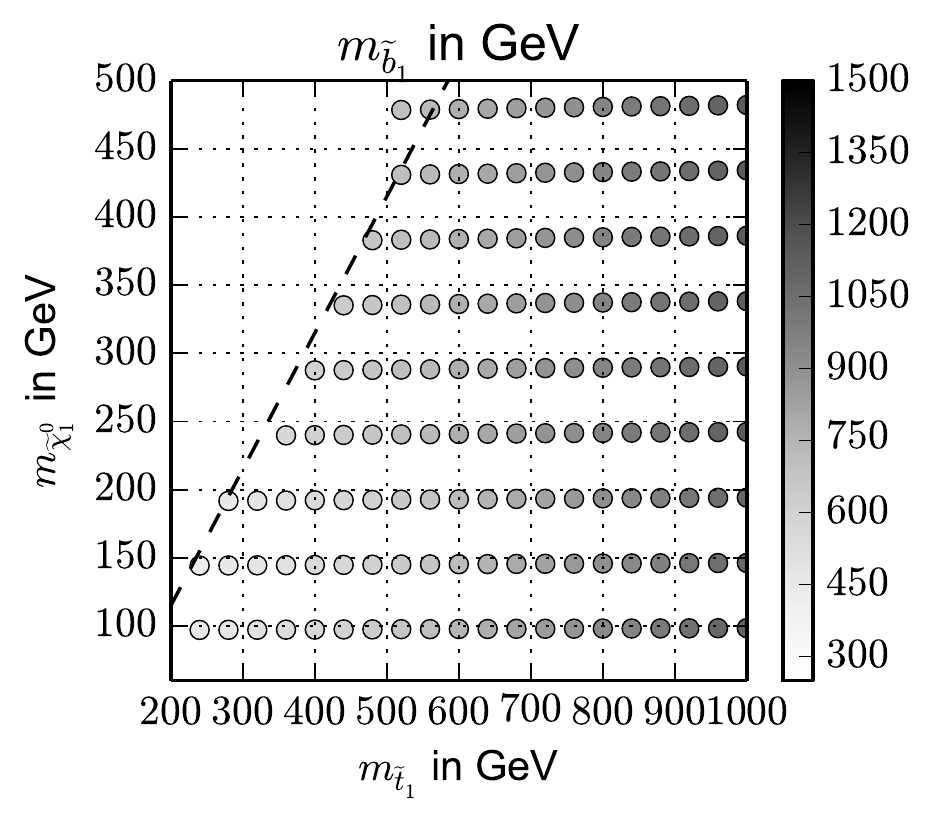}
\caption{Mass of the heavier stop and the lighter sbottom (which is very degenerate with the heavier sbottom) for a decoupled gluino and  $m_{\widetilde{\chi}_1^\pm} = 500$ GeV. Left: $\lambda_L$. Right: $\lambda_S$}
\label{fig:stopplotstop2mass}
\end{figure}
\vfill
\pagebreak
\section{Cross Section Distributions}
\begin{figure}[h!]
\includegraphics[height=0.2\textheight]{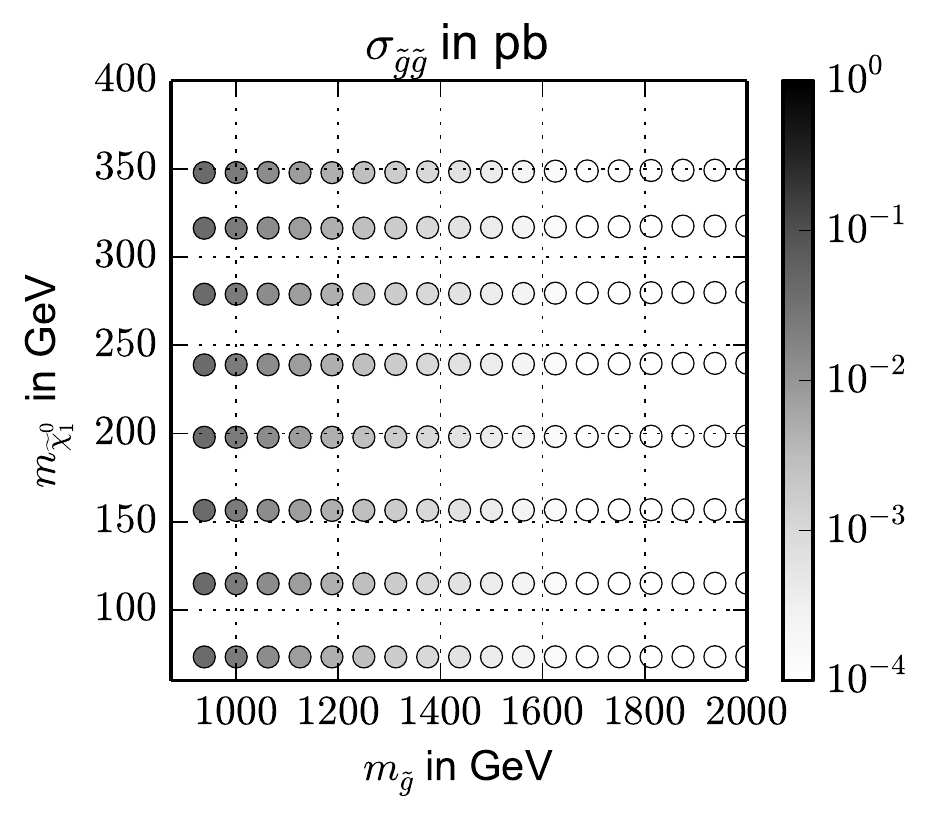}\qquad \qquad \qquad
\includegraphics[height=0.2\textheight]{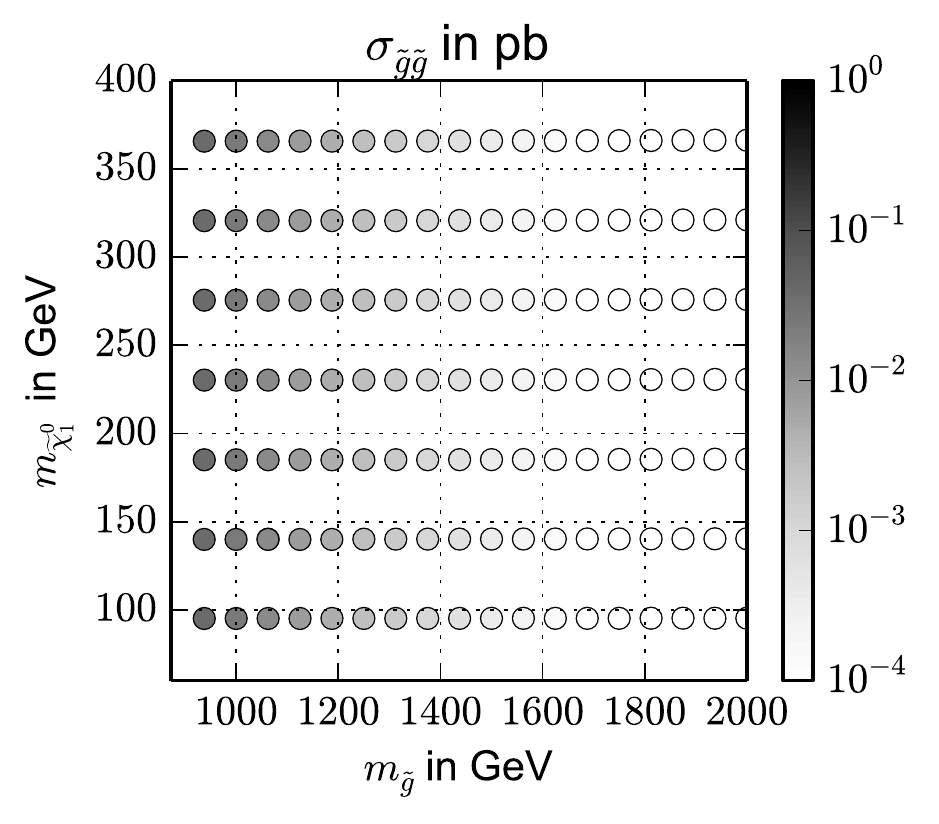}
\caption{Total production cross section for gluinos, using $m_{\tilde t_1} = 500$ GeV, $m_{\widetilde{\chi}_1^\pm} = 400 $ GeV. Left: $\lambda_L$. Right: $\lambda_S$.}
\label{fig:gluinoxsect}
\includegraphics[height=0.2\textheight]{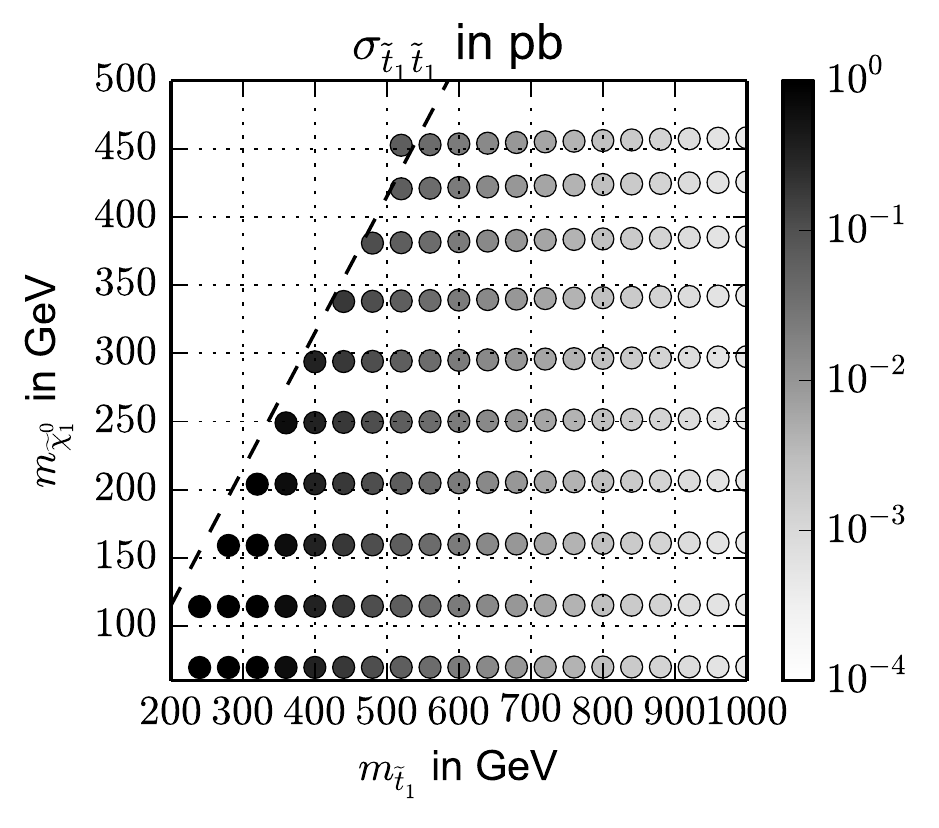}\qquad \qquad \qquad
\includegraphics[height=0.2\textheight]{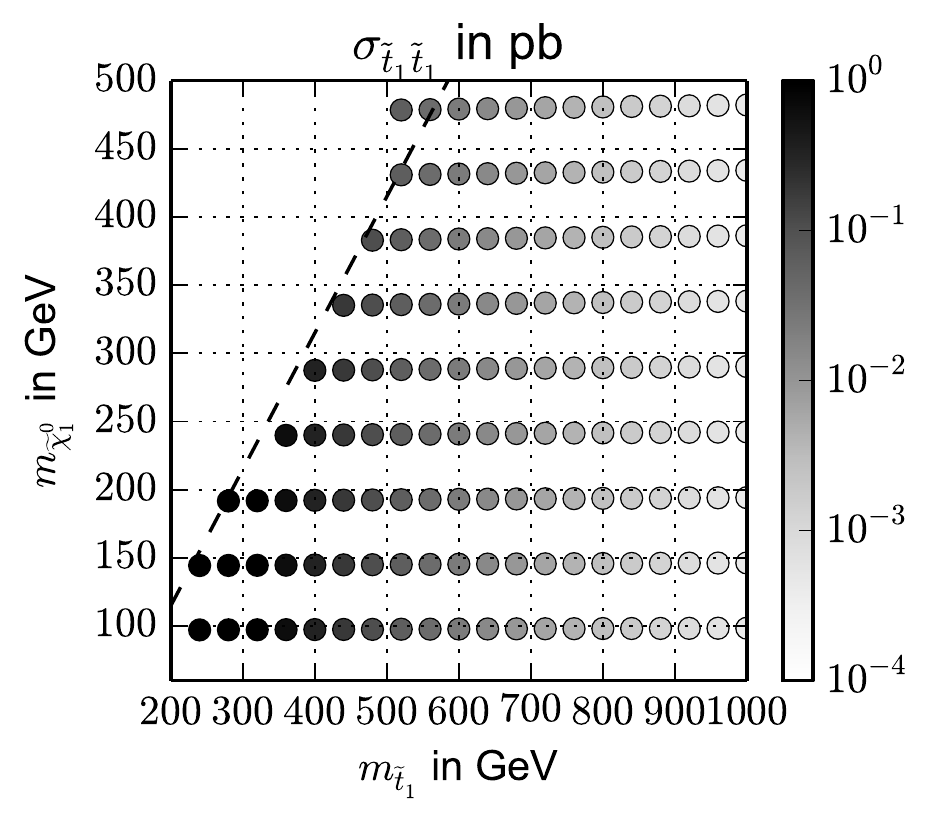}
\includegraphics[height=0.2\textheight]{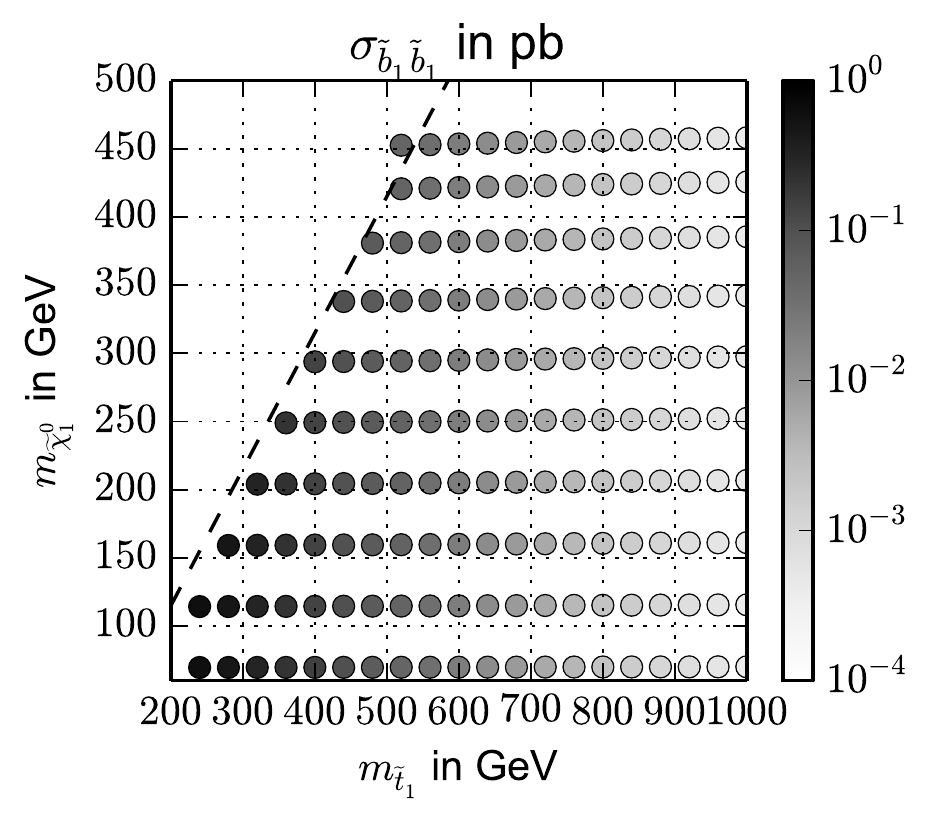}\qquad \qquad \qquad
\includegraphics[height=0.2\textheight]{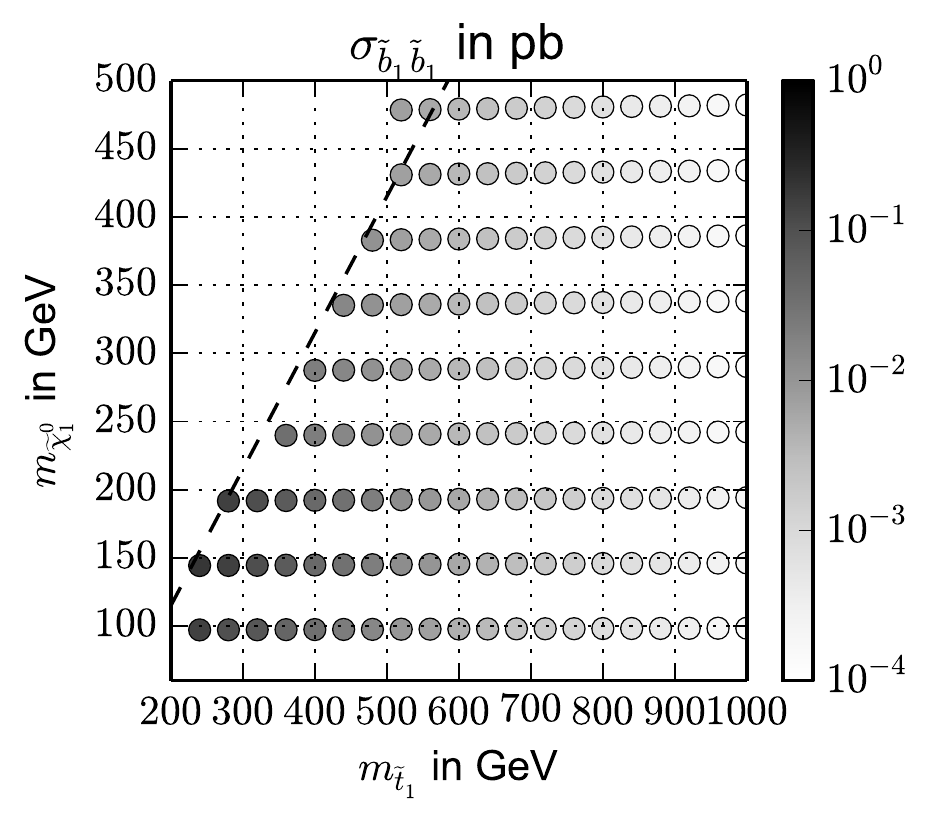}
\includegraphics[height=0.2\textheight]{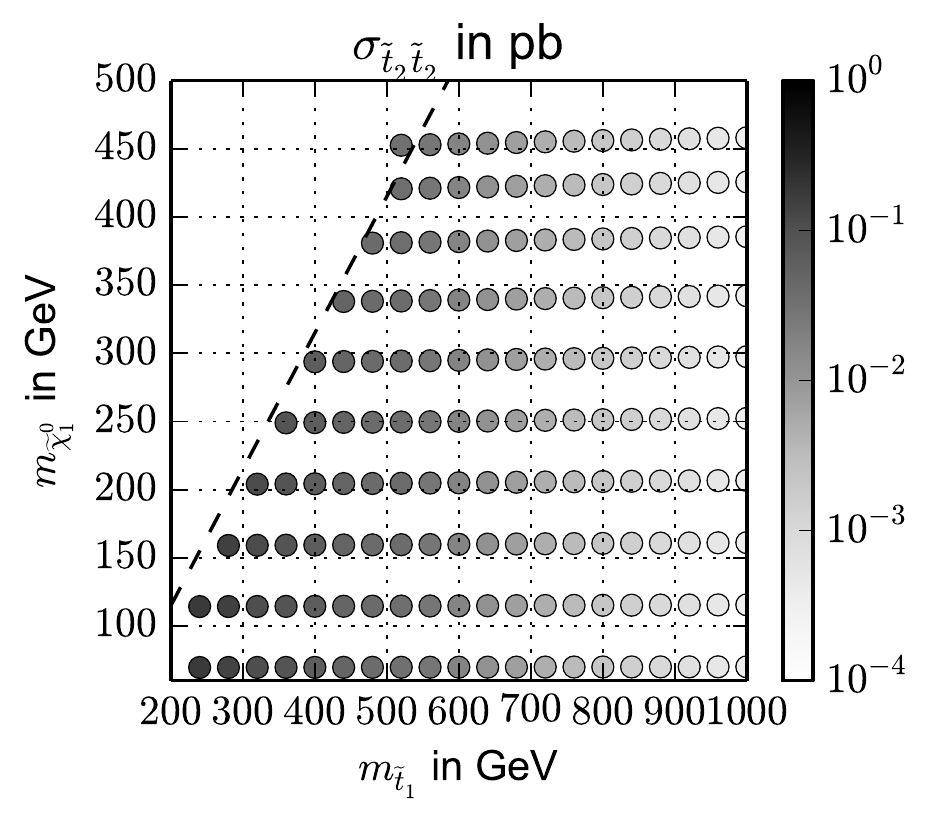}\qquad \qquad \qquad
\includegraphics[height=0.2\textheight]{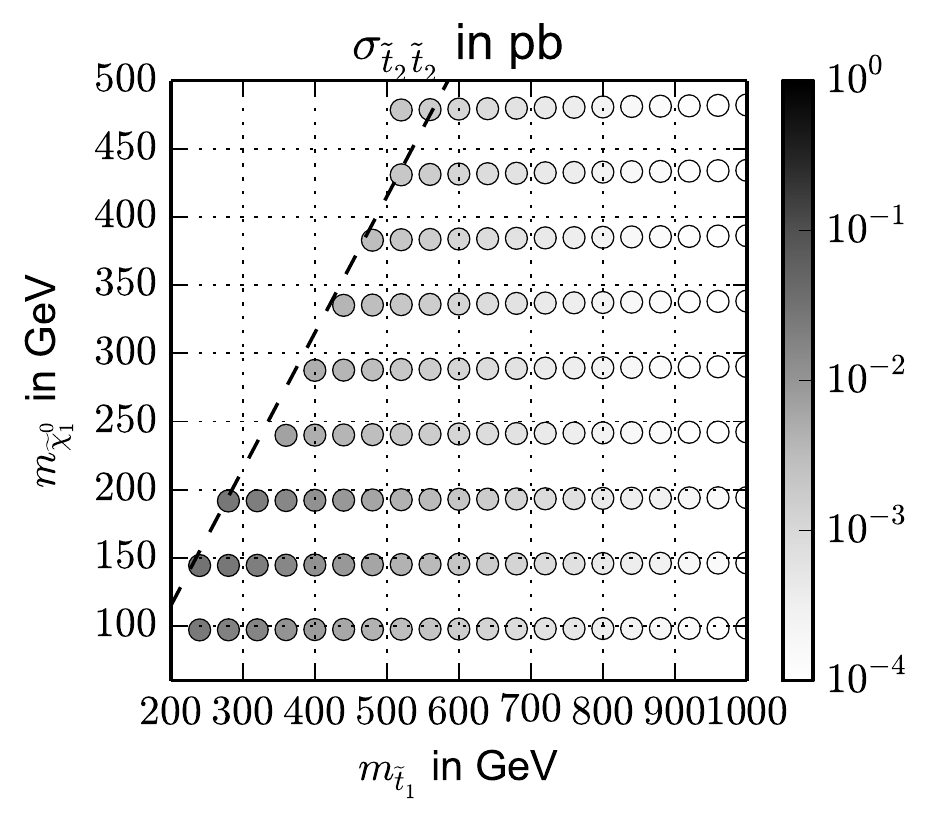}
\caption{Total production cross section for the third generation squarks for a decoupled gluino and  $m_{\widetilde{\chi}_1^\pm} = 500$ GeV. Left: $\lambda_L$. Right: $\lambda_S$}
\label{fig:stopplotstopxsect}
\end{figure}
\vfill
\pagebreak
\section{$\tilde{t}_1$ Branching Ratio Distributions}
\begin{figure}[h!]
\includegraphics[height=0.2\textheight]{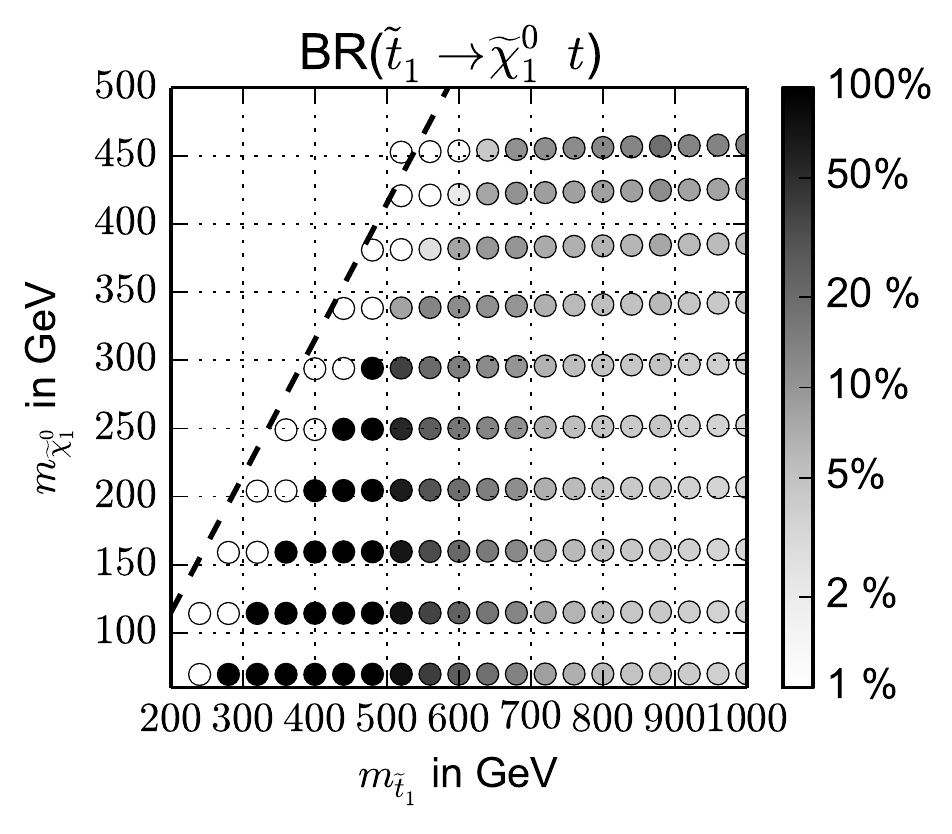}\qquad \qquad \qquad
\includegraphics[height=0.2\textheight]{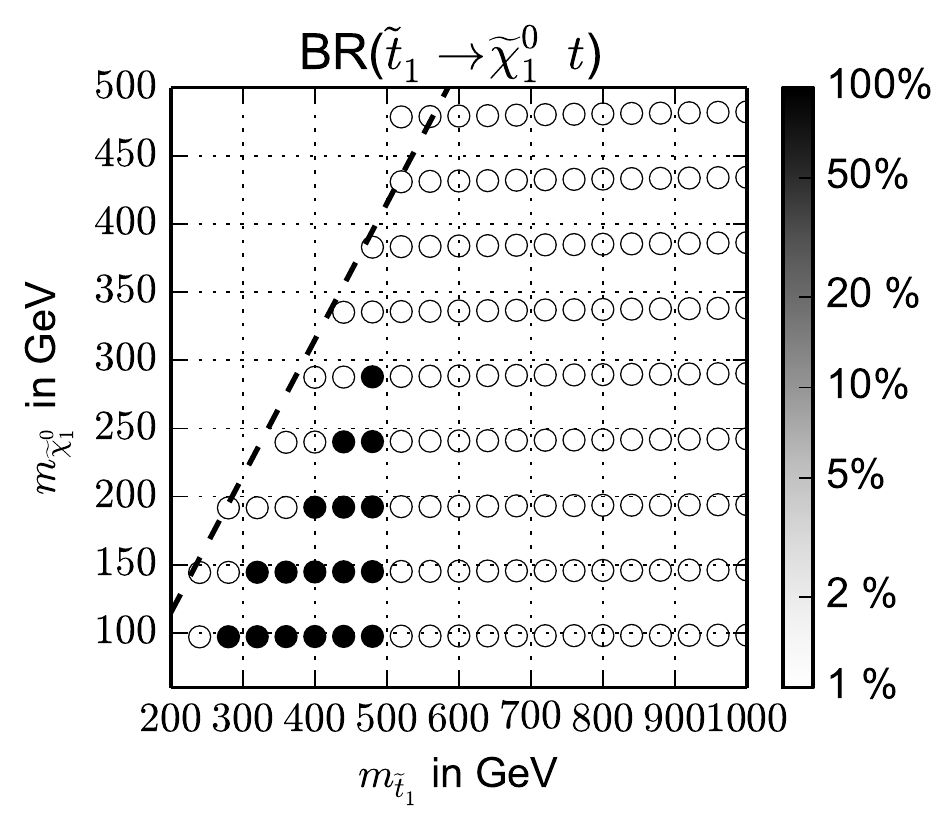}
\includegraphics[height=0.2\textheight]{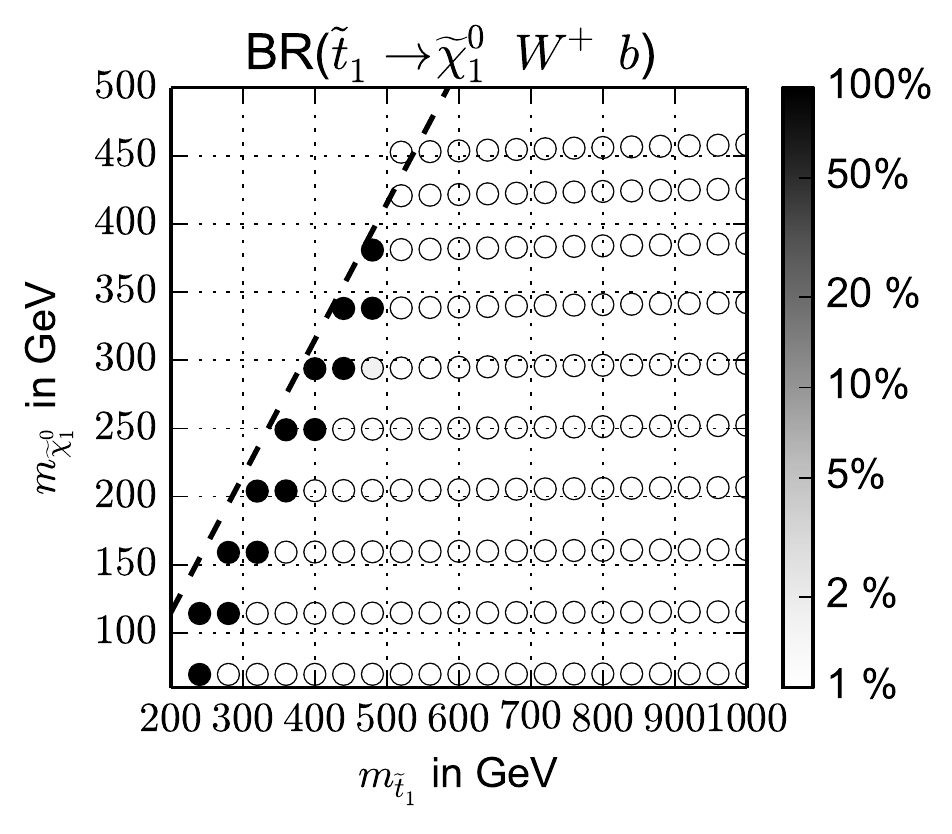}\qquad \qquad \qquad
\includegraphics[height=0.2\textheight]{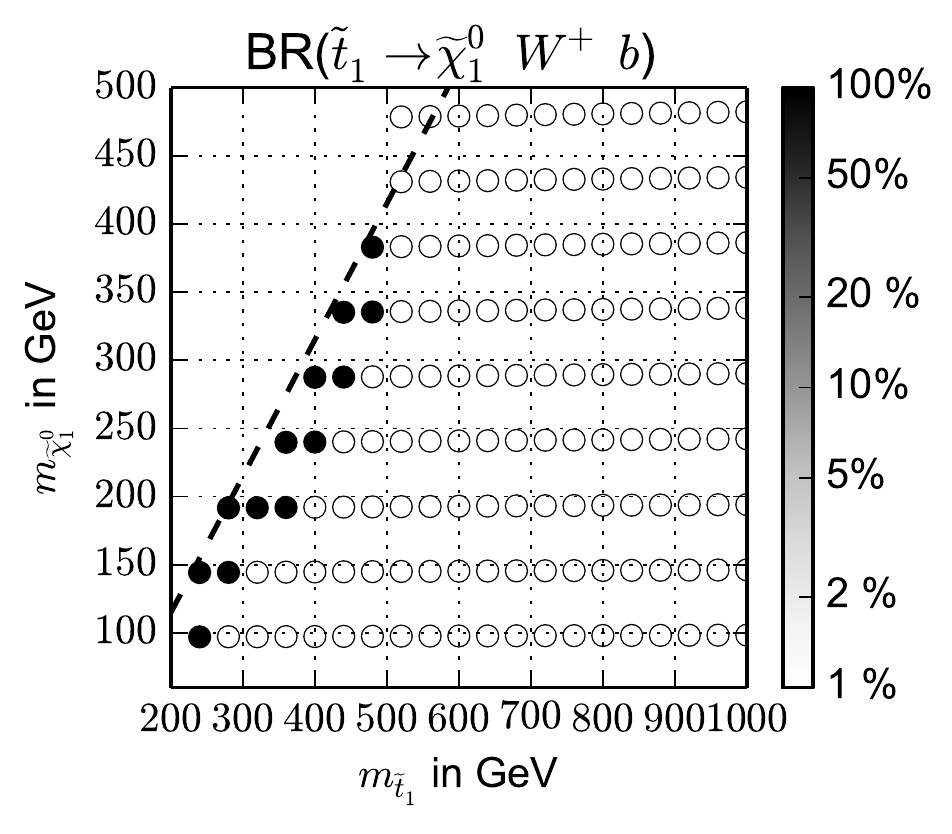}
\includegraphics[height=0.2\textheight]{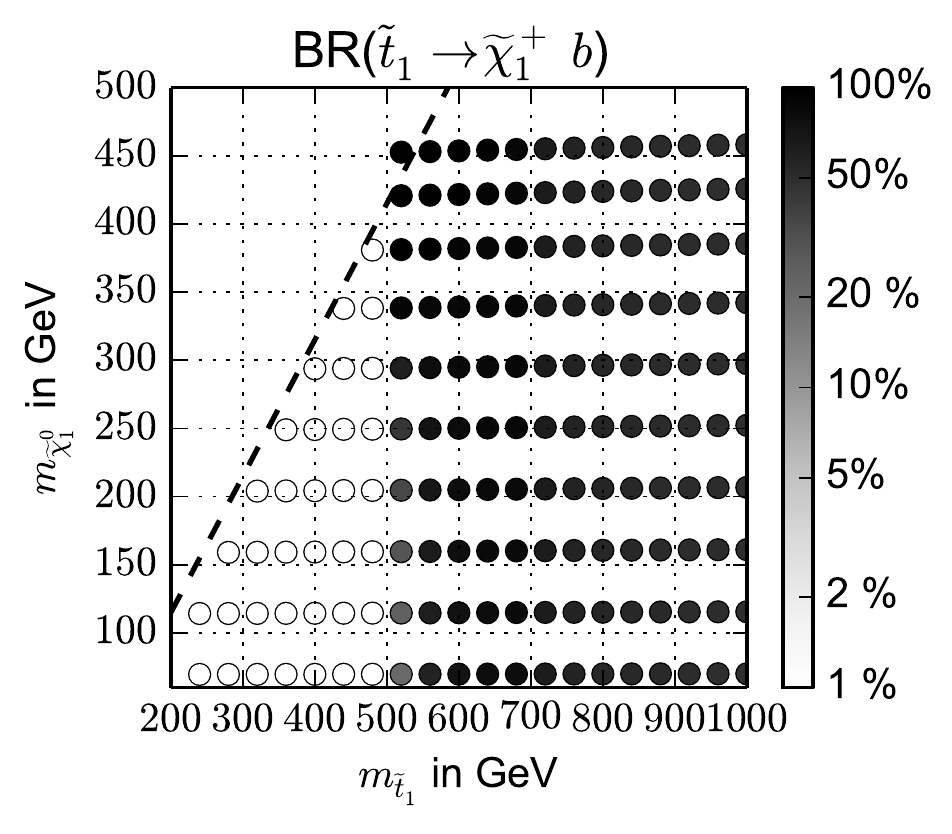}\qquad \qquad \qquad
\includegraphics[height=0.2\textheight]{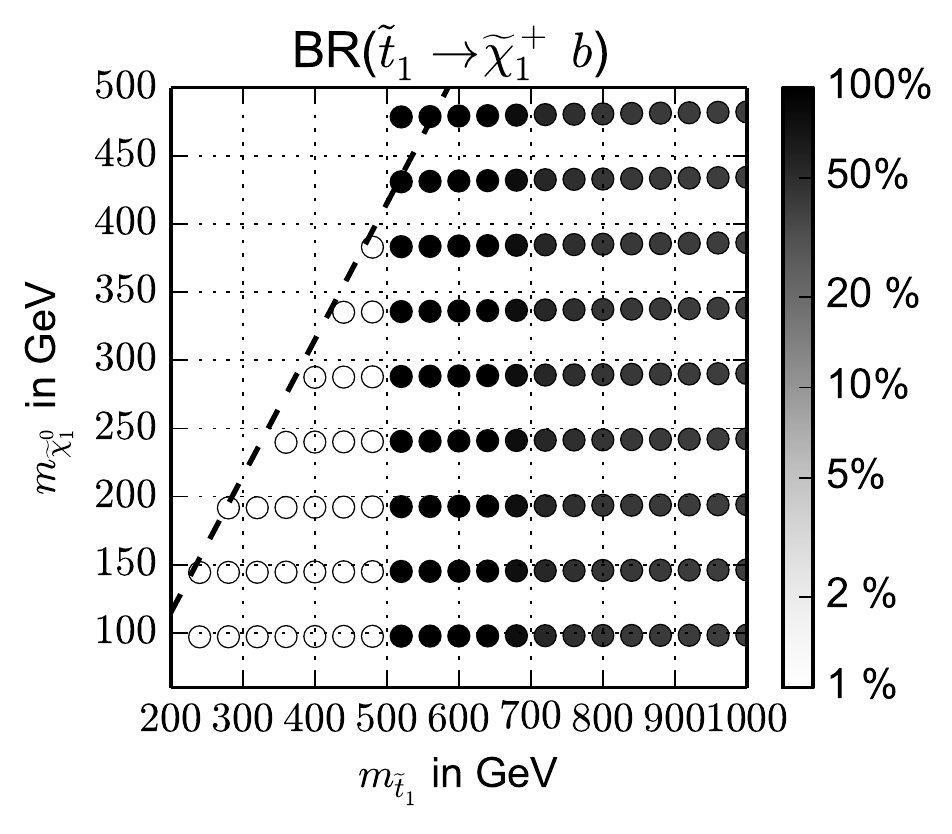}
\includegraphics[height=0.2\textheight]{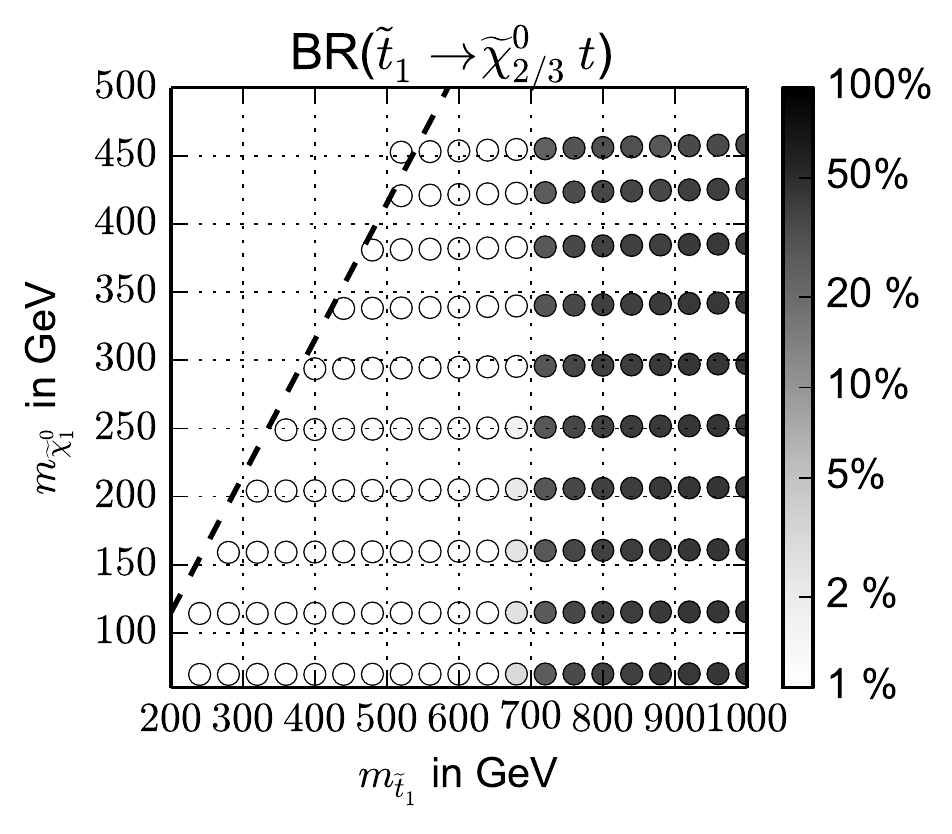}\qquad \qquad \qquad
\includegraphics[height=0.2\textheight]{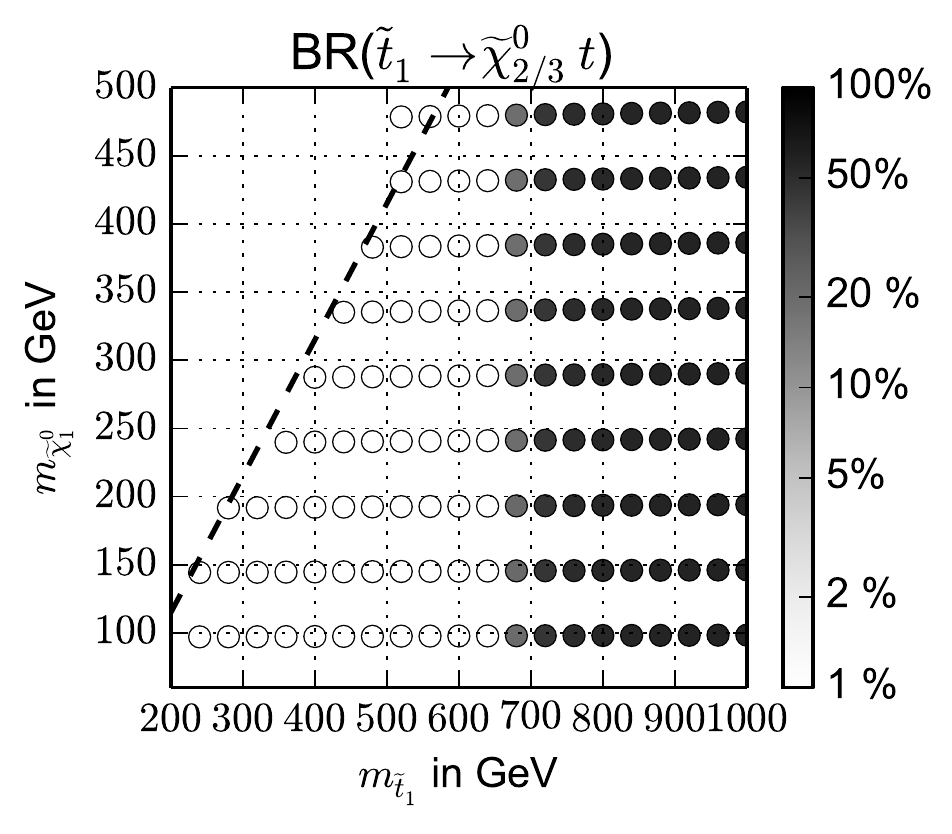}
\caption{Most significant branching ratios of the lightest stop into the the singlino LSP, the higgsino NLSPs and the chargino for a decoupled gluino and  $m_{\widetilde{\chi}_1^\pm} = 500$ GeV. Left: $\lambda_L$. Right: $\lambda_S$}
\label{fig:stopplotbrstoplsp}
\end{figure}
\vfill
\pagebreak
\begin{figure}[h!]
\includegraphics[height=0.2\textheight]{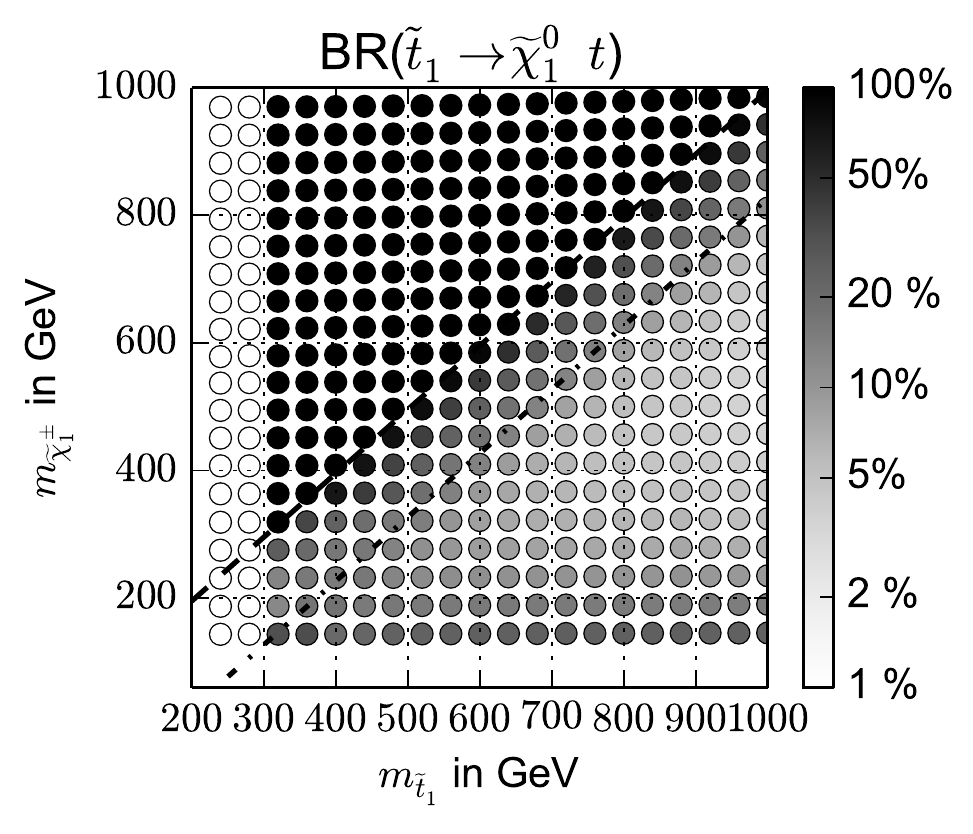}\qquad \qquad \qquad
\includegraphics[height=0.2\textheight]{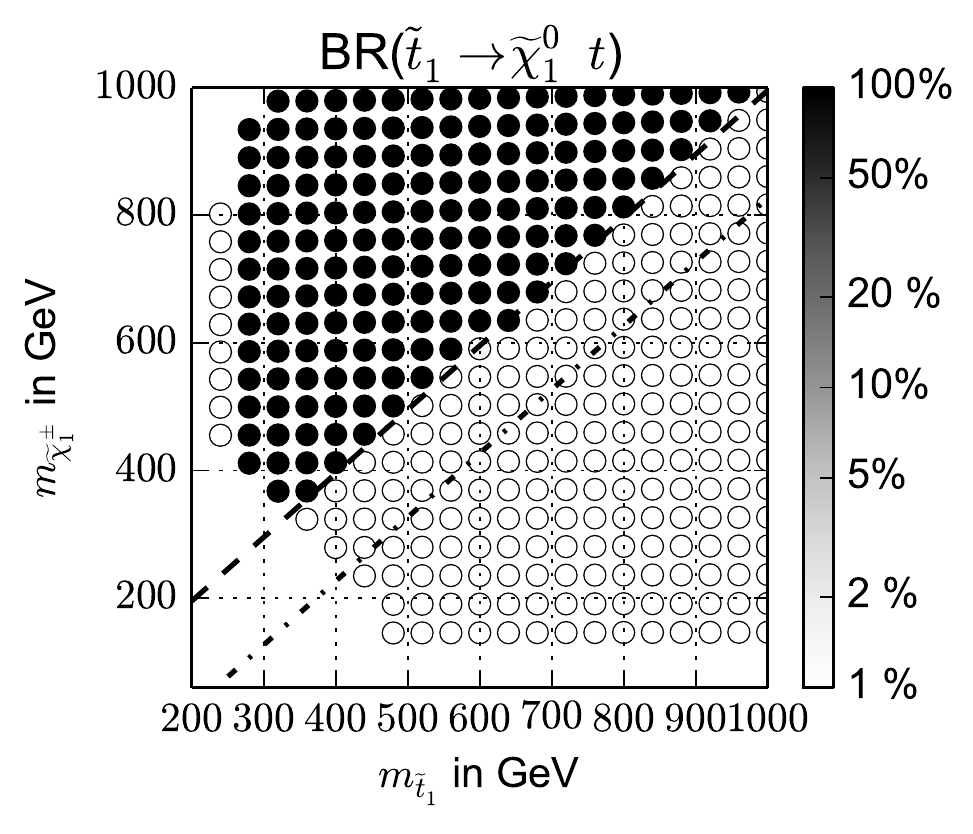}
\includegraphics[height=0.2\textheight]{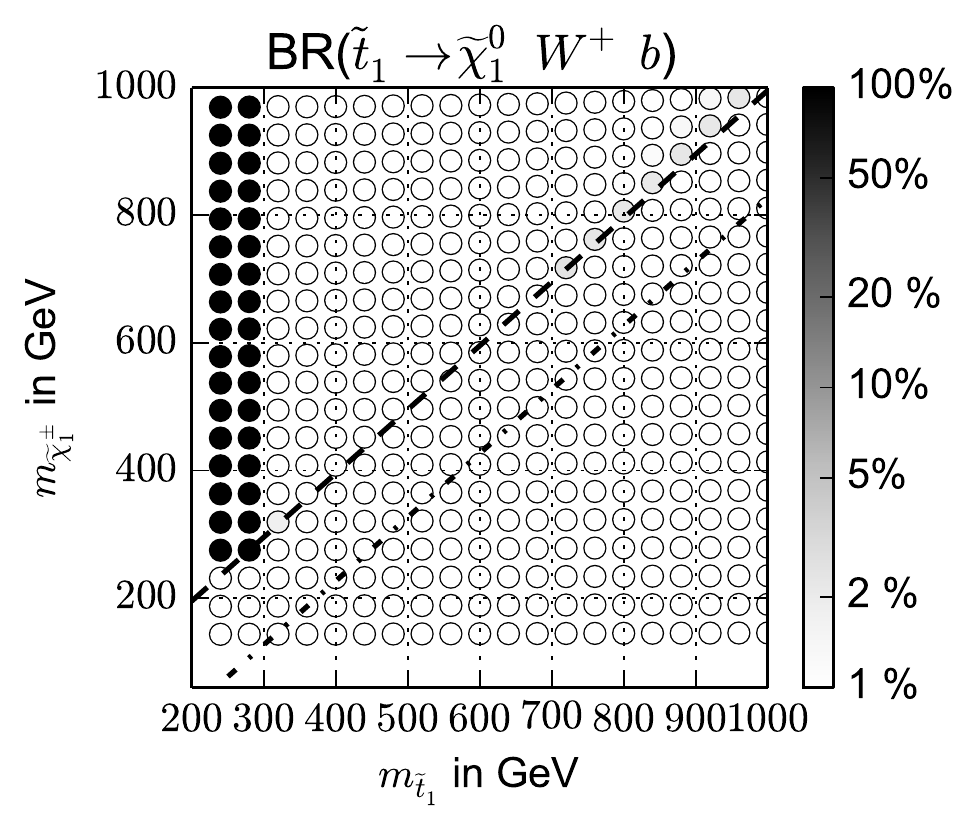}\qquad \qquad \qquad
\includegraphics[height=0.2\textheight]{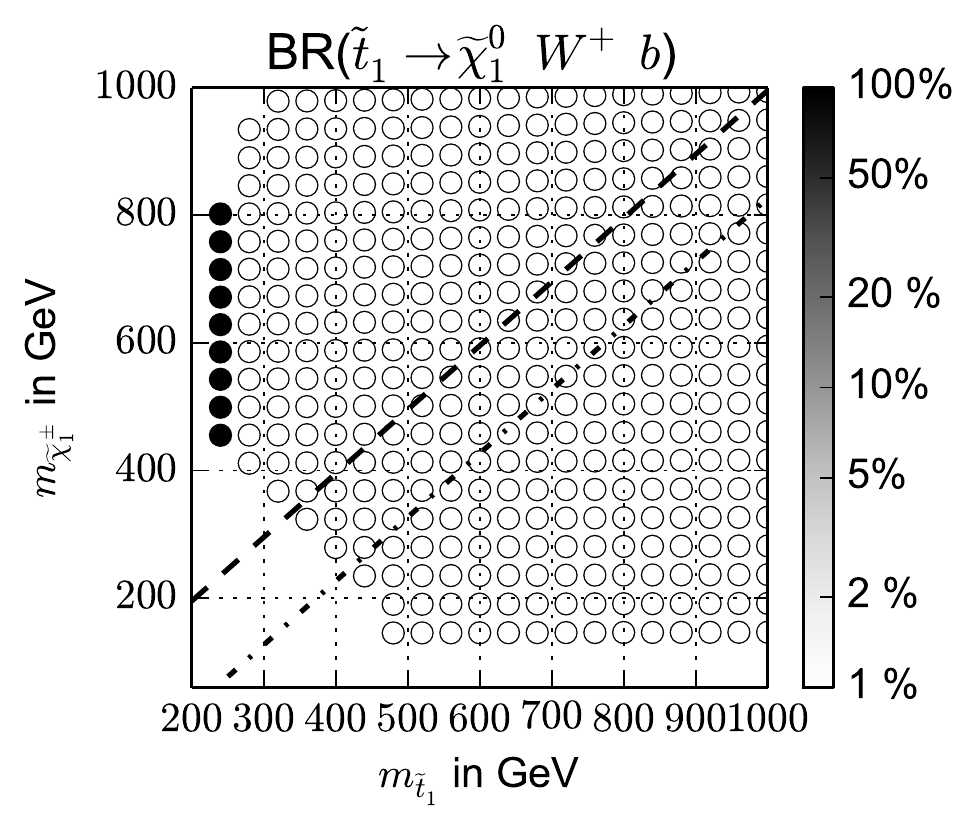}
\includegraphics[height=0.2\textheight]{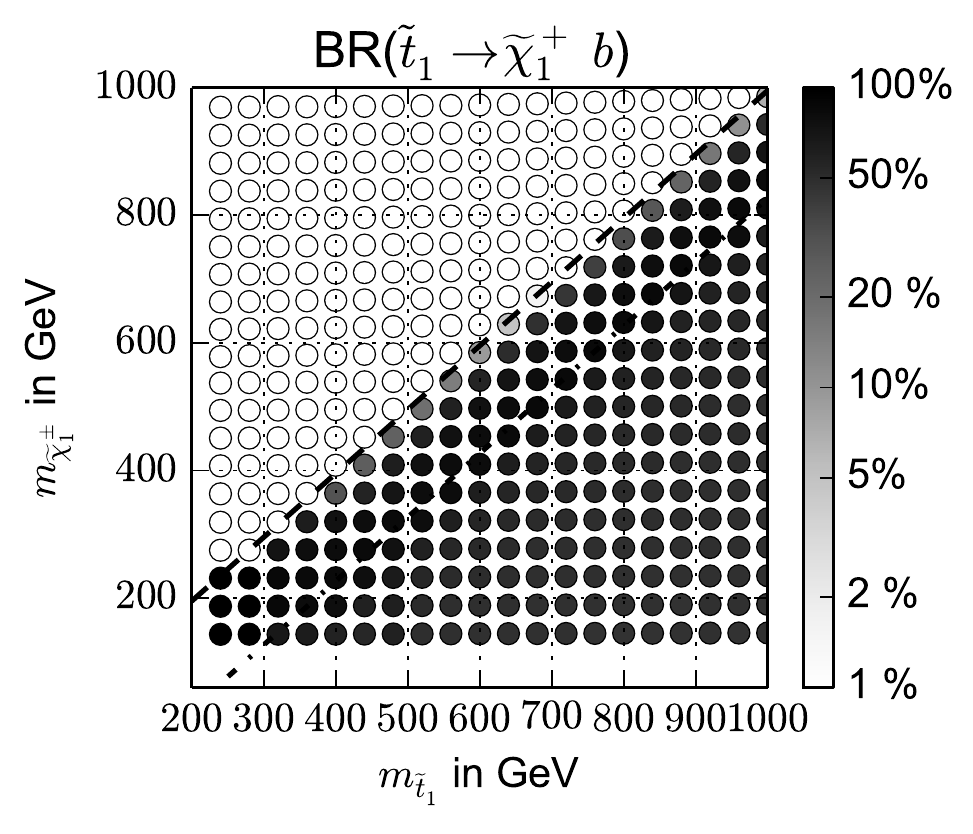}\qquad \qquad \qquad
\includegraphics[height=0.2\textheight]{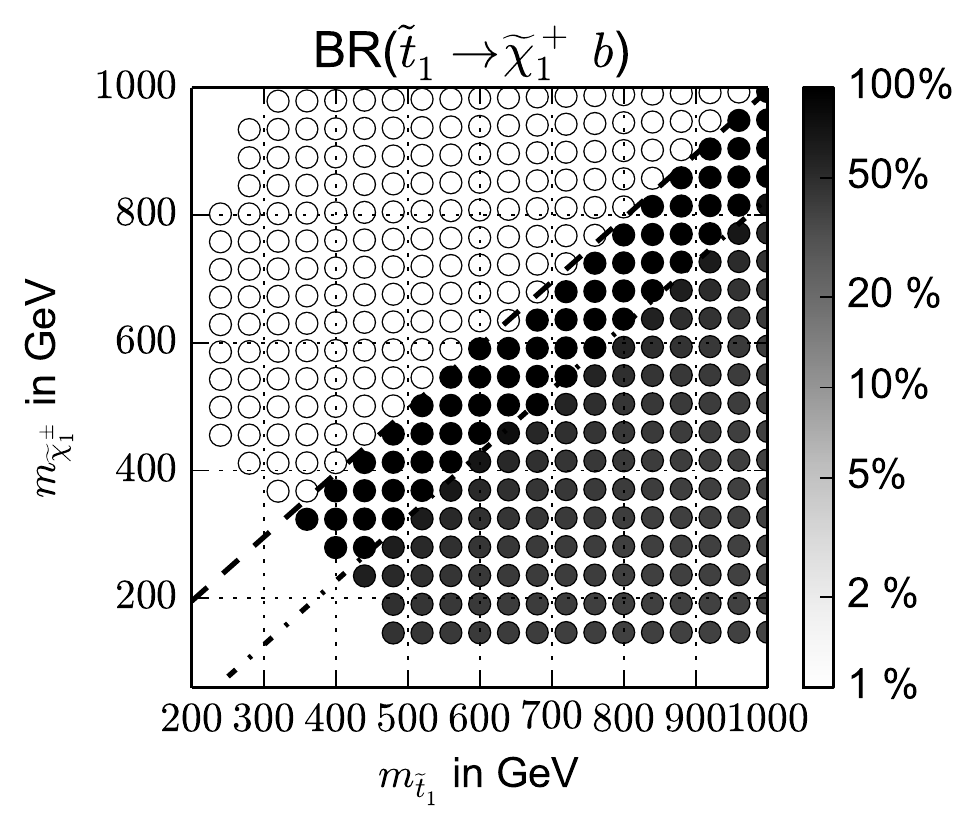}
\includegraphics[height=0.2\textheight]{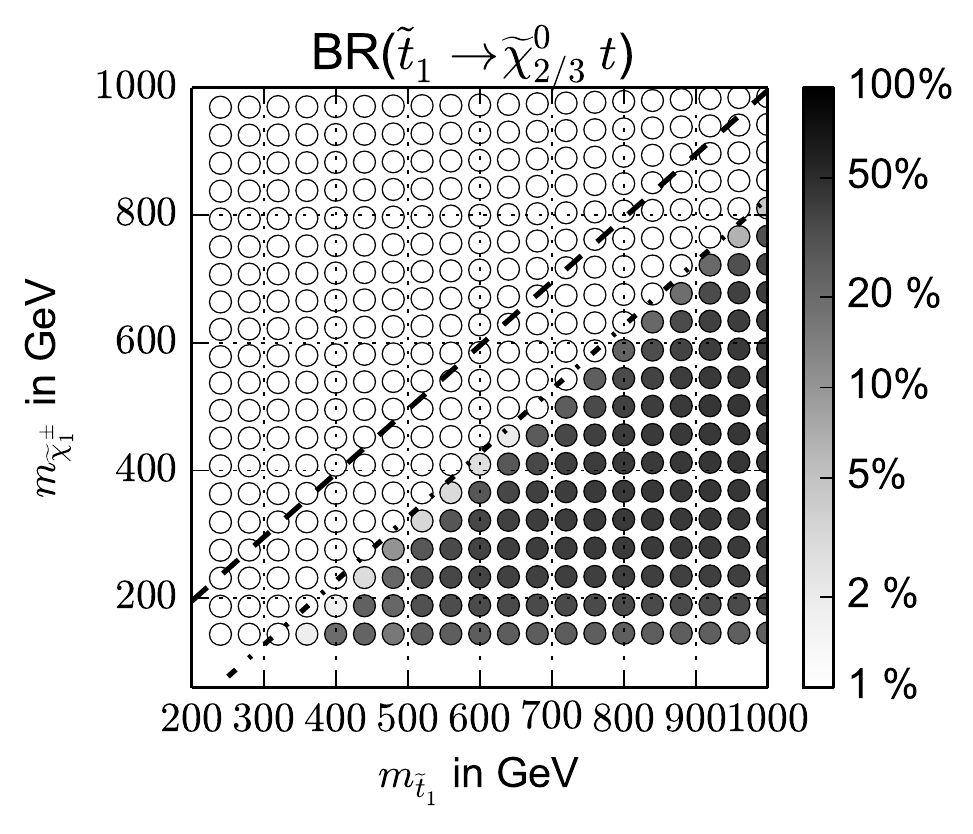}\qquad \qquad \qquad
\includegraphics[height=0.2\textheight]{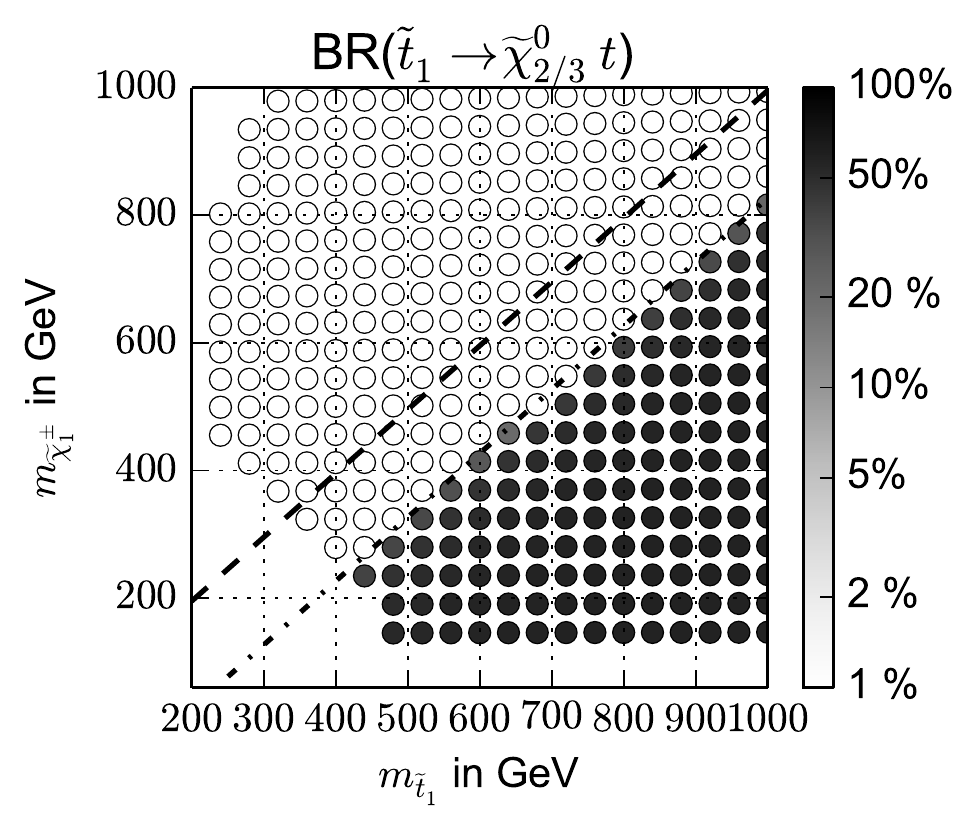}
\caption{Most significant branching ratios of the lightest stop into the the singlino LSP, the higgsino NLSPs and the chargino for a decoupled gluino and  $m_{\widetilde{\chi}^0_1} = 100$ GeV. Left: $\lambda_L$. Right: $\lambda_S$}
\label{fig:stopplotbrstoplsp2}
\end{figure}
\vfill
\pagebreak
\section{$\tilde{b}_{1/2}$ Branching Ratio Distributions}
\begin{figure}[h!]
\includegraphics[height=0.2\textheight]{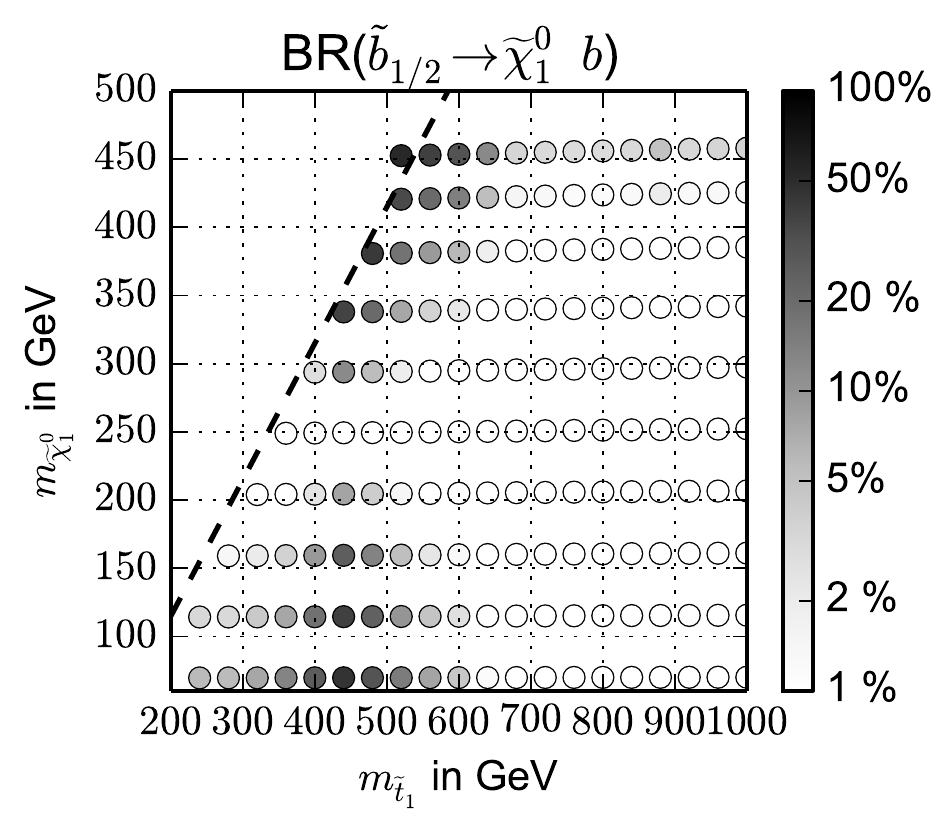} \qquad \qquad \qquad
\includegraphics[height=0.2\textheight]{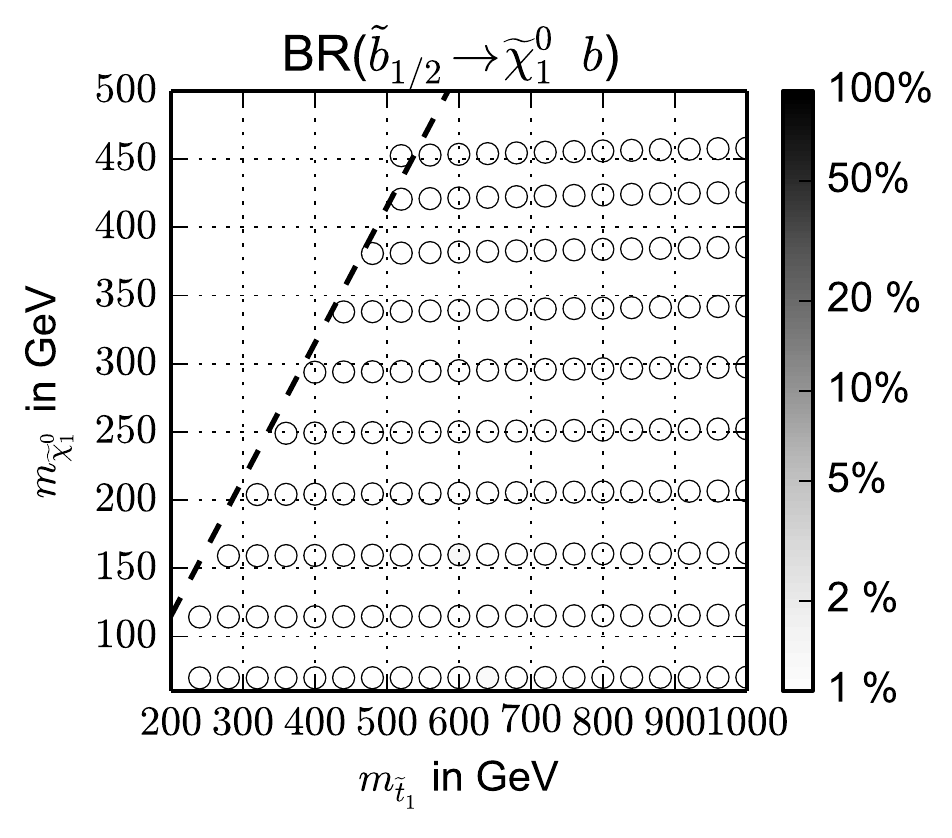} 
\includegraphics[height=0.2\textheight]{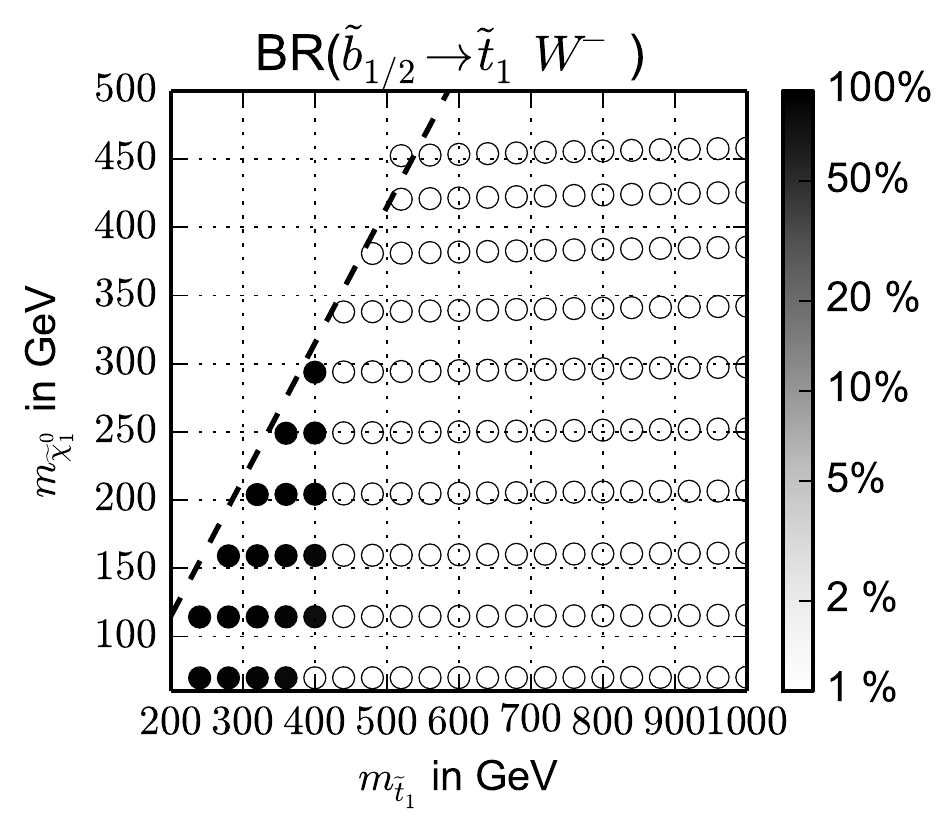}\qquad \qquad \qquad
\includegraphics[height=0.2\textheight]{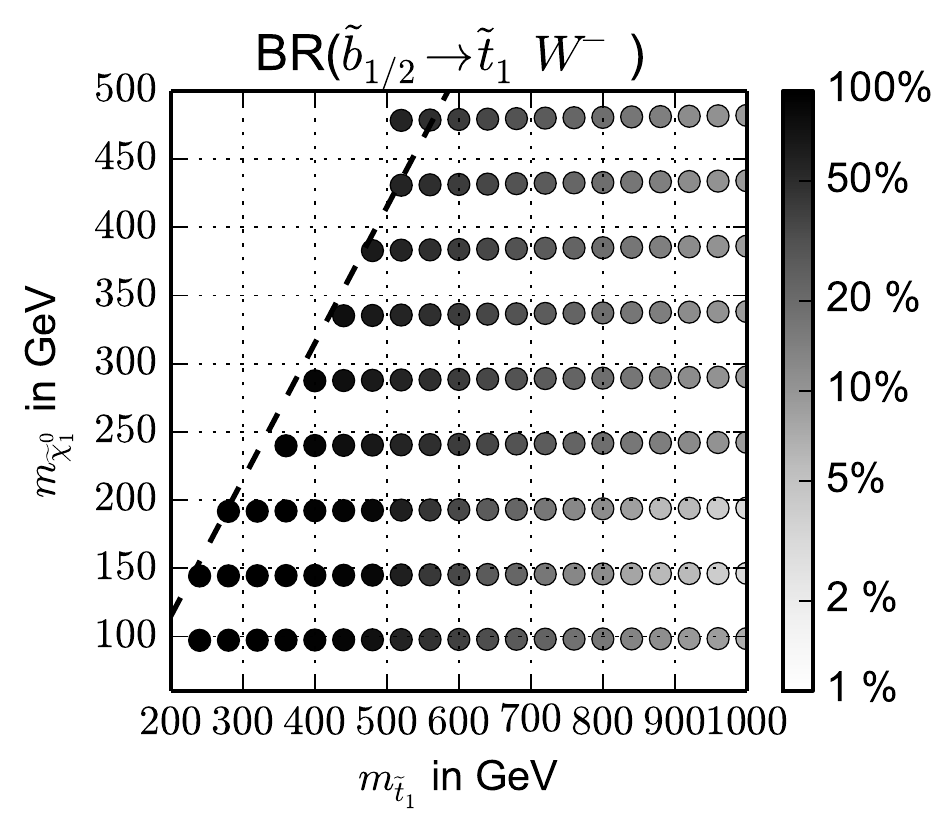}
\includegraphics[height=0.2\textheight]{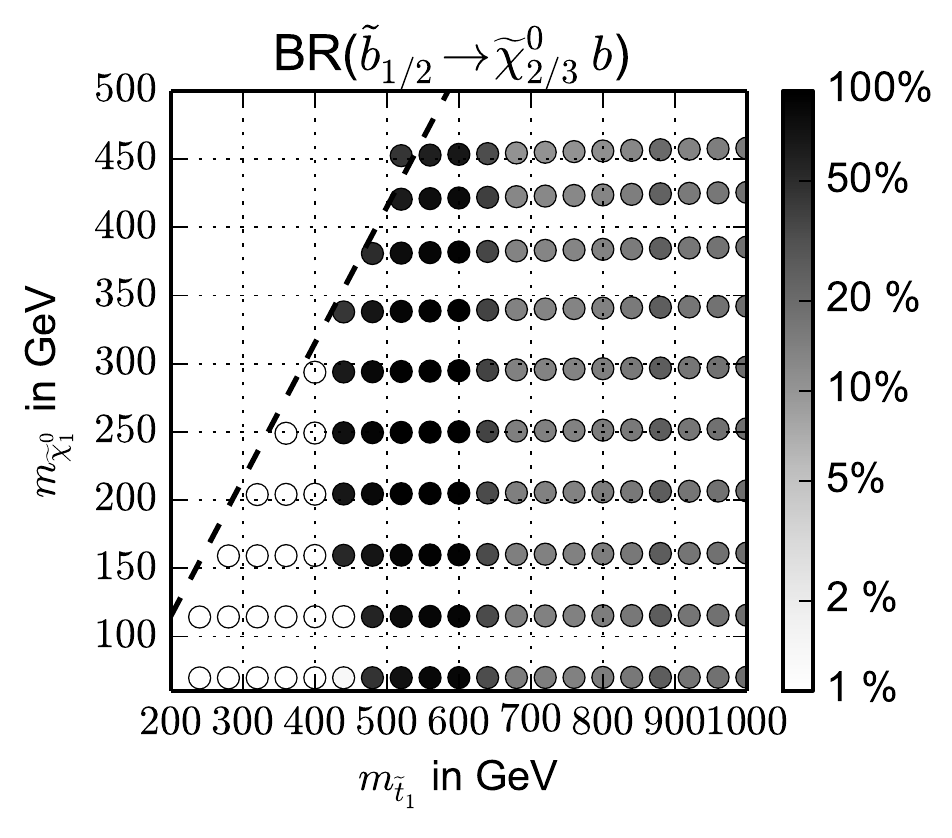}\qquad \qquad \qquad
\includegraphics[height=0.2\textheight]{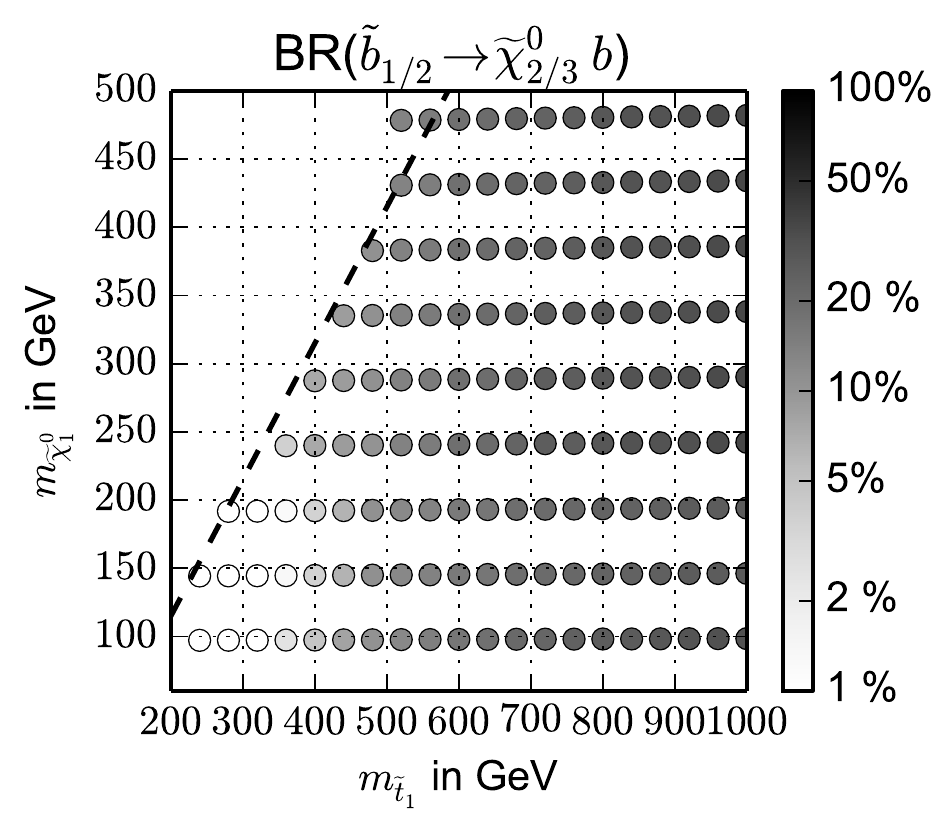}
\includegraphics[height=0.2\textheight]{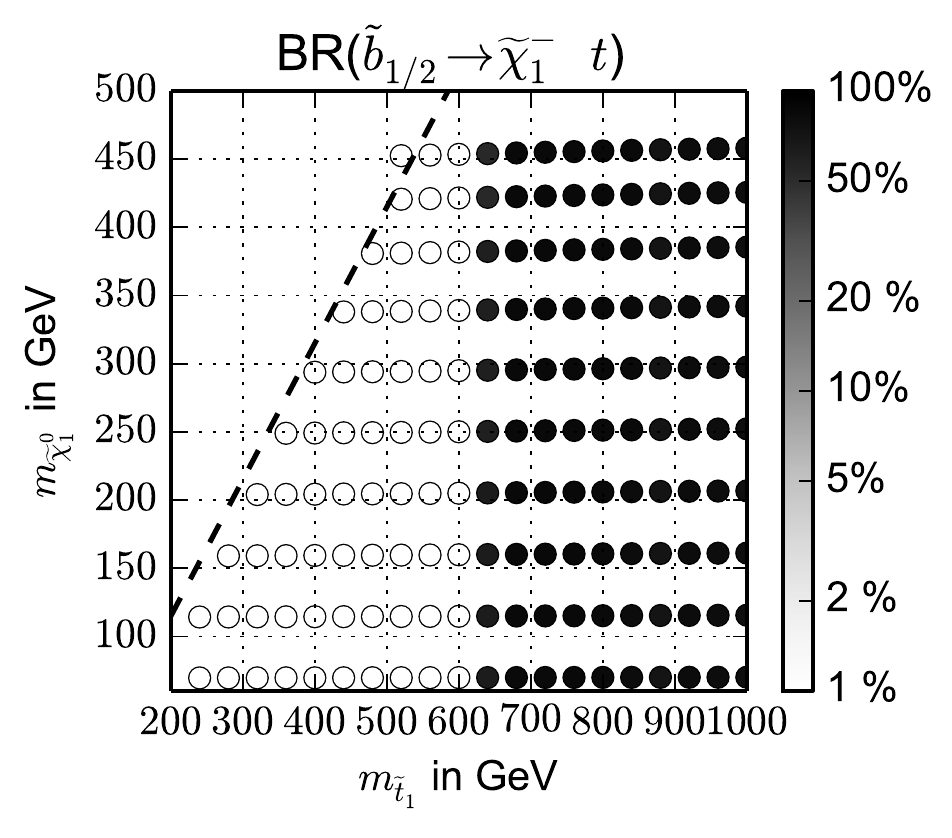}\qquad \qquad \qquad
\includegraphics[height=0.2\textheight]{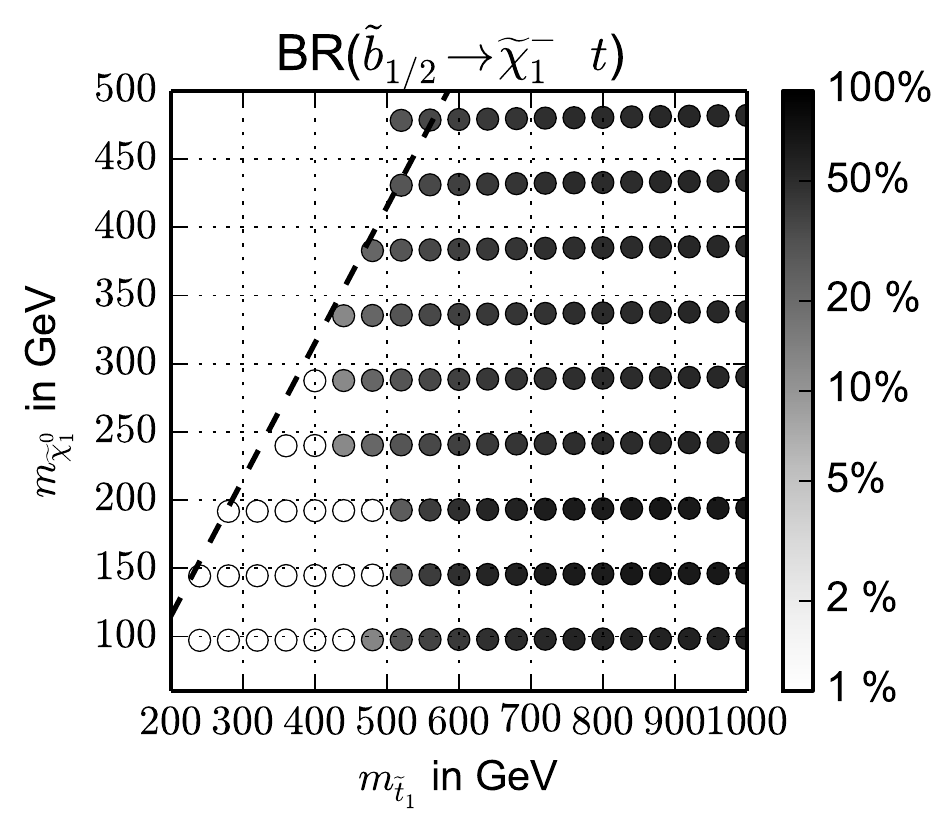}
\caption{Most significant branching ratios of the (mostly degenerate) sbottoms into the the lightest stop, the higgsino NLSPs and the chargino for a decoupled gluino and  $m_{\widetilde{\chi}_1^\pm} = 500$ GeV. Left: $\lambda_L$. Right: $\lambda_S$}
\label{fig:sbottomplotbrstoplsp}
\end{figure}
\vfill
\pagebreak
\begin{figure}[h!]
\includegraphics[height=0.2\textheight]{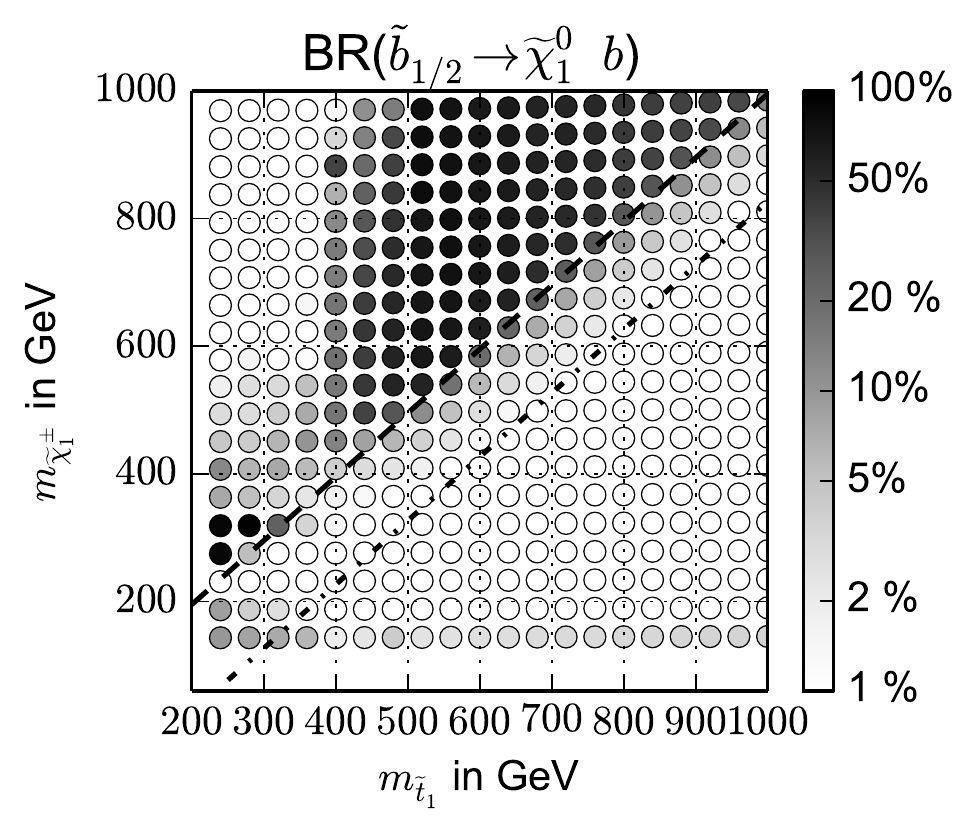} \qquad \qquad \qquad
\includegraphics[height=0.2\textheight]{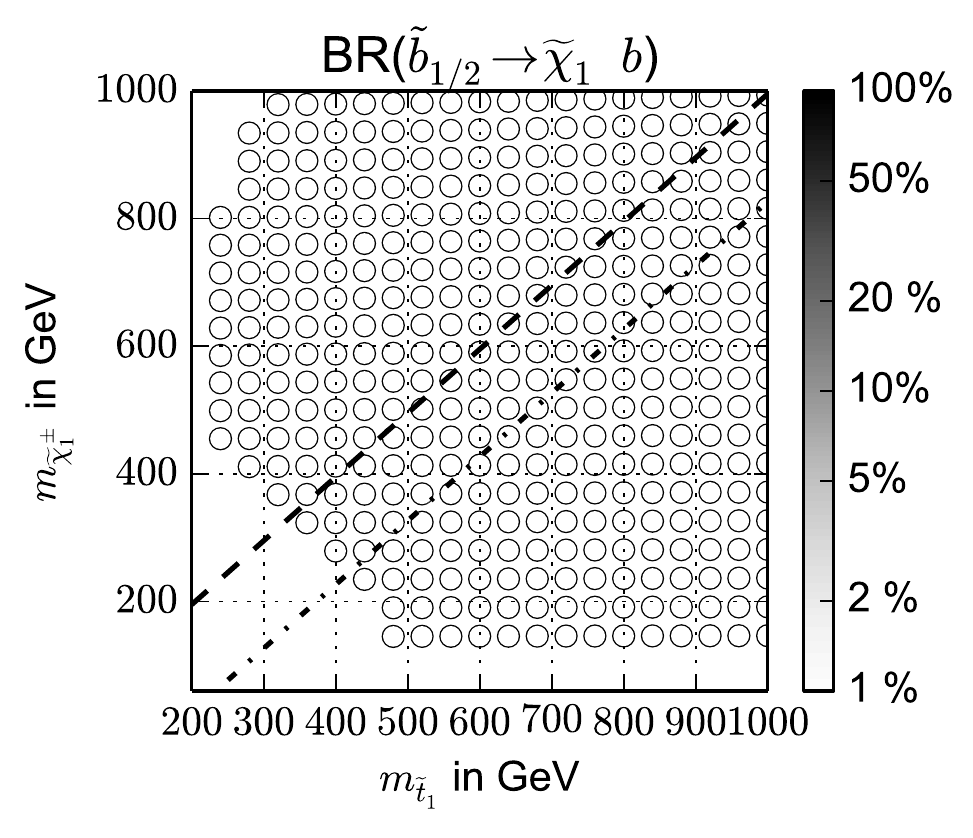} 
\includegraphics[height=0.2\textheight]{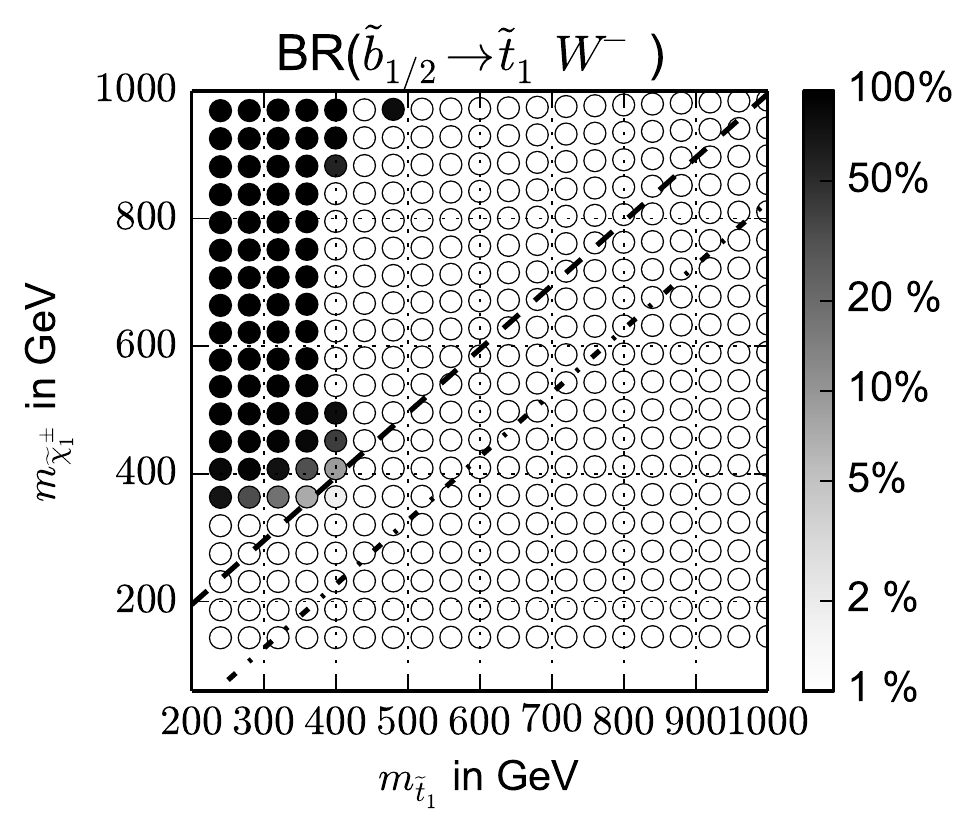}\qquad \qquad \qquad
\includegraphics[height=0.2\textheight]{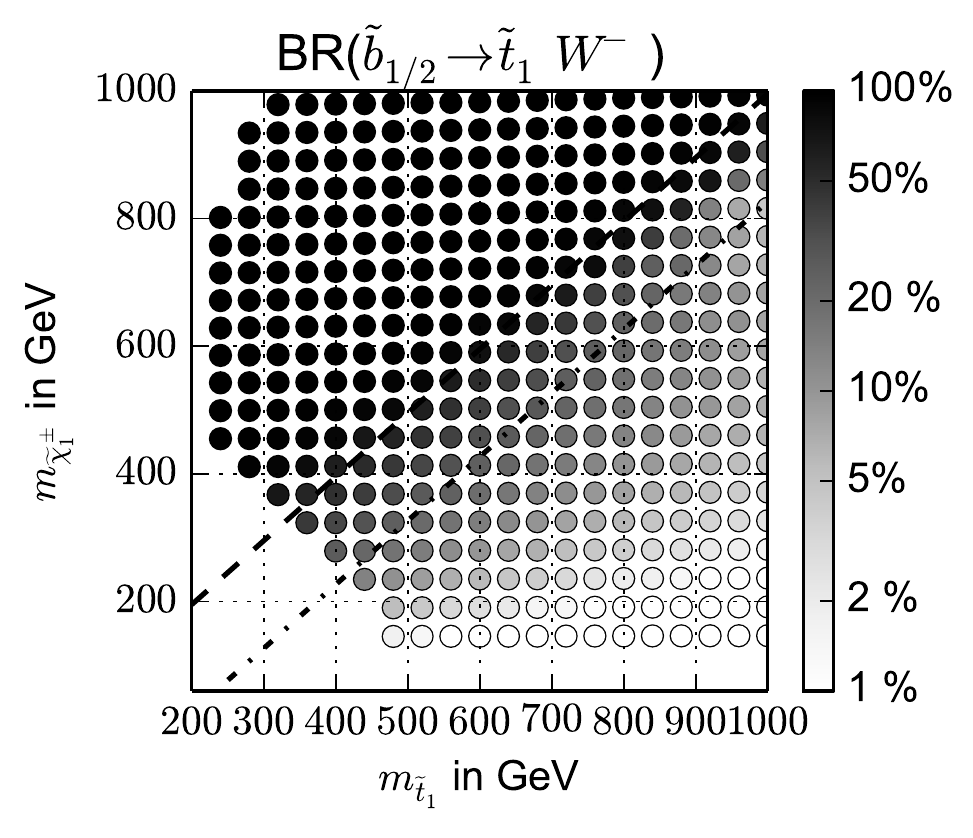}
\includegraphics[height=0.2\textheight]{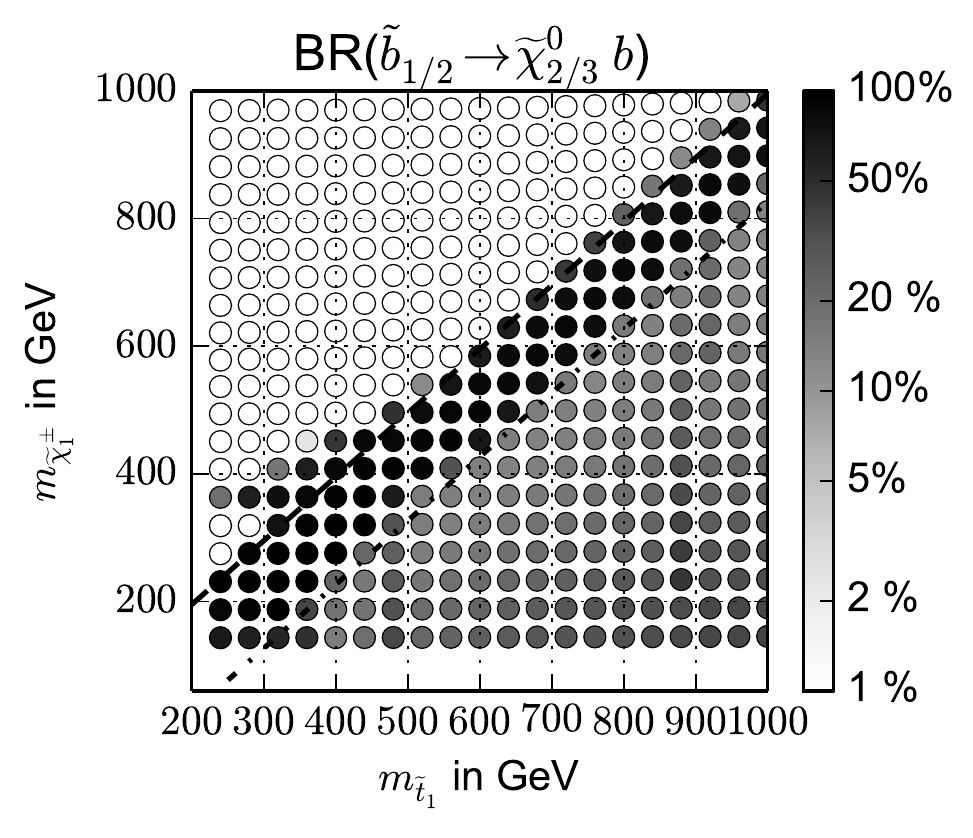}\qquad \qquad \qquad
\includegraphics[height=0.2\textheight]{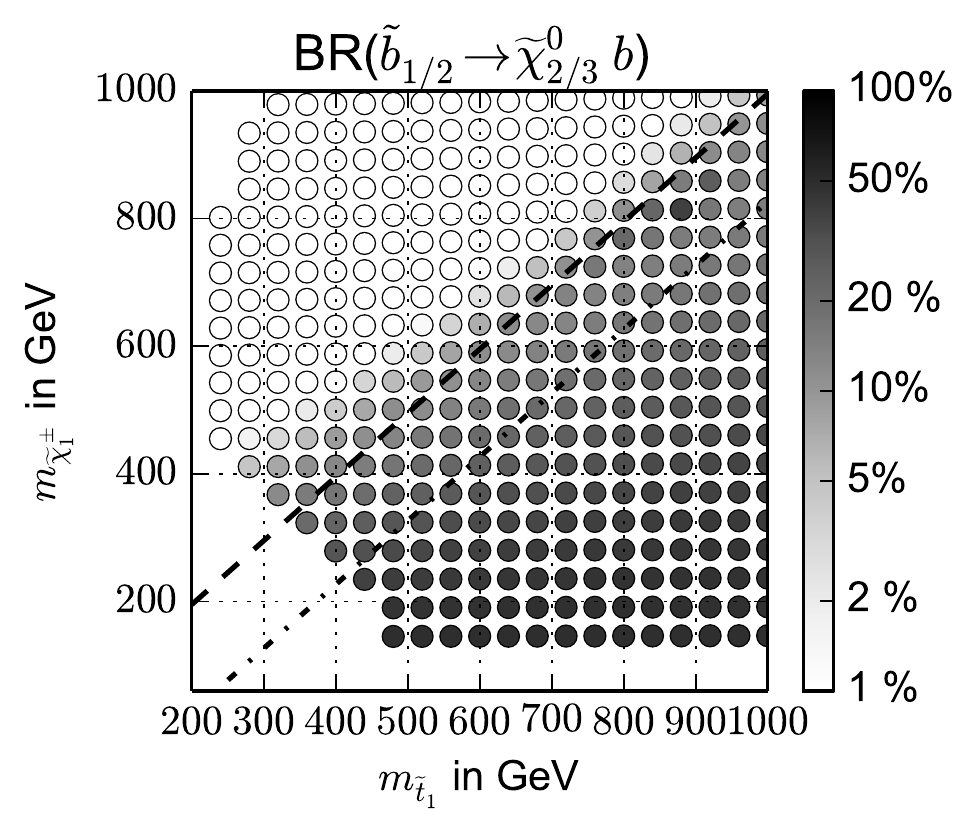}
\includegraphics[height=0.2\textheight]{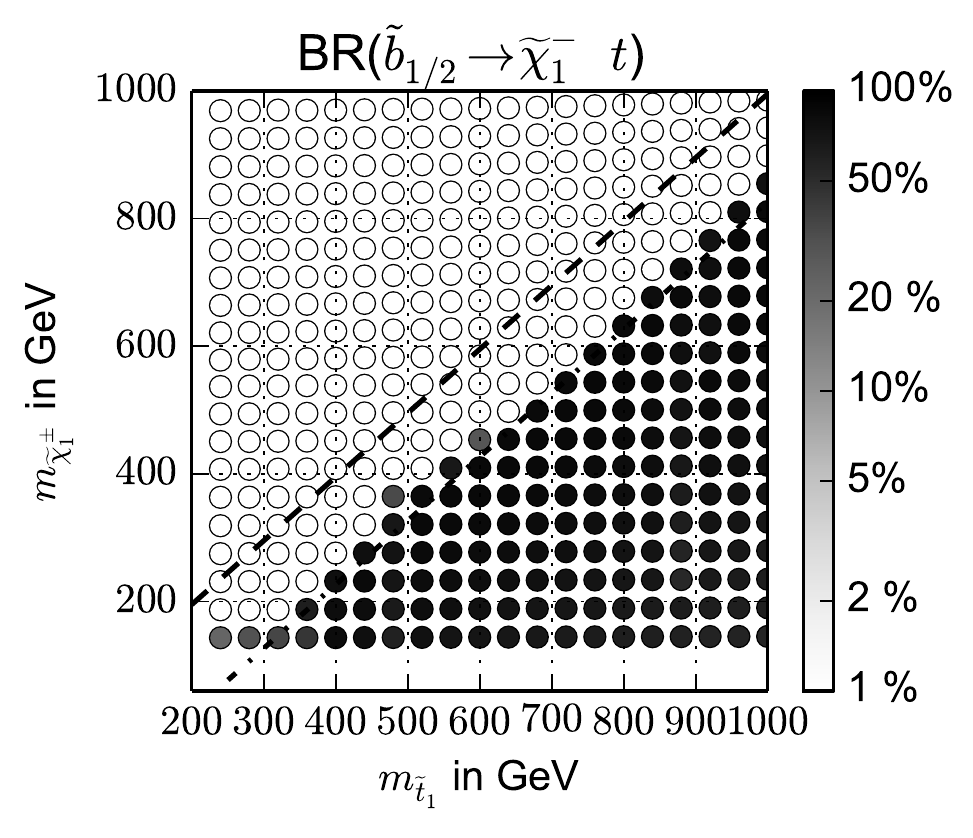}\qquad \qquad \qquad
\includegraphics[height=0.2\textheight]{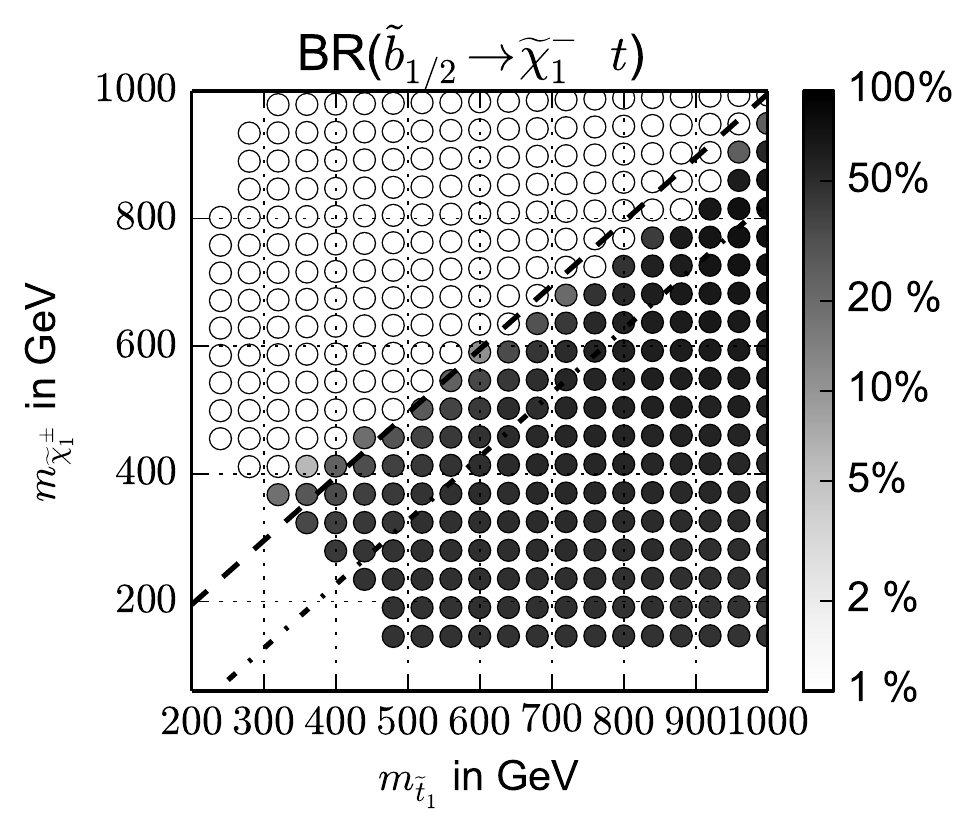}
\caption{Most significant branching ratios of the (mostly degenerate) sbottoms into the the lightest stop, the higgsino NLSPs and the chargino for a decoupled gluino and $m_{\widetilde{\chi}^0_1} = 100$ GeV. Left: $\lambda_L$. Right: $\lambda_S$}
\label{fig:sbottomplotbrstoplsp2}
\end{figure}
\vfill
\pagebreak
\section{$\tilde{t}_2$ Branching Ratio Distributions}
\begin{figure}[h!]
\includegraphics[height=0.2\textheight]{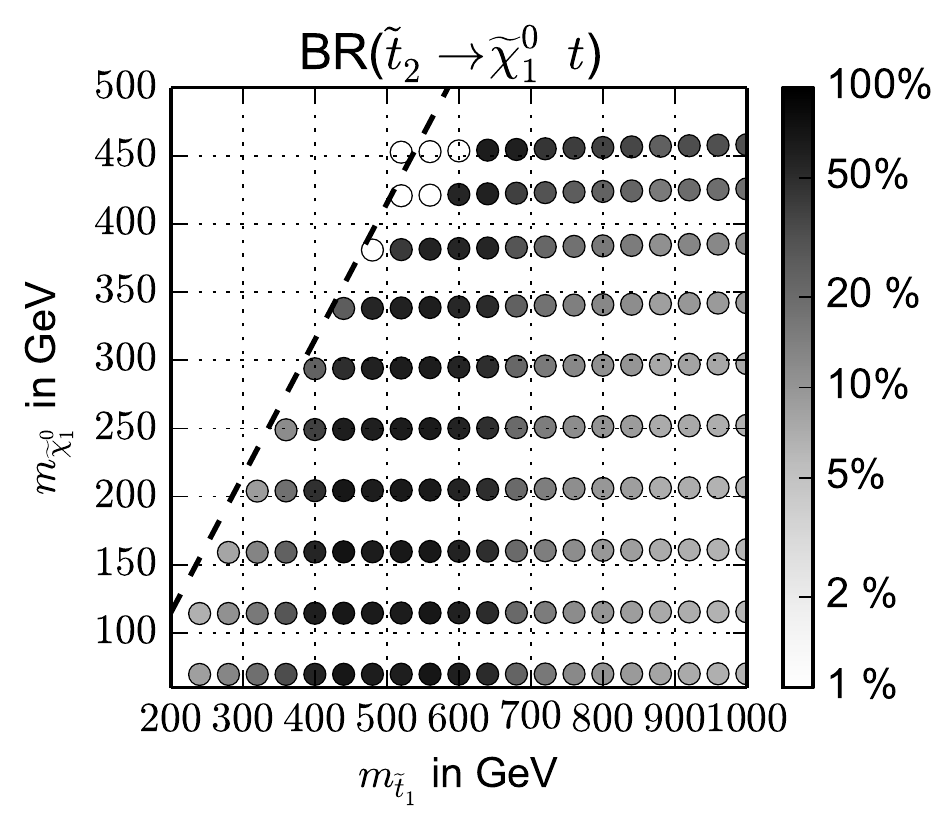}\qquad \qquad \qquad
\includegraphics[height=0.2\textheight]{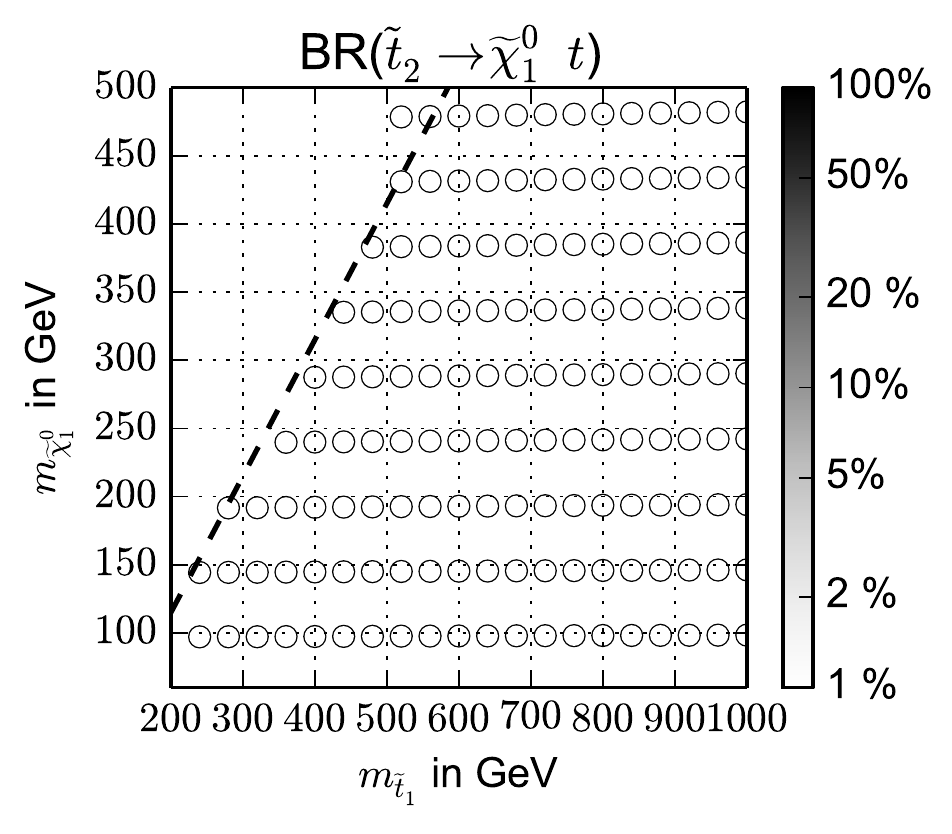}
\includegraphics[height=0.2\textheight]{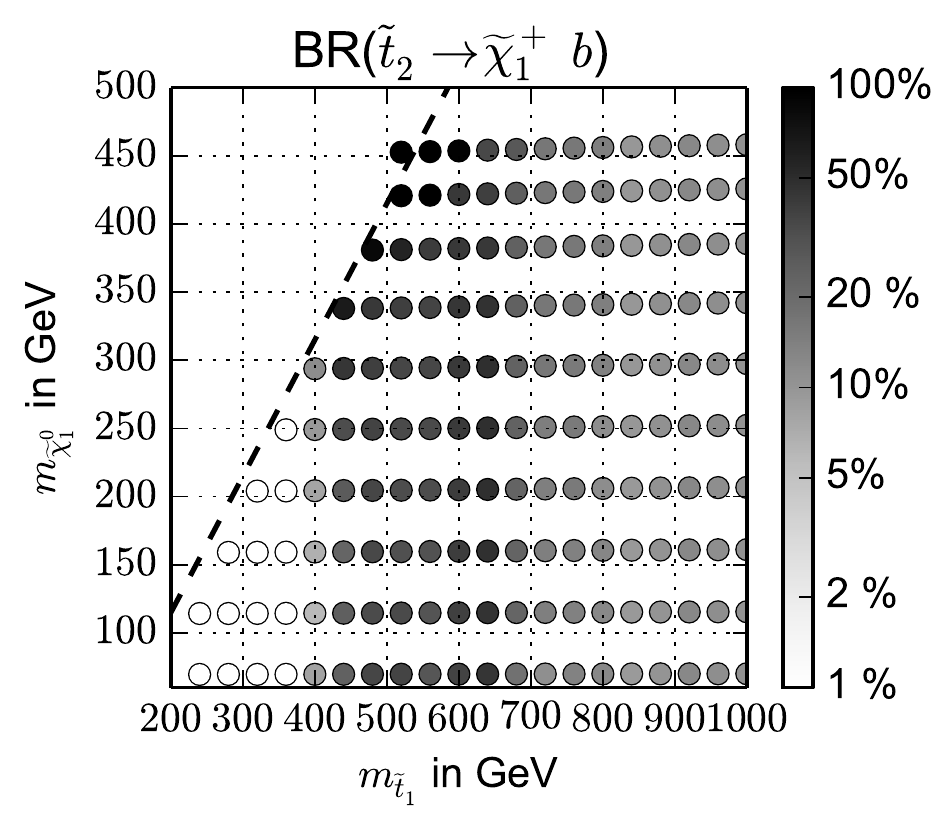}\qquad \qquad \qquad
\includegraphics[height=0.2\textheight]{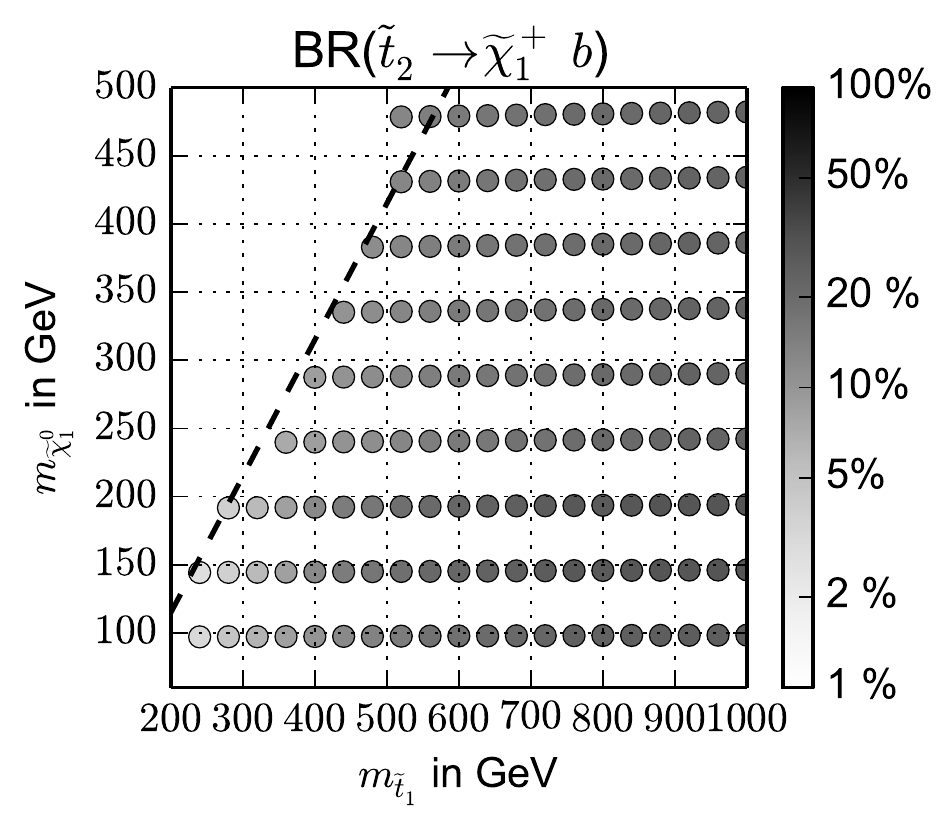}
\includegraphics[height=0.2\textheight]{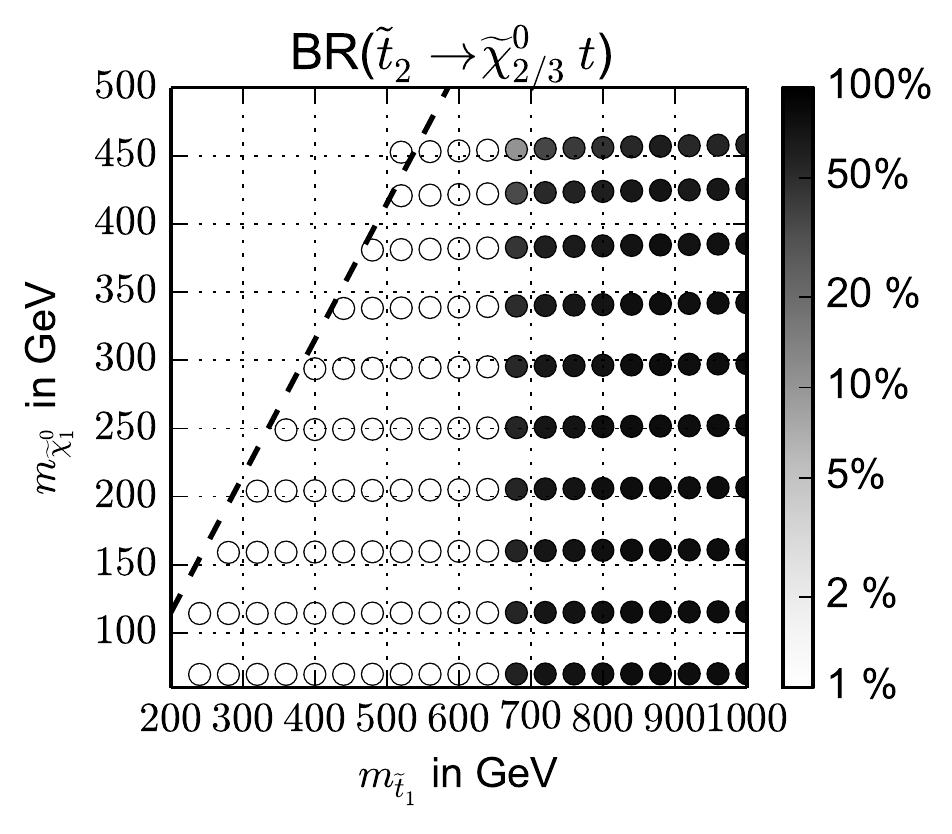}\qquad \qquad \qquad
\includegraphics[height=0.2\textheight]{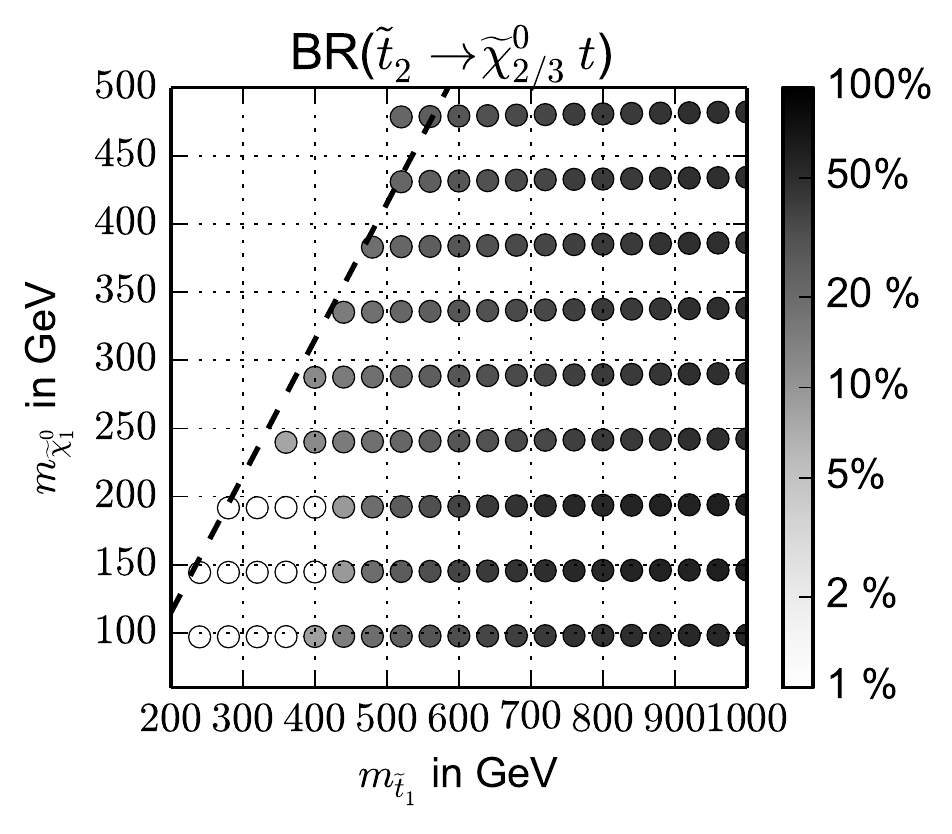}
\includegraphics[height=0.2\textheight]{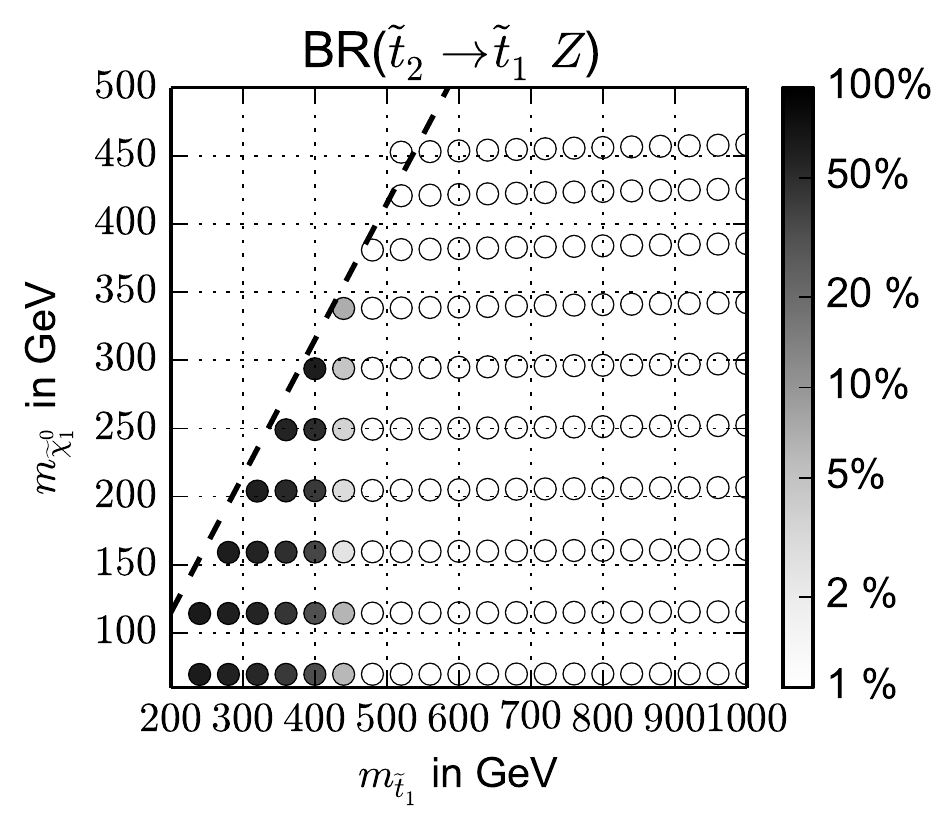}\qquad \qquad \qquad
\includegraphics[height=0.2\textheight]{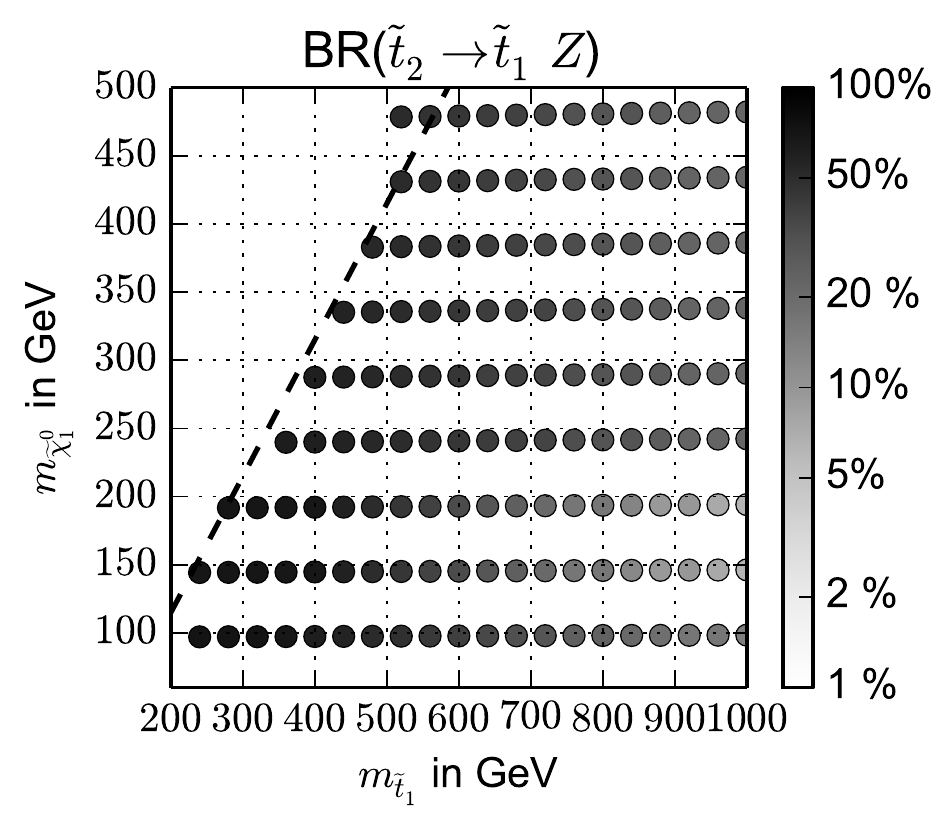}
\caption{Most significant branching ratios of the heavier stop into the the singlino LSP, the higgsino NLSPs and the chargino for a decoupled gluino and  $m_{\widetilde{\chi}_1^\pm} = 500$ GeV. Left: $\lambda_L$. Right: $\lambda_S$}
\label{fig:stopplot2}
\vspace{-5cm}
\end{figure}
\vfill
\pagebreak
\begin{figure}[h!]
\includegraphics[height=0.2\textheight]{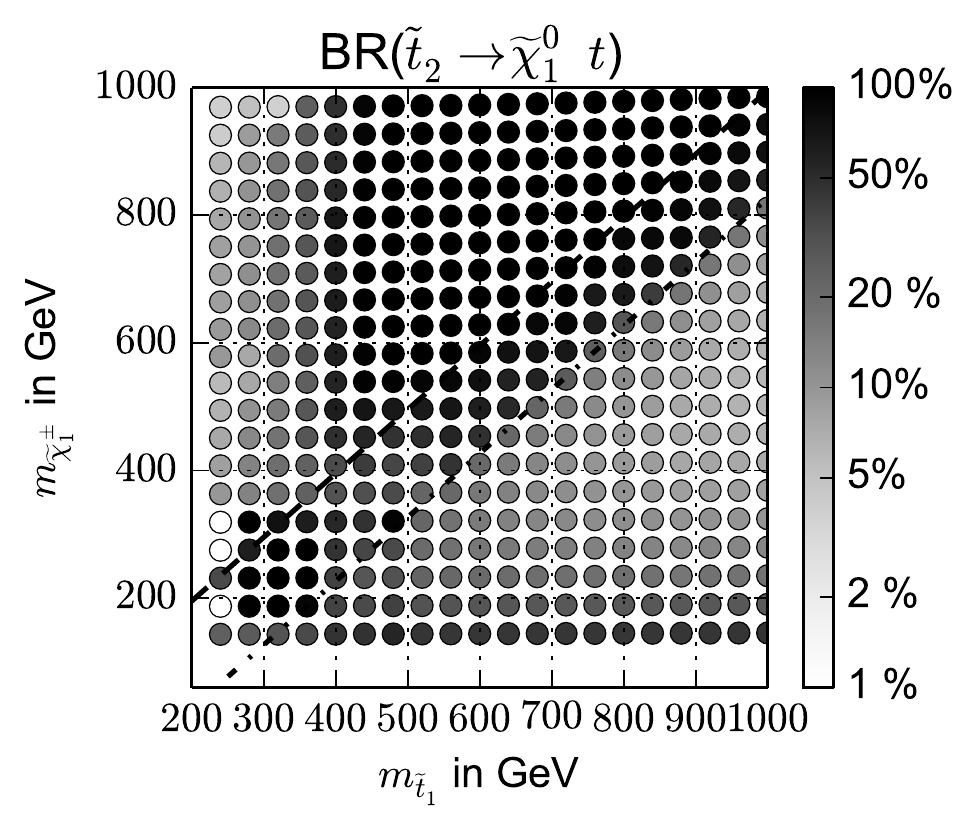}\qquad \qquad \qquad
\includegraphics[height=0.2\textheight]{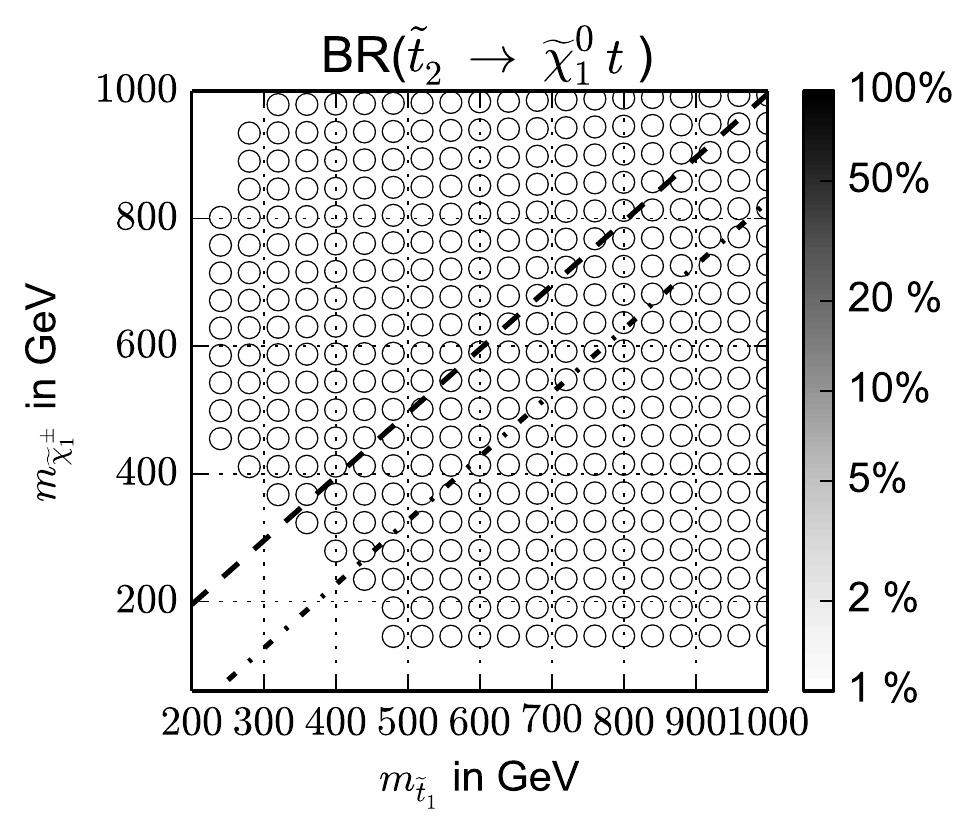}
\includegraphics[height=0.2\textheight]{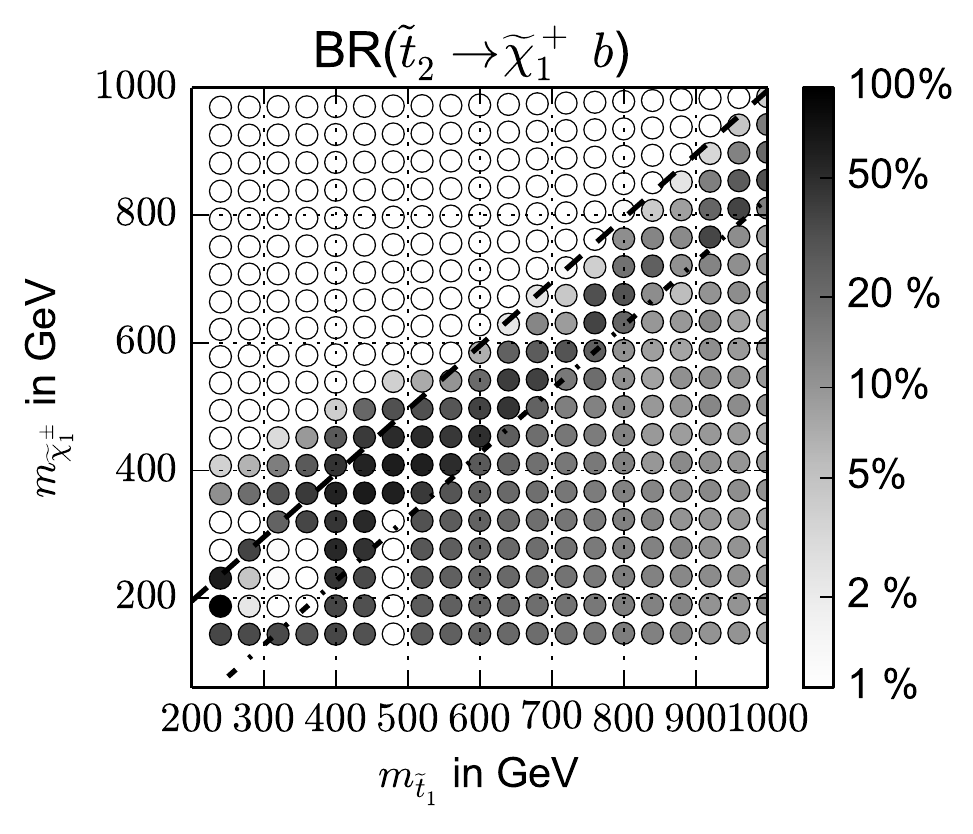}\qquad \qquad \qquad
\includegraphics[height=0.2\textheight]{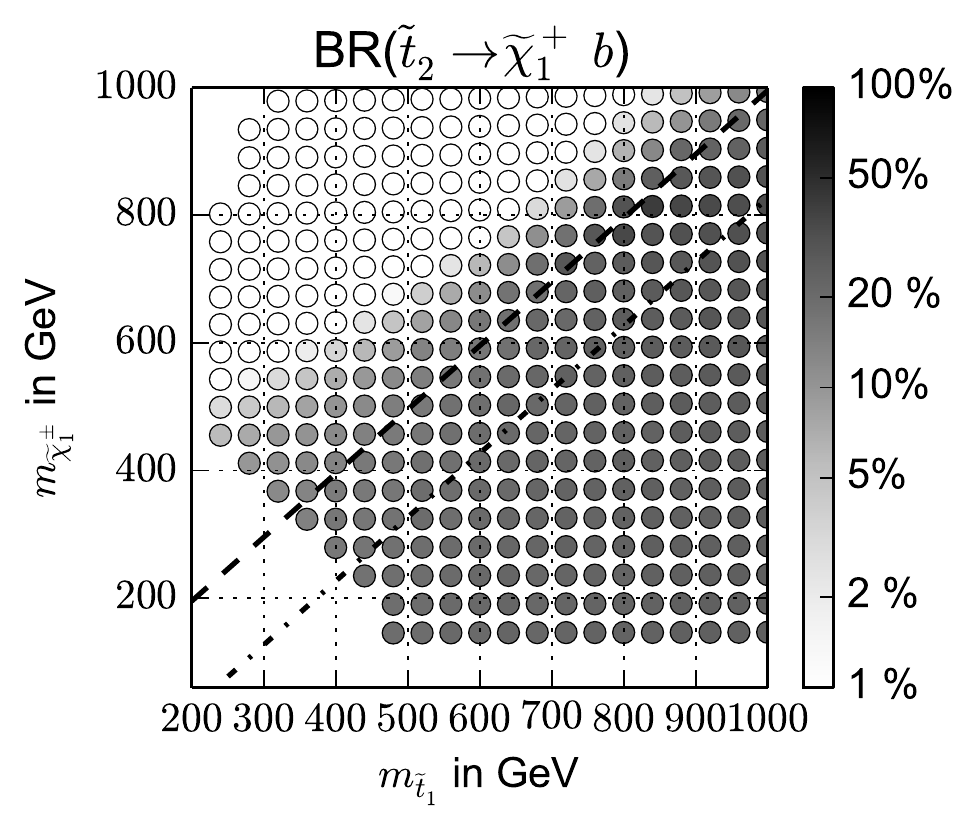}
\includegraphics[height=0.2\textheight]{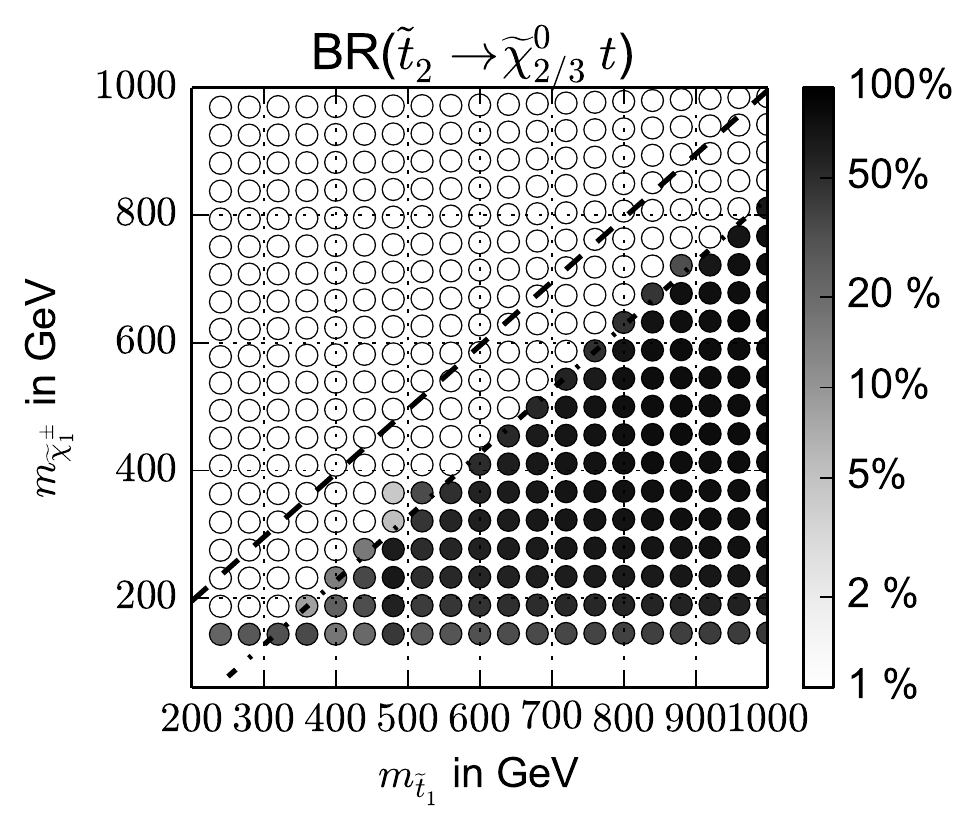}\qquad \qquad \qquad
\includegraphics[height=0.2\textheight]{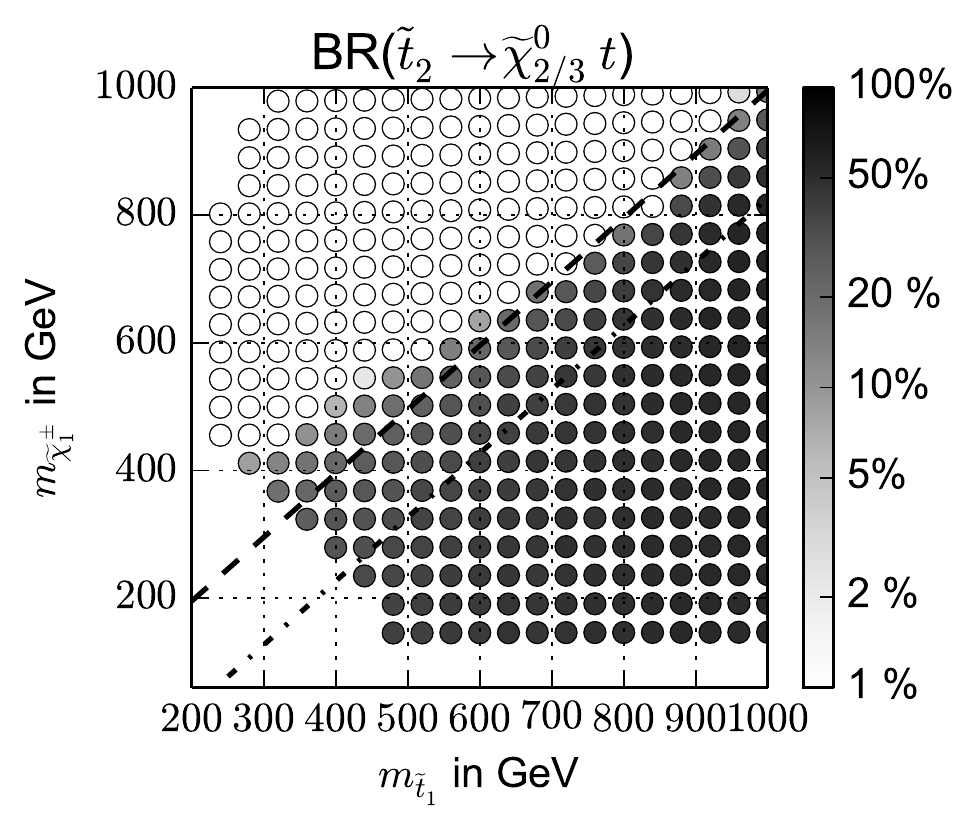}
\includegraphics[height=0.2\textheight]{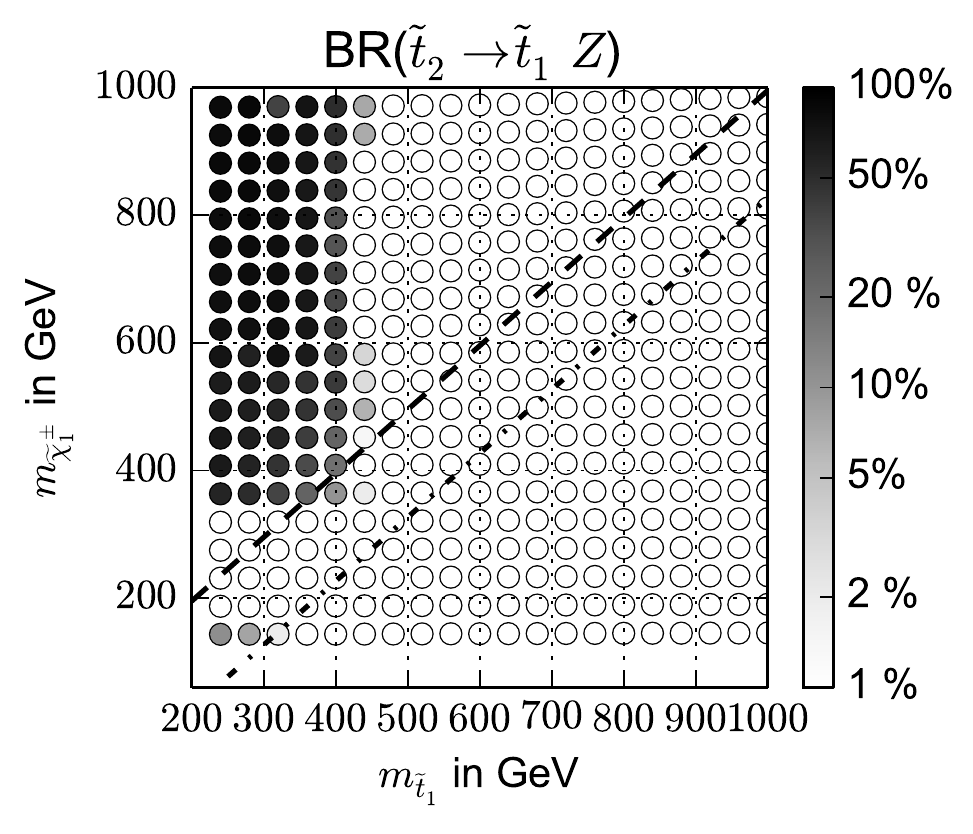}\qquad \qquad \qquad
\includegraphics[height=0.2\textheight]{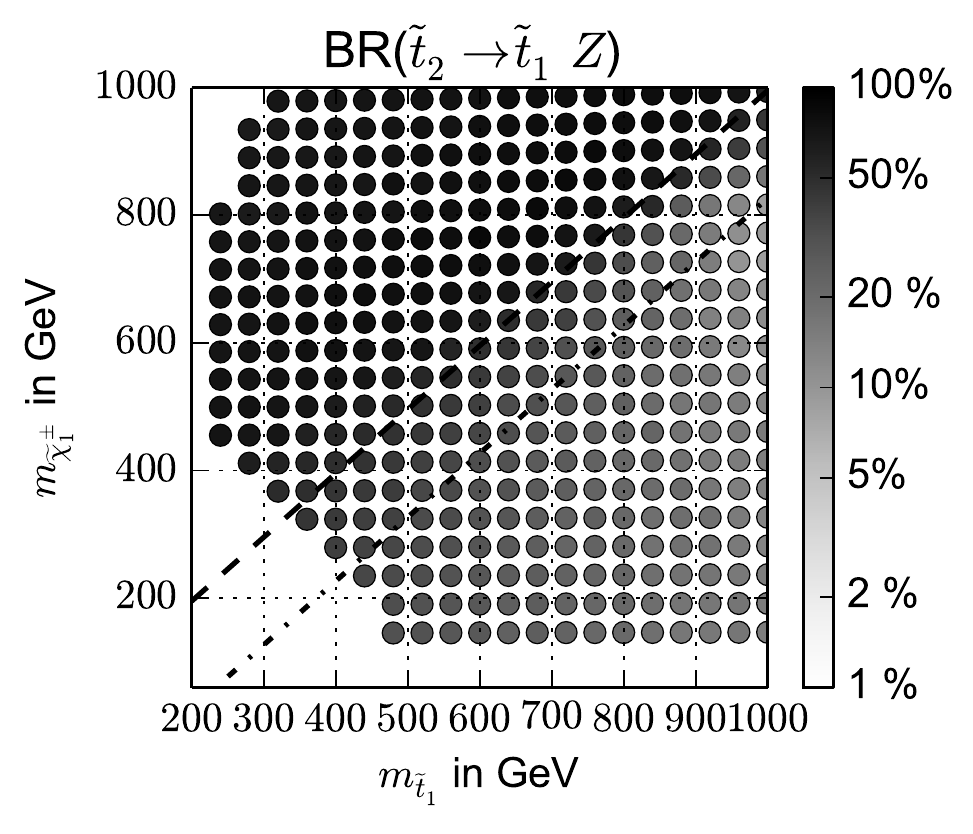}
\caption{Most significant branching ratios of the heavier stop into the the singlino LSP, the higgsino NLSPs and the chargino for a decoupled gluino and  $m_{\widetilde{\chi}^0_1} = 100$ GeV. Left: $\lambda_L$. Right: $\lambda_S$}
\vspace{-5cm}
\label{fig:stopplot22}
\end{figure}
\vfill
\pagebreak
\section{$\widetilde{\chi}^0_{2/3}$ Branching Ratio Distributions}
\begin{figure}[h!]
\includegraphics[height=0.2\textheight]{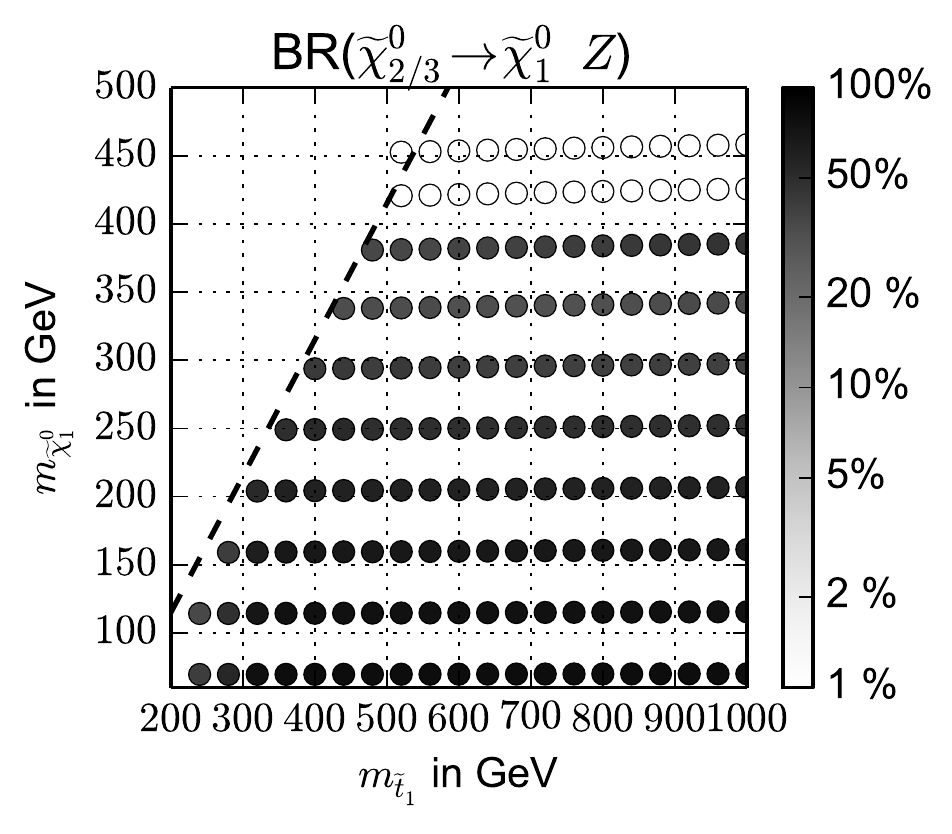}\qquad \qquad \qquad
\includegraphics[height=0.2\textheight]{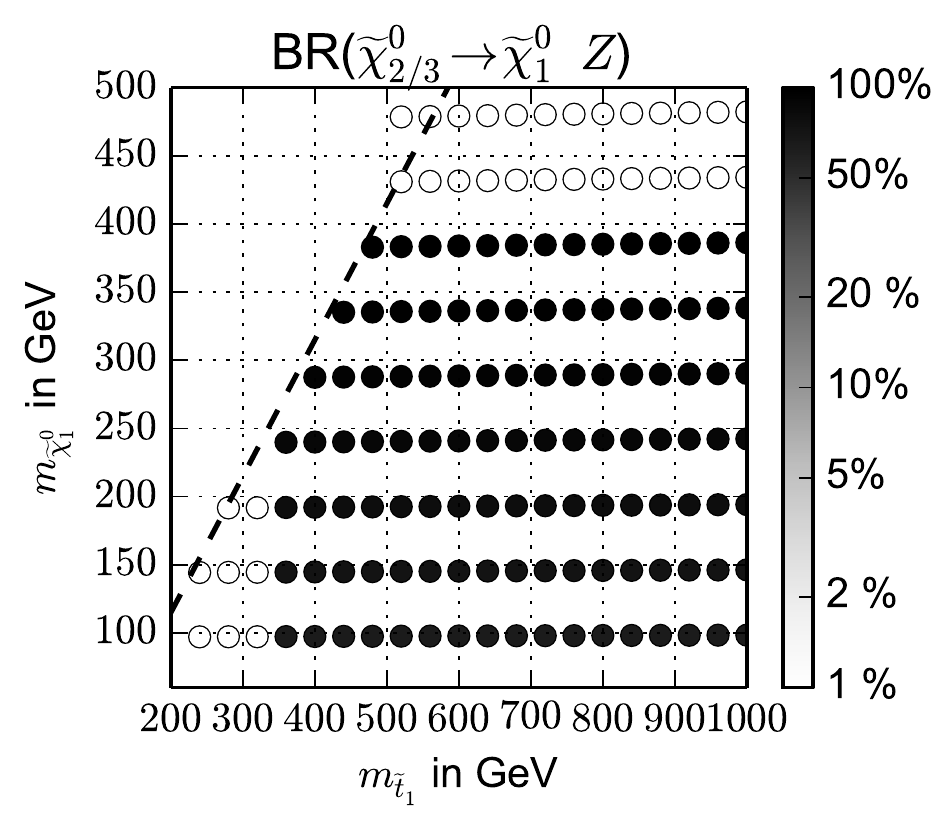}
\includegraphics[height=0.2\textheight]{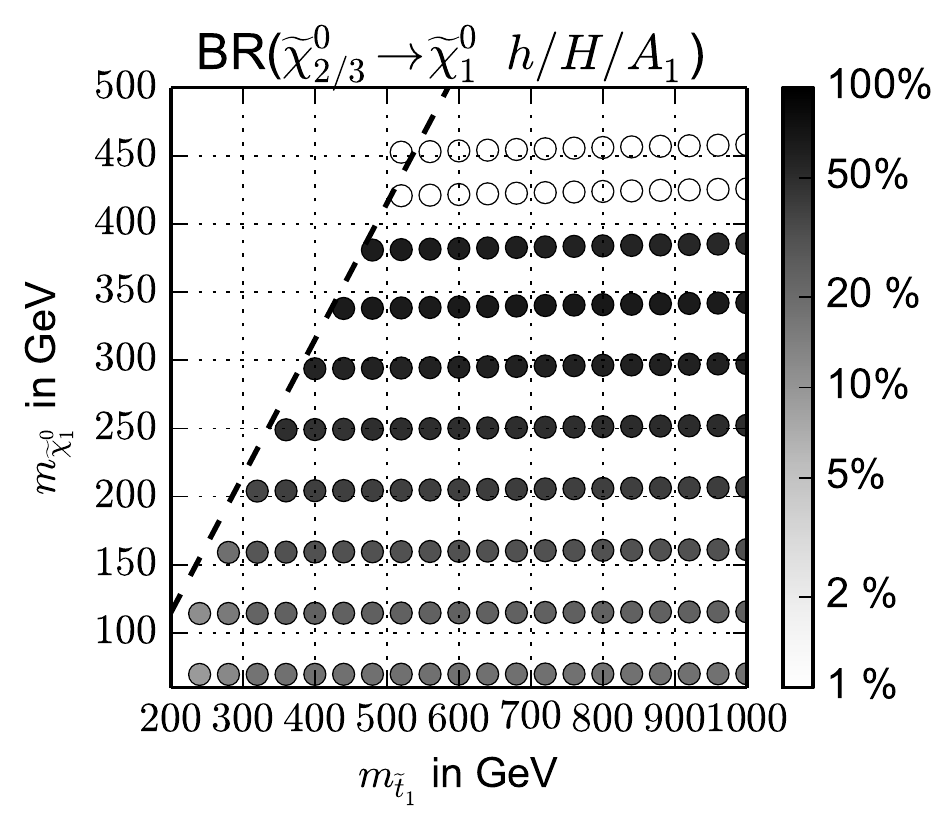}\qquad \qquad \qquad
\includegraphics[height=0.2\textheight]{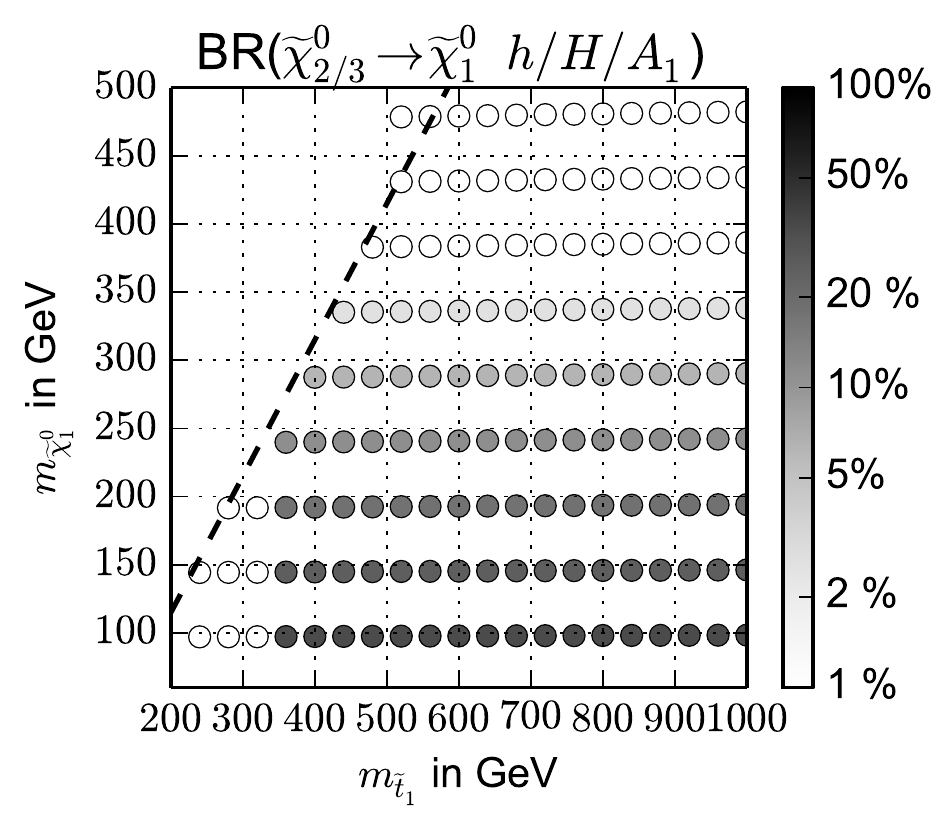}
\includegraphics[height=0.2\textheight]{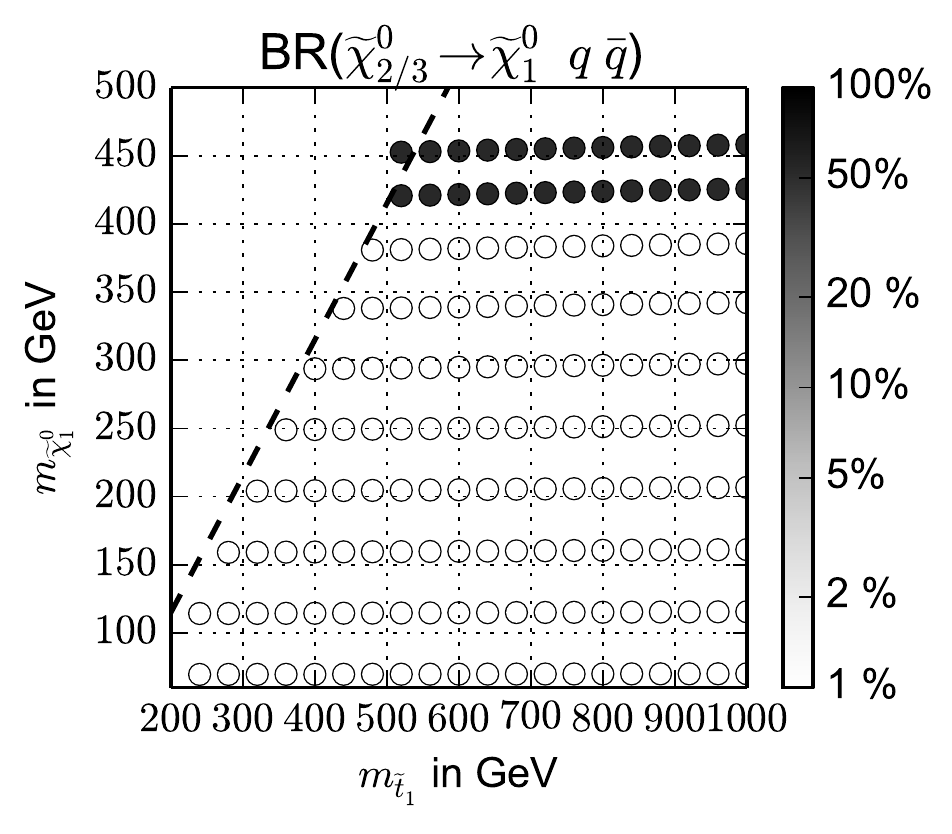}\qquad \qquad \qquad
\includegraphics[height=0.2\textheight]{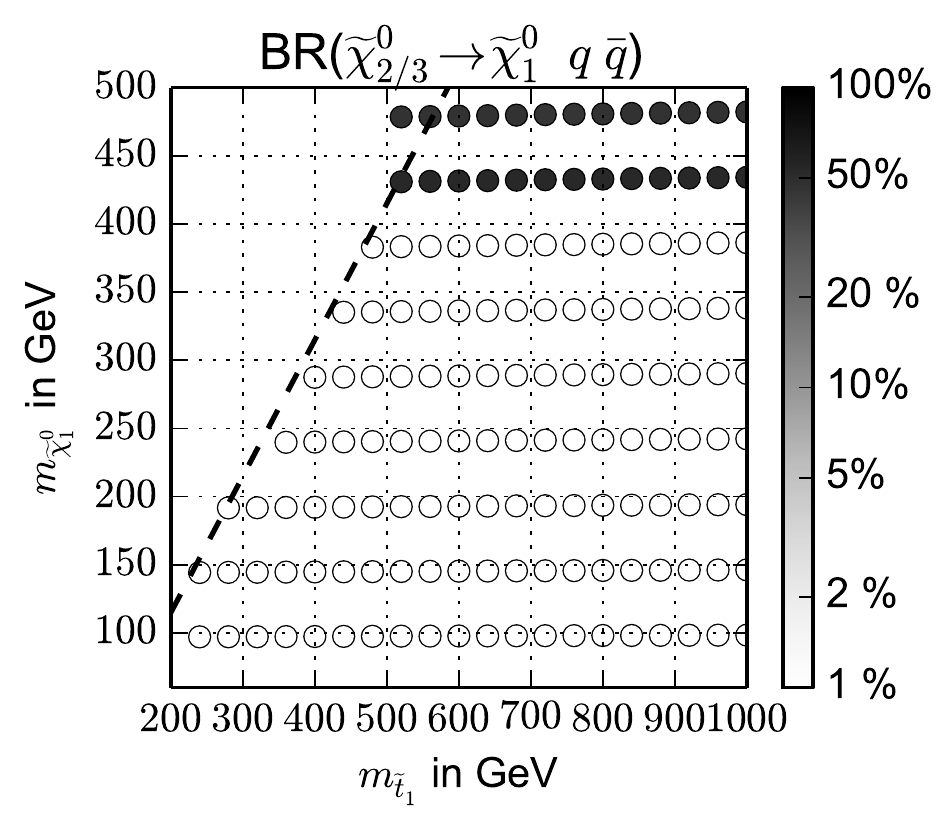}
\includegraphics[height=0.2\textheight]{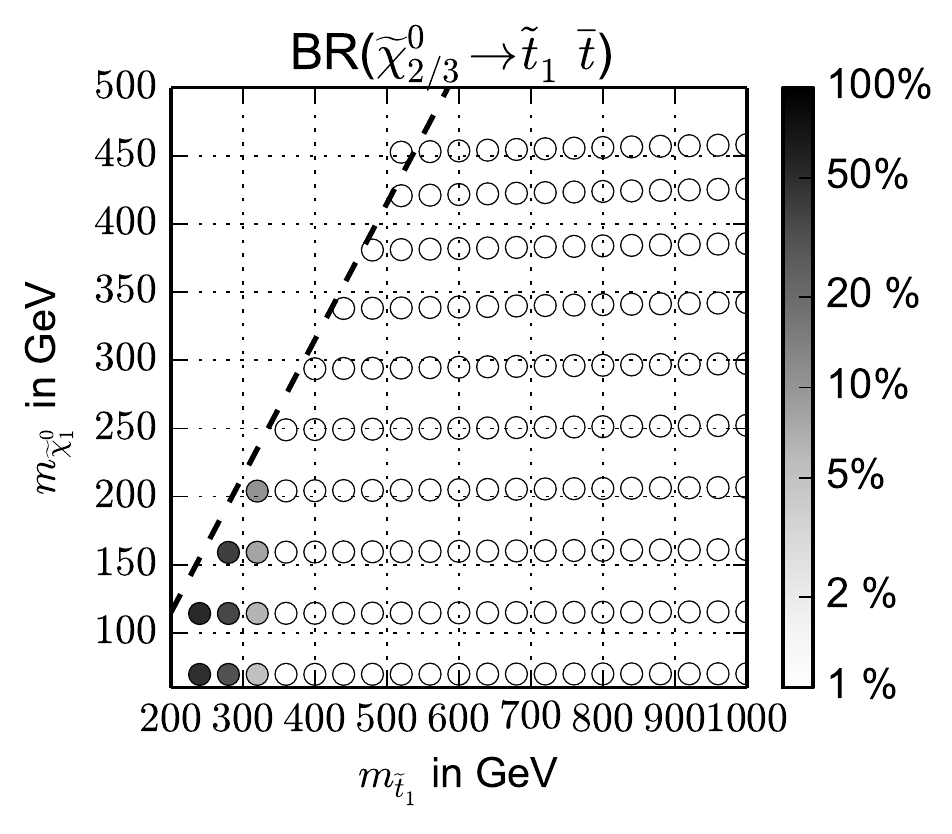}\qquad \qquad \qquad
\includegraphics[height=0.2\textheight]{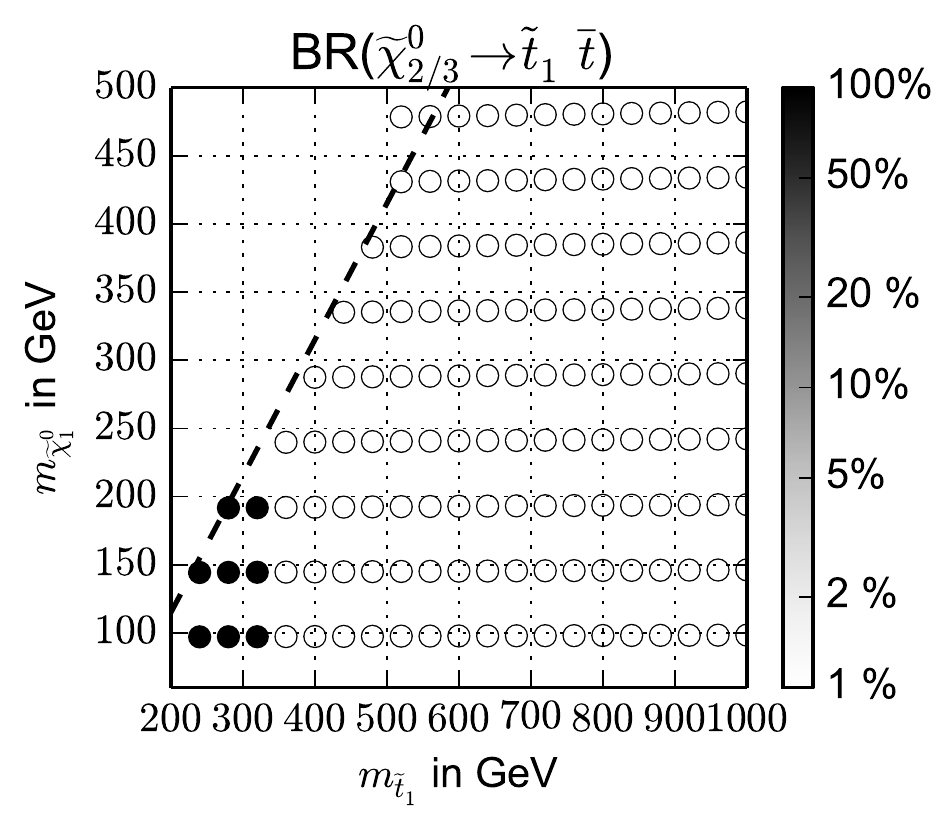}
\caption{Most significant branching ratios of the higgsino-like neutralinos for a decoupled gluino and  $m_{\widetilde{\chi}_1^\pm} = 500$ GeV. Left: $\lambda_L$. Right: $\lambda_S$.}
\label{fig:chi23branching}
\end{figure}
\vfill
\pagebreak
\begin{figure}[h!]
\includegraphics[height=0.2\textheight]{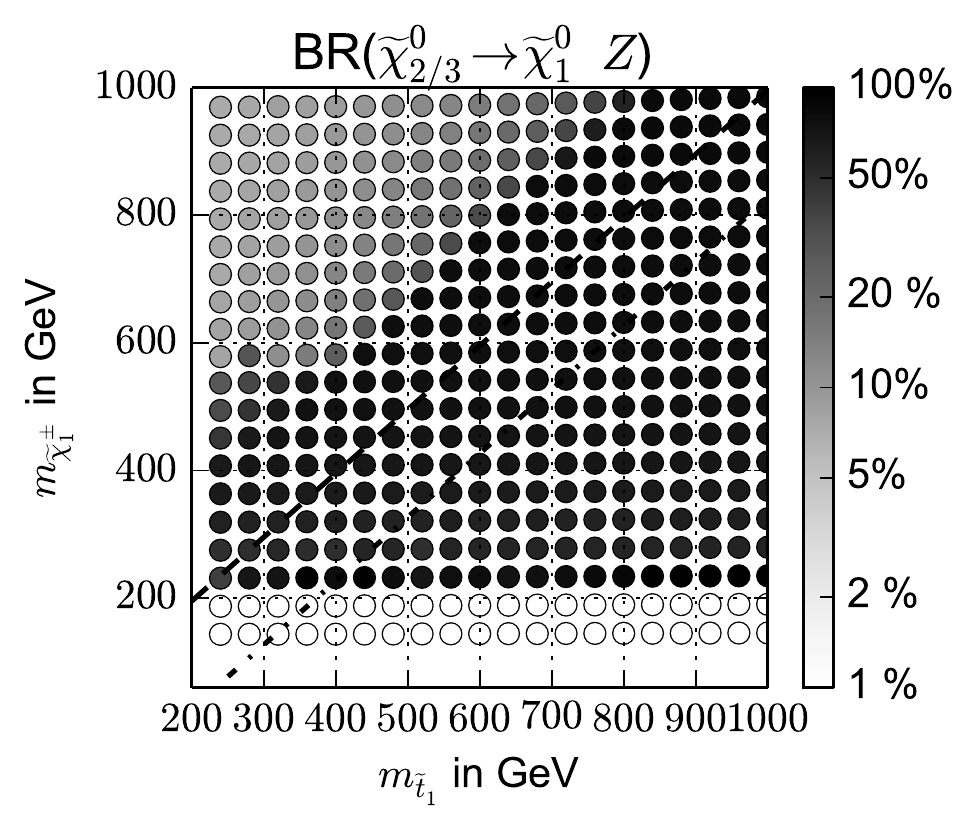}\qquad \qquad \qquad
\includegraphics[height=0.2\textheight]{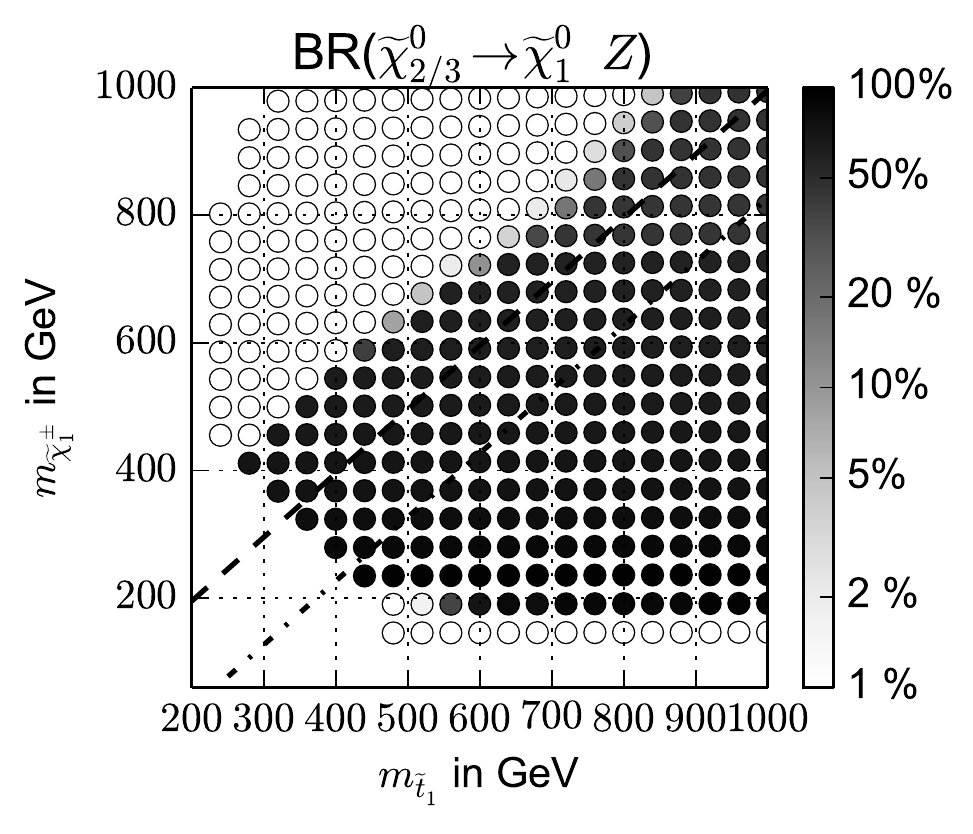}
\includegraphics[height=0.2\textheight]{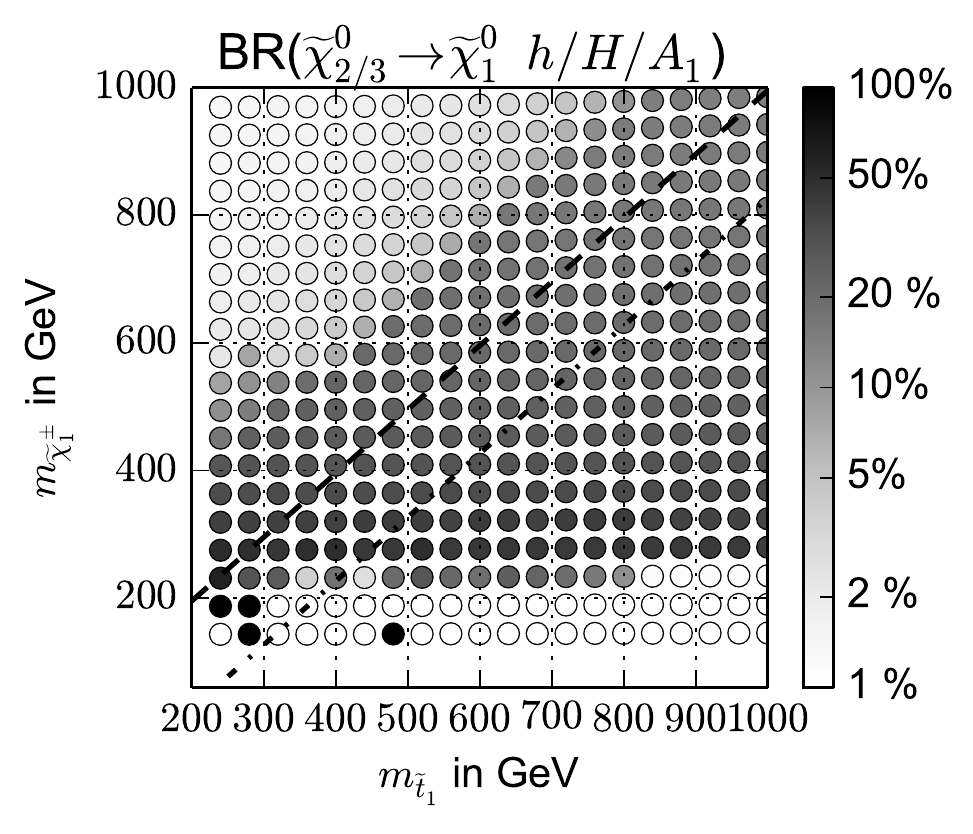}\qquad \qquad \qquad
\includegraphics[height=0.2\textheight]{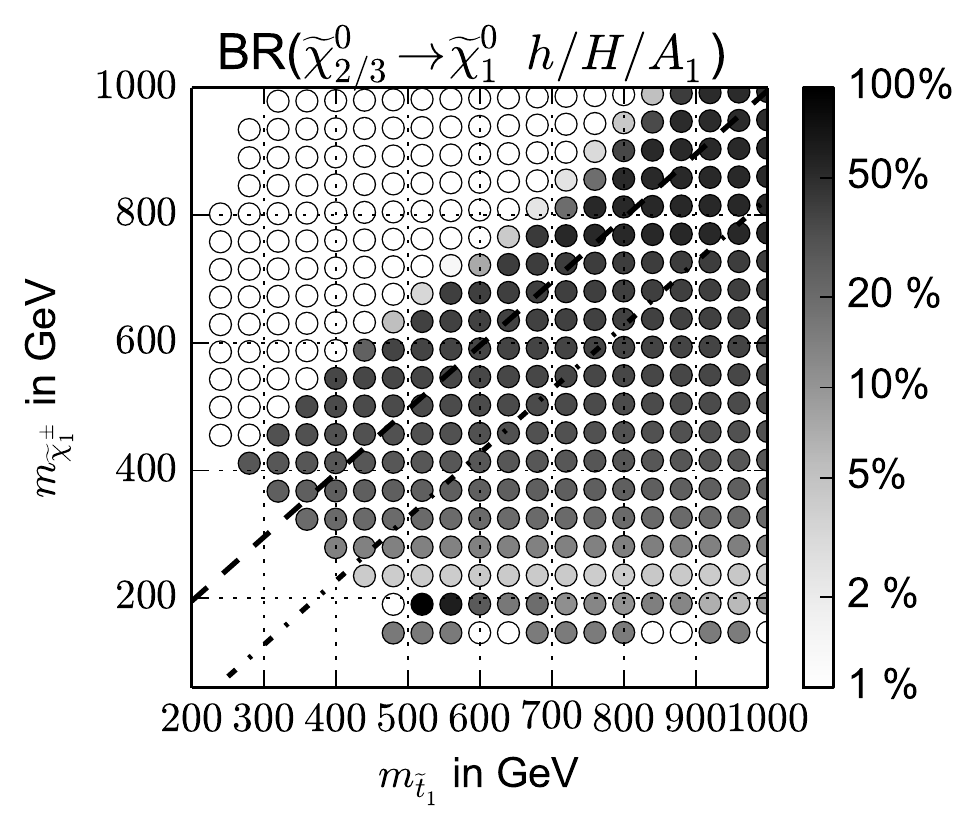}
\includegraphics[height=0.2\textheight]{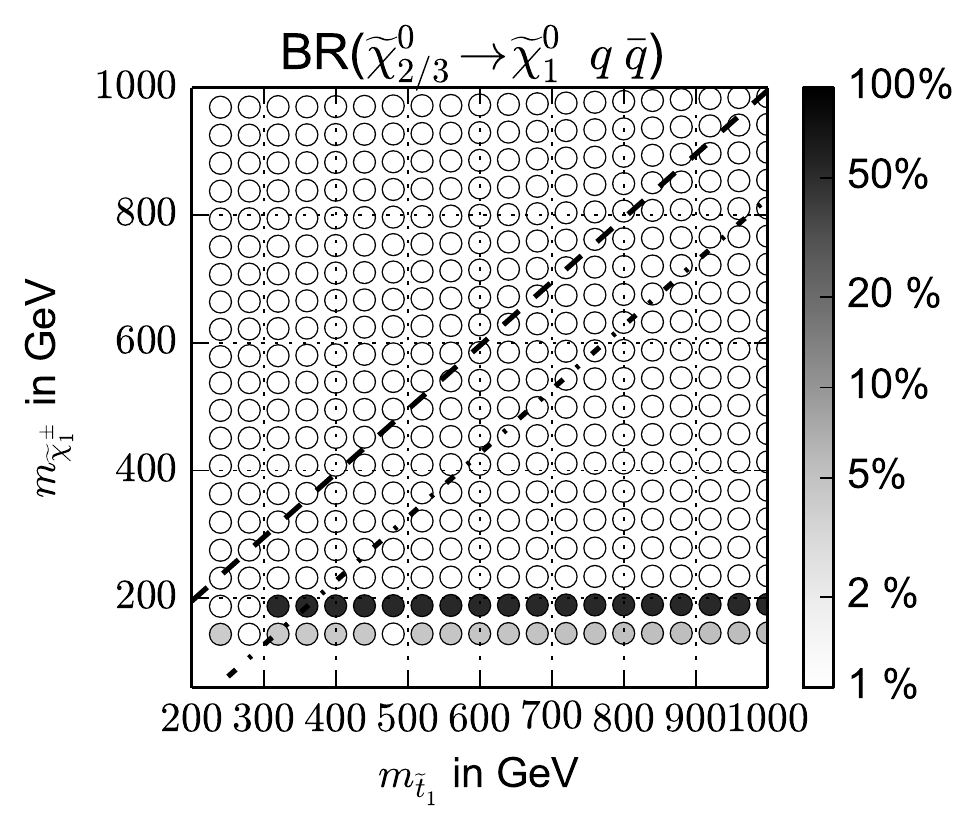}\qquad \qquad \qquad
\includegraphics[height=0.2\textheight]{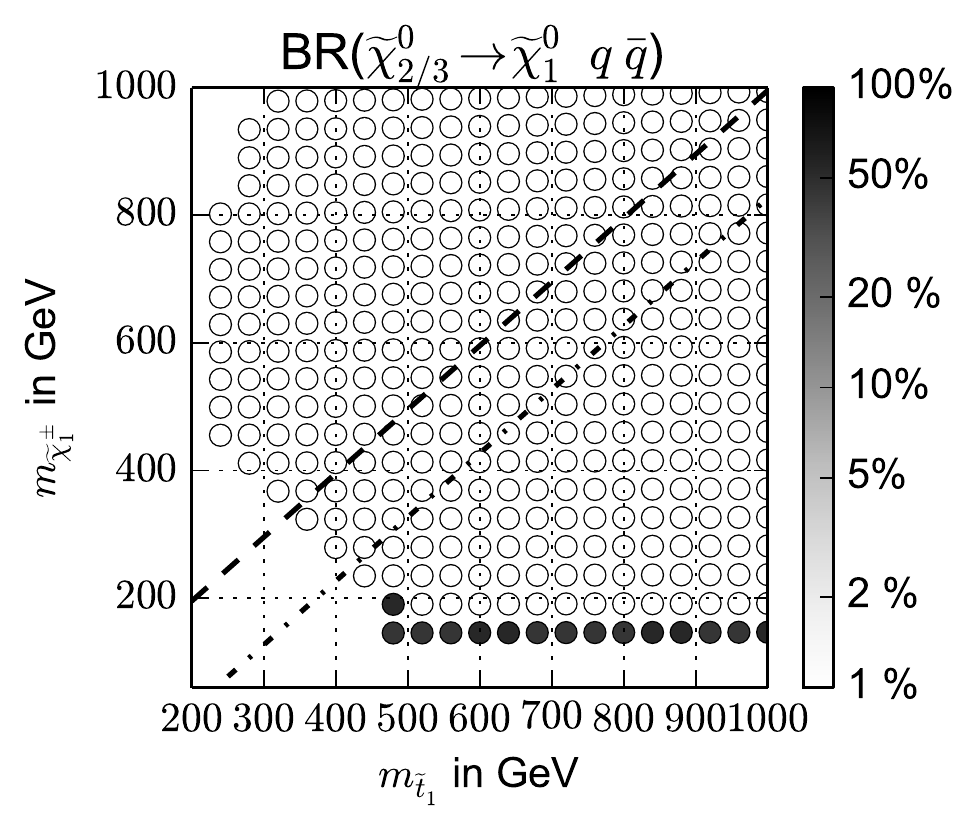}
\includegraphics[height=0.2\textheight]{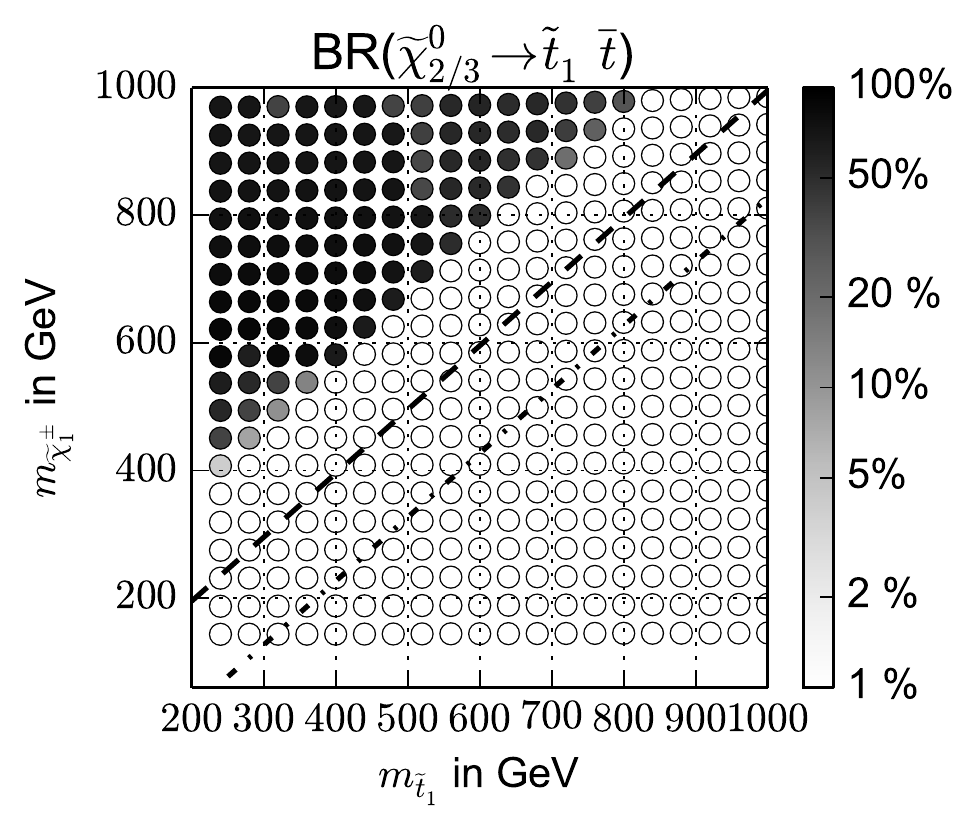}\qquad \qquad \qquad
\includegraphics[height=0.2\textheight]{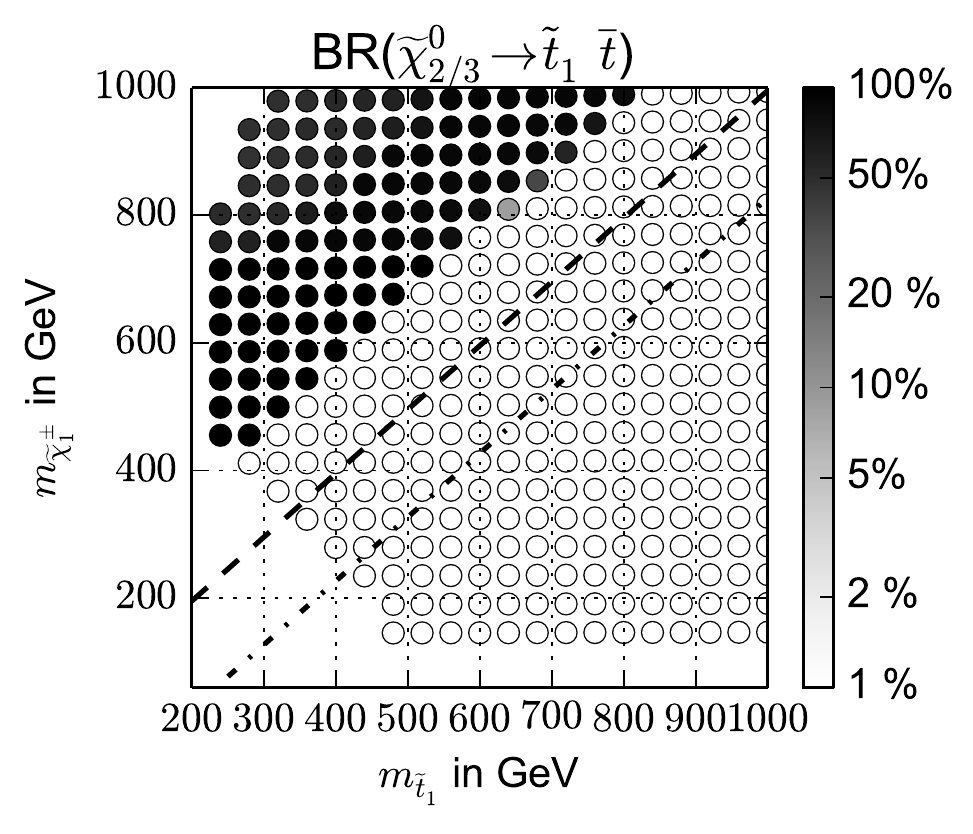}
\caption{Most significant branching ratios of the higgsino-like neutralinos for a decoupled gluino and  $m_{\widetilde{\chi}^0_1} = 100$ GeV. Left: $\lambda_L$. Right: $\lambda_S$}
\label{fig:chi23branching}
\end{figure}
\vfill
\pagebreak
\twocolumngrid
\vfill
\pagebreak
\bibliographystyle{utphys}
\bibliography{NMSSM_correct}

\end{document}